\documentstyle[amscd,12pt,amssymb,righttag]{amsart}

\newcommand{\nusquareVV}{
\unitlength=0.50mm
\makebox[18mm][l]{
\raisebox{-17mm}[20mm][19mm]{
\put(26,50){\oval(20,20)[b]}
\put(27,50){\oval(18,20)[t]}
\put(18,50){\line(0,-1){5}}
\put(18,39){\line(0,-1){8}}
\put(10,20){\oval(20,20)[t]}
\put(9,20){\oval(18,20)[b]}
\put(18,20){\line(0,1){5}}
\put(16,50){\line(0,1){20}}
\put(20,20){\line(0,-1){20}}
\put(14,0){\makebox(0,0)[cc]{$V$}}
\put(11,70){\makebox(0,0)[cc]{$V$}}
}}}

\newcommand{\dimeightV}{
\unitlength=0.50mm
\makebox[17mm][l]{
\raisebox{-10mm}[11mm][11mm]{
\put(15,30){\oval(20,20)[t]}
\put(15,10){\oval(20,20)[b]}
\put(25,30){\line(-1,-1){20}}
\put(25,10){\line(-1,1){9}}
\put(5,30){\line(1,-1){9}}
\put(31,30){\makebox(0,0)[cb]{$V$}}
\put(0,30){\makebox(0,0)[cb]{$V\pti$}}
}}}

\newcommand{\Dprime}{
\unitlength=0.75mm
\makebox[60mm][l]{
\raisebox{-13.5mm}[15mm][15mm]{
\put(0,30){\line(0,-1){30}}
\put(80,30){\line(0,-1){30}}
\put(0,34){\makebox(0,0)[cc]{$1$}}
\put(10,34){\makebox(0,0)[cc]{$k$}}
\put(30,34){\makebox(0,0)[cc]{$m$}}
\put(50,34){\makebox(0,0)[cc]{$1$}}
\put(70,34){\makebox(0,0)[cc]{$l$}}
\put(80,34){\makebox(0,0)[cc]{$n$}}
\put(50,30){\line(0,-1){30}}
\put(60,30){\line(0,-1){30}}
\put(40,15){\oval(10,10)[l]}
\put(40,14){\oval(8,8)[r]}
\put(40,20){\line(5,2){8}}
\put(70,30){\line(-3,-1){9}}
\put(58,26){\line(-4,-1){7}}
\put(10,30){\line(5,-2){26}}
\put(30,30){\line(0,-1){6}}
\put(30,20){\line(0,-1){20}}
\put(21,0){\line(0,1){24}}
\put(21,27){\line(0,1){3}}
}}}

\newcommand{\Sy}
{\text{\fontfamily{cmss}\fontseries{bx}\fontshape{n}\selectfont S}}
\newcommand{\Ex}
{\text{\fontfamily{cmss}\fontseries{bx}\fontshape{n}\selectfont\char'003}}
\newcommand{\RB}{R}
\newcommand{\di}{{\bold d}}
\newcommand{\K}{K^{\bullet\bullet}}
\newcommand{\X}{{\overset\cup{\scriptstyle\cap}}}
\newcommand{\Zp}{{\Bbb Z}_{\ge0}}

\newcommand{\C}{{\Bbb C}}

\newcommand{\Z}{{\Bbb Z}}

\newcommand{\esel}{{\frak s\frak l}}
\newcommand{\SSS}{{\frak S}}

\newcommand{\e}{\varepsilon}
\newcommand{\eps}{\varepsilon}
\newcommand{\sign}{\text{sign}}

\newcommand{\Vect}{\operatorname{-Vect}}

\newcommand{\sdim}{\operatorname{sdim}}
\newcommand{\sdet}{\operatorname{sdet}}
\newcommand{\detq}{\operatorname{det}_q}

\newcommand{\tens}{\otimes}
\newcommand{\Sym}{\operatorname{Sym}}
\newcommand{\Ant}{\operatorname{Ant}}
\newcommand{\Symshf}{\operatorname{Symshf}}
\newcommand{\Antshf}{\operatorname{Antshf}}
\newcommand{\Shuffles}{\operatorname{Shuffles}}

\newcommand{\Hom}{\operatorname{Hom}}

\newcommand{\ind}{\operatorname{ind-}}
\newcommand{\Rel}{\operatorname{Rel}}

\newcommand{\Tr}{\operatorname{Tr}}
\newcommand{\im}{\operatorname{Im}}
\newcommand{\Ker}{\operatorname{Ker}}

\newcommand{\id}{\operatorname{id}}
\newcommand{\op}{{\mathrm o\mathrm p}}
\newcommand{\tl}{\triangleleft}

\newcommand{\CC}{{\cal C}}
\newcommand{\CV}{{\cal V}}
\newcommand{\CO}{{\cal O}}
\newcommand{\CM}{{\cal M}}
\newcommand{\CR}{{\cal R}}
\newcommand{\CH}{{\cal H}}
\newcommand{\cd}{{\cal D}}
\newcommand{\DY}{{\cal D\cal Y}}

\newcommand{\be}{\begin{equation}}
\newcommand{\gr}{\Z/2\times\Z}
\newcommand{\grp}{\Z/2\times\Z_{\ge0}}
\newcommand{\md}[2]{{d^{#1}}_{#2}}
\newcommand{\mt}[2]{{t^{#1}}_{#2}}
\newcommand{\mc}[3]{{{t_{#1}}^{#2}}_{#3}}
\newcommand{\ma}[3]{{{#1}^{#2}}_{#3}}
\newcommand{\Tbar}[2]{\bar T_{#1}{@!}^{#2}}
\newcommand{\tbar}[2]{\bar t_{#1}{@!}^{#2}}
\newcommand{\tpt}[2]{t_{#1}{@!}^{#2}}
\newcommand{\tpri}[2]{t'_{#1}{@!}^{#2}}
\newcommand{\Comat}{\operatorname{Comat}}
\newcommand{\Ob}{\operatorname{Ob}}
\newcommand{\Mor}{\operatorname{Mor}}
\newcommand{\Ann}{\operatorname{Ann}}
\newcommand{\ev}{\operatorname{ev}}
\newcommand{\lev}{\operatorname{lev}}
\newcommand{\coev}{\operatorname{coev}}
\newcommand{\degree}{\operatorname{deg}}
\newcommand{\End}{\operatorname{End}}
\newcommand{\Ber}{\operatorname{Ber}}

\renewcommand{\le}{\leqslant}
\renewcommand{\ge}{\geqslant}

\newcommand{\mtwo}{_{(-2)}}
\newcommand{\mone}{_{(-1)}}
\newcommand{\nul}{_{(0)}}
\newcommand{\one}{_{(1)}}
\newcommand{\two}{_{(2)}}
\newcommand{\tre}{_{(3)}}
\newcommand{\four}{_{(4)}}

\newcommand{\qqquad}{\qquad\quad}
\newcommand{\nquad}{{\!\!\!\!\!\!}}
\newcommand{\nqquad}{\nquad\nquad}
\newcommand{\pti}{{}\spcheck}

\newcommand{\<}{\langle}
\renewcommand{\>}{\rangle}

\newcommand{\aow}{\hbox to 20pt {\rightarrowfill}}
\newcommand{\arow}{\hbox to 30pt {\rightarrowfill}}
\newcommand{\arrow}{\hbox to 45pt {\rightarrowfill}}
\newcommand{\arrrow}{\hbox to 60pt {\rightarrowfill}}
\newcommand{\arrrrow}{\hbox to 100pt {\rightarrowfill}}
\newcommand{\lfaow}{\hbox to 20pt {\leftarrowfill}}
\newcommand{\lfarow}{\hbox to 30pt {\leftarrowfill}}
\newcommand{\lfarrow}{\hbox to 45pt {\leftarrowfill}}
\newcommand{\lfarrrow}{\hbox to 60pt {\leftarrowfill}}
\newcommand{\lfarrrrow}{\hbox to 100pt {\leftarrowfill}}

\newtheorem{thm}{Theorem}[section]
\newtheorem{cor}[thm]{Corollary}

\newtheorem{prop}[thm]{Proposition}

\theoremstyle{definition}
\newtheorem{defn}{Definition}[section]
\newtheorem{example}[defn]{Example}
\newtheorem{examples}[defn]{Examples}

\theoremstyle{remark}
\newtheorem{rem}{Remark}[section]
\newtheorem{conjecture}{Conjecture}

\numberwithin{equation}{section}

\newcommand{\thmref}[1]{Theorem~\ref{#1}}
\newcommand{\secref}[1]{Section~\ref{#1}}

\newcommand{\propref}[1]{Proposition~\ref{#1}}

\begin{document}

\title [Differential forms, Koszul complexes and Berezinians]
{Quantum supergroups of $GL(n|m)$ type:
Differential forms, Koszul complexes and Berezinians}


\author[V. Lyubashenko]{Volodymyr Lyubashenko}
\address{{\it Current address\/}: (V.L.) Department of Mathematics\\
Kansas State University\\ 137 Cardwell Hall\\
Manhattan, Kansas 66506--2602\\ U.S.A.}
\email{lub@@math.ksu.edu}
\thanks{ Research was supported by EPSRC research grant GR/G 42976.}

\author[A. Sudbery]{Anthony Sudbery}
\address
{(A.S.) Department of Mathematics\\ University of York\\
Heslington, York, YO1 5DD, U.K.}
\email{as2@@york.ac.uk}


\date {26 May 1995 : \today }

\begin{abstract}
We introduce and study the Koszul complex for a Hecke $R$-matrix.
Its cohomologies, called the Berezinian, are used to define quantum
superdeterminant for a Hecke $R$-matrix. Their behaviour with
respect to Hecke sum of $R$-matrices is studied.
Given a Hecke $R$-matrix in
$n$-dimensional vector space, we construct a Hecke $R$-matrix in
$2n$-dimensional vector space commuting with a differential.
The notion of a quantum differential supergroup is derived. Its
algebra of functions is a differential coquasitriangular Hopf
algebra, having the usual algebra of differential forms as a
quotient. Examples of superdeterminants related to these algebras
are calculated. Several remarks about Woronowicz's theory are made.
\end{abstract}

\maketitle

\subsection{Short description of the paper}
\subsubsection{}
In this paper we will be concerned with differential Hopf algebras generated
by sets of matrix elements. We start (\secref{con-recon}) by giving a
construction of such algebras, generalising the construction \cite{Take,
Sud:matel} of a bialgebra generated by a single set of matrix elements
(without differential) as a universally coacting bialgebra preserving
several algebras generated by a set of coordinates. In our generalisation
the data are morphisms in the category of graded differential complexes.

\subsubsection{}
Given a Hecke $R$-matrix for a vector space $V$, we construct in this
paper another Hecke $R$-matrix $\CR$ for the space $W=V\oplus V$
equipped with the differential
$d=\bigl( \begin{smallmatrix} 0&1 \\ 0&0 \end{smallmatrix} \bigr)$
and the grading $\sigma:W \to W$,
$\sigma=\bigl( \begin{smallmatrix} 1&0 \\ 0&-1 \end{smallmatrix} \bigr)$.
The matrix $\CR$ is distinguished by the property
\[ \CR (d\tens1 + \sigma\tens d) = (d\tens1 + \sigma\tens d) \CR .\]

\subsubsection{}
The algebra $H$ of functions on the quantum supergroup constructed from
$\CR$ is a $\Z$-graded differential coquasitriangular Hopf algebra
(\secref{sec2}). In brief, it defines a differential quantum supergroup.
A quotient of $H$ is the $\Zp$-graded differential Hopf algebra $\Omega$
of differential forms defined via $R$ in
\cite{SchWatZum,Sud:supcal,Tsy:dT,Zum:calcul}.

The classical version ($q=1$) of this construction is: take a vector
space $ V$, add to it another copy of it with the opposite parity and
consider the general linear supergroup of the $\Z/2$-graded space so obtained.

\subsubsection{}
We introduce Koszul complexes for Hecke $R$-matrices in \secref{Koszul}.
They are $\Zp\times\Zp$-graded algebras with two differentials, $D$ of
degree $(1,1)$ and $D'$ of degree $(-1,-1)$. We calculate their
anticommutator, called the Laplacian. The cohomology space of $D$ is
called the Berezinian. It generalizes the determinant, coinciding in the
even case with the highest exterior power of $V$. The behaviour of Koszul
complexes and Berezinians with respect to the Hecke sum~\cite{MaMa}
is described: we prove that the Berezinian of a Hecke sum is the tensor
product of Berezinians.

\subsubsection{}
The Berezinian is used to define the quantum superdeterminant in
~\secref{CalBer}. We calculate the superdeterminant in several examples.
In particular, in the algebra $\Omega$ of differential forms on the
standard quantum $GL(n|m)$ the superdeterminant equals 1. This confirms
the idea that there are no central group-like elements in $\Omega$,
explaining why it has not been possible to construct a differential
calculus on special linear groups with the same dimension as the
classical case.

\subsubsection{}
We make several remarks on Woronowicz's theory \cite{Wor:calcul} in \secref{bimodules}.
In particular, each first order differential calculus is extended to
a differential Hopf algebra.

\subsection{Notations and conventions}
$k$ denotes a field of characteristic 0. In this paper a {\em Hopf
algebra} means a $k$-bialgebra with an invertible antipode.
Associative comultiplication is denoted by $\Delta x= x\one\tens x\two$,
the counit by $\e$, the antipode in a Hopf algebra by $\gamma$.
If $H$ is a Hopf algebra, $H^\op$ denotes the same coalgebra $H$ with
opposite multiplication, $H_\op$ denotes the same algebra $H$ with the
opposite comultiplication.

When $X$ is a graded vector space, $\hat a$ denotes the degree of a
homogeneous element $x_a \in X$.

The braiding in a braided tensor category $\CC$ \cite{JoyStr:tor} (e.g.
in
the category of representations of a quasitriangular Hopf algebra) is
denoted
by $c_{X,Y}:X\tens Y \to Y\tens X$, where $X,Y\in \Ob\CC$. By definition,
the maps $1_{V^{\tens k-1}} \tens c_{V,V} \tens 1_{V^{\tens n-k-1}} :
V^{\tens n} \to V^{\tens n}$ obey the braid group relations.
Given $\sigma\in \SSS_n$, we denote by $(c)_\sigma$ and $(c^{-1})_\sigma$
the maps $V^{\tens n} \to V^{\tens n}$ coming from the liftings of
$\sigma$ to the elements of the braid group representing reduced
expressions of $\sigma$. The first case is a word in the generators
$1\tens c\tens1$, the second is a word in $1\tens c^{-1}\tens1$.

We often use tangles to describe maps constructed from an $R$-matrix and
pairings. In the conventions of \cite{Lyu:tan} we denote
{\allowdisplaybreaks
\begin{alignat*}2
&c:X\tens Y \to Y\tens X &&\qquad \text{by} \qquad
\unitlength=1mm
\makebox[20mm][l]{
\raisebox{-7mm}[14mm][8mm]{
\put(5,0){\line(2,3){10}}
\put(5,15){\line(2,-3){4}}
\put(15,0){\line(-2,3){4}}
\put(5,17){\makebox(0,0)[cb]{$X$}}
\put(15,17){\makebox(0,0)[cb]{$Y$}}
}} ,\\
&c^{-1}:X\tens Y \to Y\tens X &&\qquad \text{by} \qquad
\unitlength=1mm
\makebox[20mm][l]{
\raisebox{-7mm}[14mm][8mm]{
\put(5,17){\makebox(0,0)[cb]{$X$}}
\put(15,17){\makebox(0,0)[cb]{$Y$}}
\put(5,15){\line(2,-3){10}}
\put(15,15){\line(-2,-3){4}}
\put(5,0){\line(2,3){4}}
}} ,\\
&\ev:X\tens X\pti \to k &&\qquad \text{by} \qquad
\unitlength=1mm
\makebox[20mm][l]{
\raisebox{-5mm}[7mm][7mm]{
\put(10,7){\oval(10,14)[b]}
\put(5,9){\makebox(0,0)[cb]{$X$}}
\put(15,9){\makebox(0,0)[cb]{$X\pti$}}
}} ,\\
&\coev:k \to X\pti\tens X &&\qquad \text{by} \qquad
\unitlength=1mm
\makebox[20mm][l]{
\raisebox{-5mm}[8mm][7mm]{
\put(10,6){\oval(10,14)[t]}
\put(5,1){\makebox(0,0)[cb]{$X\pti$}}
\put(15,1){\makebox(0,0)[cb]{$X$}}
}} .
\end{alignat*}
}

\subsection{Preliminaries}
We recall some definitions from \cite{Lyu:sym} and some results from
\cite{Lyu:dis}.

\begin{defn}
Let $T:A\tens B \to B\tens A$,
$a_i\tens b_j \mapsto T_{ij}^{kl} b_k\tens a_l$, be a linear map, written
in terms of bases $(a_i)$, $(b_j)$ of the finite-dimensional vector spaces
$A$, $B$.
Let $A^*$, $B^*$ be the spaces of linear functionals on $A, B$ with the
dual
bases $(a^i)$, $(b^j)$. Define linear maps $T^\sharp$, $T^\flat$ as
\begin{alignat*}5
&T^\sharp:B\tens A^* &&\to A^*\tens B, &&\qquad b_i\tens a^j &&\mapsto
T_{\ ij}^{\sharp kl}a^k\tens b_l \: &\overset{\text{def}}=\: &
 T_{ki}^{lj}a^k\tens b_l ,\\
&T^\flat:B^*\tens A &&\to A\tens B^*, &&\qquad b^i\tens a_j &&\mapsto
T_{\ ij}^{\flat kl}a_k\tens b^l \: &\overset{\text{def}}=\: &
 T_{jl}^{ik}a_k\tens b^l .
\end{alignat*}
These maps are characterised by the properties
\begin{equation} \label{evsharp}
(\ev\otimes 1)(1\otimes T^{\sharp}) = (1\otimes \ev)(T\otimes 1): A\otimes
B\otimes A^* \to B
\end{equation}
\begin{equation}
(1\otimes \ev)(T^\flat\otimes 1) = (\ev\otimes 1)(1\otimes T): B^*\otimes A\otimes
B \to A,
\end{equation}
where $\ev :A\otimes A^*\to k$ and $\ev :B^*\otimes B\to k$ are the usual
pairings given by evaluation.

The operations $\sharp$ and $\flat$ are inverse to each other, and $T^{\sharp
\sharp \sharp \sharp}=T$.
\end{defn}
Note that if $T^{\sharp}$ is invertible, $1 \otimes T$ can be cancelled
from an expression like $(U \otimes 1)(1 \otimes T)(S \otimes 1)$ where
$S: A\otimes C \to C\otimes A$ and $U:C\otimes B \to D$. To be precise,
suppose
\begin{eqnarray}
(U\otimes 1)(1\otimes T)(S\otimes 1) =&W:&A\otimes C\otimes B\to D\otimes
A \\  && a_i \otimes c_j \otimes b_k \mapsto w_{ijk}^{lm}d_l \otimes a_n,
\end{eqnarray}
where $(c_i)$ and $(d_i)$ are bases of $C$ and $D$, and define
\[
W^{\sharp}:C\otimes B\otimes A^* \to A^* \otimes D \quad \text{by} \quad
b_j \otimes c_k \otimes a^m \mapsto w_{ijk}^{lm}a^i \otimes d_l.
\]
Then

\begin{equation}\label{cancel}
(1 \otimes U)(S^{\sharp}\otimes1)=W^{\sharp}(1 \otimes T^{\sharp -1})\,:\,
C\otimes A^*\otimes B\to A^*\otimes B\otimes C.
\end{equation}

\begin{defn}
Let $\CC$ be an abelian monoidal category with an exact monoidal functor
to $k$-vect, e.g. $\CC=H$-mod for some Hopf algebra $H$. Let $X\in\Ob\CC$.
An object $X\pti\in\CC$ with a non-degenerate pairing and co-pairing
\[ \ev: X\tens X\pti \to k, \qqquad \coev: k\to X\pti\tens X \]
is called a {\em right dual} of $X$ if $\ev,\coev\in\Mor\CC$ are
compatible in the standard sense (see e.g. \cite{JoyStr:tor}).
Similarly, a {\em left dual} is an object $\pti X\in\CC$ with
\[ \ev: \pti X\tens X \to k, \qqquad \coev: k\to X\tens\pti X. \]
When $X$ is an $H$-module, $X\pti$ and $\pti X$ are two different
$H$-module structures on the space of linear functionals $X^*$. The
category is called {\em rigid} if each object has left and right duals.
\end{defn}

The following theorem was proven in the symmetric monoidal case in
~\cite{Lyu:sym}.

\begin{thm}[\cite{Lyu:dis}] \label{sharpsflats}
Let $\CC$ be a rigid braided abelian monoidal category with an exact
monoidal functor to $k$-vect, e.g. $\CC=H$-mod for some quasitriangular
Hopf algebra $H$. Denote the braiding for some $A,B\in\CC$ by
\[ R = c_{A,B} : A\tens B \to B\tens A .\]
Then the following braiding isomorphisms are uniquely determined by $R$:
\begin{alignat*}3
c_{A,B\pti} &= (R^{-1})^\sharp &:& A\tens B\pti &&\to B\pti\tens A ,\\
c_{A\pti,B} &= (R^\sharp)^{-1} &:& A\pti\tens B &&\to B\tens A\pti ,\\
c_{A\pti,B\pti} &= R^{\sharp\sharp} &:&
A\pti\tens B\pti &&\to B\pti\tens A\pti ,\\
c_{A,\pti B} &= (R^\flat)^{-1} &:& A\tens\pti B &&\to\pti B\tens A ,\\
c_{\pti A,B} &= (R^{-1})^\flat &:& \pti A\tens B &&\to B\tens\pti A ,\\
c_{\pti A,\pti B} &= R^{\flat\flat} &:&
\pti A\tens\pti B &&\to\pti B\tens\pti A .
\end{alignat*}
\end{thm}

\begin{thm}[\cite{Lyu:dis}] \label{invertible}
If $\RB$ is an $\check R$-matrix, i.e. a map $\RB:V\tens V\to V\tens V$
for finite-dimensional $V$ satisfying the Yang-Baxter relation (braid
relation)
\begin{equation}\label{eq:braid}
 (\RB\tens1)(1\tens\RB)(\RB\tens1) = (1\tens\RB)(\RB\tens1)(1\tens\RB):
V\tens V\tens V \to V\tens V\tens V ,
\end{equation}
with invertible $\RB$ and $\RB^\sharp$, then $\RB^{-1\sharp}$,
$\RB^{\sharp\sharp}$, $\RB^\flat$, $\RB^{-1\flat}$, $\RB^{\flat\flat}$
are also invertible.
\end{thm}

\begin{pf} We regard $V^*$ as a right dual of $V$, with the pairing
ev: $V\tens V^* \to k$ given by the usual evaluation, which we also write
as $\ev(x \tens \xi) = \langle x, \xi \rangle$ for $x \in V, \xi \in V^*$.
We define a pairing between $V \tens V$ and $V^* \tens V^*$ by
\[
\langle x \tens y, \xi \tens \eta \rangle = \langle x, \eta \rangle \langle
y, \xi \rangle. \]
Then $R ^{\sharp \sharp}$ is the transpose of $R$ with respect to this
pairing, which is non-degenerate, so $R^{\sharp\sharp}$ is invertible
if $R$ is, and $R^{\flat}=(R^{\sharp})^{\sharp\sharp}$ is invertible if
$R^{\sharp}$ is. Since $R^{\flat\flat}=R^{\sharp\sharp}$ while
$R^{-1\flat}=R^{-1 \sharp\sharp\sharp}=(R^{\sharp\sharp})^{-1\sharp}$
and $R^{\sharp\sharp}$ satisfies the same conditions as $R$, it remains
only to prove that $R^{-1\sharp}$ is invertible.

According to Theorem \ref{sharpsflats}, $R^{\sharp -1}:V^*\tens V\to
V\tens V^*$ is a braiding isomorphism, and we can regard $V^*$ as a left
dual of $V$ with the pairing $\lev:\, V^* \tens V \to k$ given by $\lev
= \ev \circ R^{\sharp -1}$. The braid relation \eqref{eq:braid} in $V
\tens V \tens V$ implies
\[
(R^{\sharp -1}\tens 1)(1 \tens R)(R^{-1\sharp} \tens 1) = (1 \tens
R^{\sharp -1})(R \tens1)(1\tens R^{\sharp -1}): V\tens V^* \tens V
\to V\tens V^* \tens V
\]
Multiplying on the left by $\ev\otimes 1$ and using \eqref{evsharp} with
$T=R^{-1}$,
\begin{equation}\label{levRR}
(\lev\otimes 1)(1\otimes R)(R^{-1}\otimes 1)=1\otimes \lev: V\otimes
V^*\otimes V\to
V.
\end{equation}
Using \eqref{cancel} with $T=R$, $S=R^{-1\sharp}$ and $U=\lev$, so
that $W=1\otimes \lev$ and $W^{\sharp}=\lev\otimes 1$,
\[
(1\otimes\lev) (R^{-1\sharp\sharp}\otimes 1) = (\lev\otimes 1) (1\otimes
R^{\sharp -1}: V^*\otimes V^*\otimes V\to V^*.
\]
Let $N\subset V$ be the null space of lev, so that $\lev(V^*\otimes N)=0$.
Then $V^*\otimes V^*\otimes N$ is annihilated by the left-hand side of
the above equation, so from the right-hand side we obtain $R^{\sharp -1}(V^*\otimes
N)\subseteq N\otimes V^*$. Since $R^{\sharp -1}$ is non-singular, we
have equality. Hence
\[
\ev(N\otimes V^*)=\ev\circ R^{\sharp -1}(V^*\otimes N)=\lev(V^*\otimes
N)=0.
\]
But ev is a non-degenerate pairing, so $N=0$; hence lev is a non-degenerate
pairing.

Now return to \eqref{levRR} and, by tensoring on the right with $V^*$
and applying ev, write it in the form
\[
\ev(\lev\otimes 1\otimes 1)(1 \otimes R\otimes 1)(R^{-1\sharp}\otimes
1\otimes 1)=\ev(1\otimes \lev\otimes 1):V\otimes V^*\otimes V\otimes V^*\to
k.
\]
Since ev and lev are non-degenerate, the right-hand side is a non-degenerate
bilinear form on $V\otimes V^*$ and therefore $R^{-1\sharp}$ is non-singular.
\end{pf}
\begin{defn}
A {\em Hecke} $\check R$-matrix is an $\check R$-matrix $\RB$ satisfying
the quadratic equation
\[ (\RB-q) (\RB+q^{-1}) =0 \]
for some $q\in k^\times$, $q^2\ne-1$, and such that $\RB^\sharp$ is
invertible.
\end{defn}

\section{A construction for graded Hopf algebras}
\label{con-recon}
\subsection{Universally coacting bialgebras} \label{reconbialg}
Let $\CO$ be a collection of finite-dimensional
$\gr$-graded vector spaces, and let
$\CM = \{ f: X_{i_1}\tens\dots\tens X_{i_k} @>>>
X_{j_1}\tens\dots\tens X_{j_m},  \}$ ($X_{i_l},X_{j_n}\in\CO$)
be a family of linear maps, preserving degree or changing it by $(1,0)$.

Let $X_0 =k$ be a chosen one dimensional space of degree $(0,0)$.

Consider a category $\CH$ of $\gr$-bialgebras $H$ together with grading-preserving
coactions $(\delta_X: X \to X\tens H)_{X\in\CO}$ such that

(a) the coaction on $X_0$ is given by $\delta_0 :k\to k\tens H$,
$1\mapsto 1\tens1$;

(b) any $f\in\CM$ is a $H$-comodule homomorphism.

\noindent The morphisms $(H,\delta_X)_{X\in\CO} @>>> (H',\delta_X')_{X\in\CO}$
are $\gr$-graded bialgebra maps $g:H\to H'$ such that
\begin{equation} \label{del'del}
\delta_X' = (X @>\delta_X>> X\tens H @>1\tens g>> X\tens H').
\end{equation}

\begin{thm}[Bialgebra construction] \label{constr1}
There exists a universal coacting bialgebra $(H,\delta_X)$ = initial object
of the category $\CH$. That is, for any bialgebra $(H',\delta'_X) \in
\CH$
there is exactly one bialgebra map $g:H\to H'$ with the property
\eqref{del'del}.
\end{thm}

\begin{pf} Choose a graded basis $(x_a)$ in each space $X\in\CO$.
Introduce a $\gr$-graded coalgebra
$C=\oplus_{X\in\CO} (\End_kX)^* = k\{\mc{X}ab \}$ with coaction
\[ X\to X\tens C,\quad x_b \mapsto x_a \tens \mc{X}ab \]
on $X_i$. Then the tensor algebra $T(C)$ is a $\gr$-graded bialgebra \cite{Lar:ord}
coacting on each $X_i$. With each map
\[ f: X^1\tens\dots\tens X^k @>>> Y^1\tens\dots\tens Y^m \in\CM , \]
$X^i\in\CO$, $Y^j\in\CO$, is associated a subspace $\Rel(f)\subset T(C)$
as follows. Choosing graded bases $(x_a^i)\subset X^i$, $(y_b^j)\subset
Y^j$ and writing
corresponding matrix elements as $\mc{X^i}ca$, $\mc{Y^j}db$ we define
this
$\gr$-graded subspace as
\begin{multline*}
\Rel(f) = k\left\{ (-1)^{\sum_{i<j} (\hat a_i-\hat c_i)\hat c_j}
f_{c_1\dots c_k}^{d_1\dots d_m}
\mc{X^1}{c_1}{a_1} \dots \mc{X^k}{c_k}{a_k} \right. \\
\left. - (-1)^{\sum_{n<l} (\hat b_n-\hat d_n)\hat d_l}
\mc{Y^1}{d_1}{b_1} \dots \mc{Y^m}{d_m}{b_m}
f_{a_1\dots a_k}^{b_1\dots b_m} \right\} .
\end{multline*}
where $\hat g\in\gr$ denotes the degree of a basic vector indexed by $g$
(the product $\hat g\hat h$ is the inner product
$(\gr)\times(\gr) \to \Z/2$). A more readable form for $\Rel(f)$
will be given later (\thmref{Hascoend}).

The bialgebra $T(C)$ is a universal bialgebra coacting on each $X_i$.
In
particular, for any $H'\in \CH$ there is unique bialgebra map
$h:T(C) \to H'$, preserving coactions in the sense of \eqref{del'del}.
For
any such $H'$, we get $h(\Rel(f)) =0$ and this equation is equivalent
to
the condition that $f$ is a $H'$-comodule morphism.

Define now an algebra
\[ H= T(C)/(\mc k11 -1, \Rel(f))_{f\in\CM} .\]
As the subspaces $k\{\mc k11 -1\}$ and $\Rel(f)$ are coideals of $T(C)$
for any
$f\in\CM$, the algebra $H$ is a bialgebra. By the above consideration
$H$ belongs to $\CH$ and is an initial object.
\end{pf}

This theorem is due essentially to Takeuchi \cite{Take} (except for the
grading) where the case $\CO= \{X\}$, $\CM = \{P_k: X\tens X \to X\tens
X\}$
is considered ($P_k$ are  projections and $\sum P_k =1$).

\subsection{Hopf algebra case}
Let $\CO$ and $\CM$ be as in \secref{reconbialg}. Now we discuss the
question when the bialgebra $H$ constructed in  \thmref{constr1} is
a Hopf algebra.

\begin{thm}[Hopf algebra construction] \label{constr2}
Assume that for each $X\in\CO$ there exist compositions of tensor monomials
in maps from $\CM$, i.e.
\[ f= 1\tens f_1\tens1 \cdot\dots\cdot 1\tens f_n\tens1 :
X\tens (Y_1\tens\dots\tens Y_k)  @>>> k ,\]
\[ g= 1\tens g_1\tens1 \cdot\dots\cdot 1\tens g_m\tens1 :
(W_1\tens\dots\tens W_l) \tens X  @>>> k ,\]
with $f_i\in\CM$, which are pairings, non-degenerate in the argument $X$. 
Assume also existence of compositions of tensor monomials in maps from 
$\CM$
\[ h= 1\tens h_1\tens1 \cdot\dots\cdot 1\tens h_a\tens1 : k @>>>  
(U_1\tens\dots\tens U_c) \tens X ,\]
\[ j= 1\tens j_1\tens1 \cdot\dots\cdot 1\tens j_b\tens1 : k @>>> 
X\tens (V_1\tens\dots\tens V_d) , \]
which are non-degenerate in $X$ in the sense that the induced 
linear maps $(U_1\tens\dots\tens U_c)\pti @>>> X$ and 
$(V_1\tens\dots\tens V_d)\pti @>>> X$ are surjective. Then the 
universal bialgebra from \thmref{constr1} has an invertible antipode.
\end{thm}

\begin{pf}
Extend $\CO$ and $\CM$, adding for each $X$ new spaces $A,B$ to $\CO$ 
and
new maps
\[ Y_1\tens\dots\tens Y_k \to Y_1\tens\dots\tens Y_k/\Ann^{\text{right}} 
f
\equiv A ,\qquad X\tens A\to k, \]
\[ W_1\tens\dots\tens W_l \to W_1\tens\dots\tens W_l/\Ann^{\text{left}} 
g
\equiv B ,\qquad B\tens X\to k. \]
It is easy to see that the resulting bialgebra $H$ will not change. Also 
we add
minimal subspaces together with inclusions
\[ C \hookrightarrow U_1\tens\dots\tens U_c ,\qquad 
   D \hookrightarrow V_1\tens\dots\tens V_d \]
such that $h$ and $j$ factorize through 
\[ k \hookrightarrow C\tens X, \qquad k \hookrightarrow X\tens D \]
(and we add also these maps to $\CM$). Again the algebra $H$ will not 
change.

But now for each $X\in\CO$ there are $X_1\pti,X_2\pti,X_3\pti,X_4\pti 
\in\CO$
together with non-degenerate pairings and co-pairings
\[ \ev : X\tens X_1\pti \to k ,\qquad \ev : X_2\pti \tens X \to k ,\]
\[ \coev:k\to X_3\pti \tens X ,\qquad \coev:k\to X\tens X_4\pti . \]
The theorem follows from a proposition, proven in~\cite{Lyu:dis} in the 
even
case and easily generalized to the $\Z/2$-graded case:

\begin{prop}
Let data $\CO,\CM$ be such that for each $X\in\CO$ there are $X_1\pti$, 
$X_2\pti$, $X_3\pti$, $X_4\pti \in\CO$ as above. Then the universal
bialgebra $H$ has an invertible antipode $\gamma$. Denote the coaction 
of $H$
by $x_i\mapsto x_j\tens \mc Xji$, $x^k\mapsto x^l\tens \tpt{X\pti l}k$,
${}^mx\mapsto {}^nx\tens \tpt{\pti Xn}m$, where $(x_i)$, $(x_k)$, 
$({}^mx)$ are bases of $X$, $X\pti=X_1\pti$ or $X_3\pti$, $\pti X=X_2\pti$
or $X_4\pti$. Then the antipode and its inverse satisfy
\begin{align*}
\gamma(\mc Xji) &= (-1)^{\hat j(\hat i-\hat j)} \tpt{\pti Xi}j ,\\
\gamma^{-1}(\mc Xji) &= (-1)^{\hat j(\hat i-\hat j)} \tpt{X\pti i}j .
\end{align*}
\end{prop}
\end{pf}

\subsection{Bialgebras in rigid monoidal categories}\label{monocat}
The ideas of \secref{reconbialg} can be applied not only to vector 
spaces, or graded vector spaces, or to vector spaces with additional 
structures. It can be generalized to symmetric (or even braided) 
closed monoidal categories. See e.g. \cite{Maj:rec,Sch:recon,Yet:recon}. 
Having in mind applications to quantum differential calculus, we 
describe one such scheme following \cite{Lyu:tan,Lyu:mod}.

Let $\CV$ be a noetherian abelian symmetric monoidal rigid category. (We 
can
even assume it braided instead of symmetric, but this generalization will 
not be used here. See \cite{Lyu:mod} for applications in the braided case.) 
Let $k=\End_\CV I$ be a field, where $I\in \Ob \CV$ is a unit object; 
then 
$\CV$ is $k$-linear. Consider the category $\hat\CV=\ind\CV$ of $k$-linear
left exact functors $\CV^{\op}\to k\Vect$. It is known that $\hat\CV$ 
is a
$k$-linear abelian symmetric monoidal closed category. The category $\CV$ 
can be viewed as a full subcategory of $\hat\CV$ via the representing 
functor $h:\CV\to\hat\CV$, $X\mapsto h_X$, $h_X(Y)=\Hom_{\CV}(Y,X)$.

Take a family $\CO$ of $\Z/2$-graded objects of $\CV$ and a family 
$\CM = \{ f: X_{i_1}\tens\dots\tens X_{i_k} @>>> 
X_{j_1}\tens\dots\tens X_{j_m} \}$ of morphisms of $\CV$, preserving the 
grading or changing it by 1. Denote by $\CC$ the subcategory of $\CV$ 
consisting of all tensor products $X_{i_1}\tens\dots\tens X_{i_k}$ with 
all tensor polynomials of $f\in\CM$ as morphisms. $\CC$ is a monoidal 
subcategory of $\CV$. The bifunctor $B:\CC^{\op}\times\CC\to\CV$, 
$X\times Y\mapsto X\pti\tens Y$ has a coend 
$H=\int^{Y\in\CC}Y\pti\tens Y\in\hat \CV$ which 
can be found from the exact sequence in $\hat\CV$
\[ \bigoplus_{f:Y\to Z\in\Mor\CC} Z\pti\tens Y @>1\tens f-f^t\tens1>>
   \bigoplus_{Y\in\Ob\CC} Y\pti\tens Y \to H \to 0 . \]
Here $f^t:Z\pti\to Y\pti\in\CV$ is the dual morphism to $f:Y\to Z\in\CV$. 
As could be expected from \cite{Maj:rec,Lyu:mod}, the object $H\in\hat\CV$ 
turns out to be a superbialgebra in $\hat\CV$.

\begin{thm}\label{Hascoend}        
(a) Let  $Z=\oplus_{Y\in\Ob\CC} Y\pti\tens Y\in\hat\CV$ with the 
operations of comultiplication
\begin{multline*}
\Delta:\oplus_Y Y\pti\tens Y=\oplus_Y Y\pti\tens I\tens Y
   @>1\tens\coev_Y\tens 1>> \oplus_Y Y\pti\tens Y\tens Y\pti\tens Y \\
\subset (\oplus_Y Y\pti\tens Y)\tens (\oplus_Y Y\pti\tens Y);
\end{multline*}
counit 
\[ \eps:\oplus_Y Y\pti\tens Y @>\ev>> \oplus_Y I 
@>{\scriptscriptstyle\sum}>> I ;\]
multiplication
\begin{multline*}
m:(\oplus_X X\pti\tens X)\tens(\oplus_Y Y\pti\tens Y)=
   \oplus_{X,Y} X\pti\tens X\tens Y\pti\tens Y
   @>1\tens S\tens 1\cdot S\tens 1\tens 1>> \\
\oplus_{X,Y} Y\pti\tens X\pti\tens X\tens Y @>\sigma\tens 1\tens 1>>
\oplus_{X,Y}(X\tens Y)\pti\tens (X\tens Y) \to \oplus_Z Z\pti\tens Z,
\end{multline*}
where $\sigma:Y\pti\tens X\pti\tens X\to Y\pti\tens X\pti\tens X$ equals
$(-1)^{i(k-j)}$ when restricted to a homogeneous component 
$Y_i\pti\tens X_j\pti\tens X_k$ with $i,j,k\in\Z/2$, and $S$ is the symmetry;

and unit
\[ \eta: I\simeq I\pti\tens I \hookrightarrow \oplus_Y Y\pti\tens Y .\]
Then $Z$ is a superalgebra in $\hat\CV$.

(b) $\sum_f \im(1\tens f - f^t\tens1)$ is a bi-ideal in 
$\oplus_Y Y\pti\tens Y $, therefore $H$ is a superbialgebra in $\hat\CV$.

(c) The morphisms
\[\delta_X:X= I\tens X @>\coev_{X}\tens1>> X\tens X\pti\tens X
\hookrightarrow X\tens(\oplus_Y Y\pti\tens Y) \to X\tens H \]
are coactions of $H$ on $X\in\CO$.

(d) The bialgebra $H$ together with the coactions $\delta_X$ is universal 
in a
sense similar to that of \secref{reconbialg}.
\end{thm}

The proof is straightforward. New in this theorem in comparison with, 
say, 
\cite{Lyu:mod} is the $\Z/2$-grading. It is valid also for braided 
categories $\CV$.

\begin{example}
The case of $\CV=\Z$-grad-vect, the category of $\Z$-graded finite 
dimensional vector spaces with grading-preserving linear maps, is already 
considered in \secref{reconbialg}. Here $\hat\CV=\Z$-grad-Vect. The 
symmetry is given by $S(x\tens y)= (-1)^{p(x)p(y)} y\tens x$ on 
homogeneous vectors $x,y$, where $p$ is the degree.
The projection map $j:\oplus_Y Y\pti\tens Y \to H$ onto the universal
algebra from \secref{reconbialg} is given by $y^a\tens y_b \mapsto \mc 
Yab$. 
The statement that $\sum_f\im(1\tens f-f^t\tens 1)=\Ker j$ is equivalent 
to 
the statement from \thmref{reconbialg} that $\Rel(f),f\in \CM$, generate 
the ideal of relations of $H$.
\end{example}

\thmref{constr2} holds also in this setting with maps replaced by morphisms.

\begin{thm}[Hopf algebra construction] \label{constrmono}
Assume that for each $X\in\CO$ there exist compositions of tensor monomials
in morphisms from $\CM$
\[ f= 1\tens f_1\tens1 \cdot\dots\cdot 1\tens f_n\tens1 : 
X\tens (Y_1\tens\dots\tens Y_k)  @>>> k ,\]
\[ g= 1\tens g_1\tens1 \cdot\dots\cdot 1\tens g_m\tens1 : 
(W_1\tens\dots\tens W_l) \tens X  @>>> k ,\]
which are pairings, non-degenerate in the argument $X$. Assume also 
existence of compositions of tensor monomials in maps from $\CM$
\[ h= 1\tens h_1\tens1 \cdot\dots\cdot 1\tens h_a\tens1 : k @>>>  
(U_1\tens\dots\tens U_c) \tens X ,\]
\[ j= 1\tens j_1\tens1 \cdot\dots\cdot 1\tens j_b\tens1 : k @>>> 
X\tens (V_1\tens\dots\tens V_d) , \]
which are non-degenerate in $X$ in the sense that the  induced 
morphisms $(U_1\tens\dots\tens U_c)\pti @>>> X$ and 
$(V_1\tens\dots\tens V_d)\pti @>>> X$ are epimorphisms. Then the 
universal bialgebra $H\in\hat\CV$ has an invertible antipode.
\end{thm}

\begin{pf} Repeating the proof of \thmref{constr2}, we can assume that 
$\CC$ 
is closed under duality $-\pti$. Then we define an antiendomorphism
\[ \gamma= (-1)^{j(i+j)}\  
\unitlength=1mm
\makebox[20mm][l]{
\raisebox{-18mm}[19mm][20mm]{
\put(5,13){\line(3,5){10.17}}
\put(5,30){\line(1,-2){4}}
\put(7,15){\line(1,-6){1.83}}
\put(5,16){\oval(10,6)[l]}
\put(10,20){\line(1,-2){8}}
\put(5,34){\makebox(0,0)[cc]{$Y_i\pti$}}
\put(15,34){\makebox(0,0)[cc]{$Y_j$}}
\put(9,0){\makebox(0,0)[cc]{$Y_j\pti\pti$}}
\put(18,0){\makebox(0,0)[cc]{$Y_i\pti$}}
}}
:\bigoplus_{Y,i,j} Y_i\pti\tens Y_j \to
\bigoplus_{Y\pti,i,j} Y_j\pti\pti\tens Y_i\pti .\]
It preserves the ideal of relations of $H$ and the antipode of $H$ comes 
as
its quotient.
\end{pf}

\subsection{Differential bialgebra case}
Some methods of constructing differential bialgebras were described by
Maltsiniotis~\cite{Mal:Gqdif,Mal:lang}, Manin~\cite{Man:deRham} and one 
of us \cite{Sud:diffor}. We
present here a general framework for such constructions. The results of 
the previous section are applied to the category $\CV=\cd$ of $\Z$-graded 
 differential finite-dimensional vector spaces, $(V,d:V\to V)$,
$V=\oplus_{i\in\Z} V_i$, $d^2=0$. The differential $d$ has degree 1.
The category
$\hat\CV= \hat\cd$ consists of all $\Z$-graded differential vector spaces
and their grading-preserving linear maps commuting with the differential.

So, our data are a family of differential $\gr$-graded spaces $\CO$
with differentials of degree (0,1) and a family of linear maps $\CM=\{f\}$
of degree (0,0) or (1,0) commuting with the differential. The output of 
\thmref{Hascoend} is a $\gr$-graded bialgebra $H$ in the category $\hat\cd$, 
that is, a differential $\gr$-graded bialgebra.

We have a forgetful functor $\Phi:\cd\to \Z$-grad-vect forgetting the 
differential. The constructions of \secref{reconbialg} with 
$\CV'=\Z$-grad-vect and above with $\CV=\cd$ give bialgebras $H'$ and 
$H$. 
They are identified, $\Phi(H)=H'$. To give explicitly the differential 
in 
the bialgebra $H'$ making it into $H$, consider the coalgebra
$C= \oplus_{X\in\CO} \Comat X = \oplus_{X\in\CO} X\pti\tens X$. It has
a unique map $d: C\to C$ of degree $(0,1)$ which makes the diagram
\[ \begin{CD}
X @>\delta_X>> X\tens C \\
@VdVV            @VVd\tens1+\sigma\tens dV \\
X @>\delta_X>> X\tens C 
\end{CD} \]
commute. In our conventions $\sigma:X\to X$ is the map whose 
eigenvectors are the homogeneous elements $x$ of $X$, the eigenvalue
associated with such an element being  $(-)^x=(-1)^{\hat x\cdot (0,1)}$. 
Indeed, $d: C\to C$ is given by
\[ d\mc{X_i}fa = (-)^f (\mc{X_i}fc \md ca - \md fb \mc{X_i}ba ) ,\]
where $\md ca$ denotes the matrix of the map $d: X_i \to X_i$, 
$x_a \mapsto x_c \md ca$, in a chosen basis.

\begin{prop} The map $d:C\to C$ is a codifferential of degree 
$(0,1) \in\gr$, that is, $d$ is a graded coderivation 
\[ \begin{CD}
C @>\Delta>> C\tens C \\
@VdVV        @VVd\tens1+\sigma\tens dV \\
C @>\Delta>> C\tens C 
\end{CD} \]
and $d^2=0$.
\end{prop}

\begin{pf}
Straightforward.
\end{pf}

\subsubsection{Construction for differential bialgebras} 
\label{pracondifbial}
The differential bialgebra $H$ constructed in \secref{monocat} usually
does not correspond to a geometric type of object. However, using previous
results we can construct more realistic examples.

Assume given the following data (comp. \cite{Sud:diffor,Tsy:dT}). Let 
$X^\bullet_1, X^\bullet_2, \dots, X^\bullet_r$  be 
$\gr\times\Z_{\ge0}$-graded algebras (where $\bullet$ denotes the 
$\Zp$-degree) with multiplication
$m_u: X^\bullet_u \tens X^\bullet_u \to X^\bullet_u$, $1\le u\le r$,
generated by finite-dimensional spaces $X^1_u$ and $X^0_u=k$. Let 
$d_u: X^\bullet_u \to X^\bullet_u$ be differentials of degree $(0,1,0)$. 
Assume bijective linear maps $f_u: X^1_u \to X^1_{u+1}$ of degree 
$\degree f_u = (p_u,0) \in \gr$ are given for $1\le u\le r-1$ such that
$f_u d_u = d_{u+1} f_u$.

These data are transformed to the family 
$\CO=\{ X^i_u\}_{i\ge0,1\le u \le r}$ of differential $k$-spaces 
and a family of maps 
$\CM= \{m_u: X^i_u\tens X^j_u \to X^{i+j}_u, f_v:X^1_v \to X^1_{v+1} \mid 
i,j\ge0, 1\le u\le r, 1\le v<r \}$ of degree $(*,0) \in\gr$. 
We have already proved

\begin{thm}[Differential bialgebra construction] \label{difbicon}
The bialgebra $H$ constructed from these data in \thmref{constr1} 
is a 
 differential bialgebra. 
\end{thm}

\begin{rem}
We can always reduce such data to the case of algebras 
$Y^\bullet_1, Y^\bullet_2, \dots, Y^\bullet_r$ generated by the same space
$Y^1_u = Y^1$ with $f_u = \id_{Y_1}$. For instance, having 
$X^\bullet_1, X^\bullet_2$ and $f_1: X^1_1 \to X^1_2$, we define
$Y^\bullet_1 = X^\bullet_1$, and $(Y_2^\bullet, \tilde m_2)$ generated 
by
$Y^1_2 = X^1_1$ is defined as a graded algebra ``isomorphic'' to 
$X_2^\bullet$ via a bijective map 
$f\equiv f^\bullet : Y_2^\bullet \to X_2^\bullet$, such that $f^1 = f_1$ 
and
$\degree f^k = (k p_1,0) \in\gr$ (recall that $\degree f_1 = (p_1,0)$ 
). 
Precisely we require  the diagram
\[ \begin{CD}
Y^\bullet_2 \tens Y^\bullet_2 @>\tilde m_2>> Y^\bullet_2 \\ 
@Vf\tens fVV                                 @VVfV \\
X^\bullet_2 \tens X^\bullet_2 @>m_2>> X^\bullet_2 
\end{CD} \]
to commute, where as usual $(f\tens f^\bullet) (y\tens y') = 
(-1)^{\hat f^\bullet\cdot \hat y} f(y)\tens f^\bullet(y')$. 
The map $f$ will be an isomorphism of graded algebras only if $p_1=0$.

All such $(Y_2^\bullet, \tilde m_2)$ are naturally isomorphic as graded 
associative algebras. One can verify that 
$Y^\bullet_2 = k \oplus X_1^1 \oplus X_2^2 \oplus X_2^3 \oplus X_2^4 \dots$ 
(with appropriately changed grading if $p_1=1$) satisfies these conditions.
\end{rem}

\subsubsection{Differential forms on quantum semigroups} 
Let $X^\bullet_1, X^\bullet_2, \dots, X^\bullet_r$  be 
$\gr\times\Z_{\ge0}$-graded algebras with additional structures
as in \secref{pracondifbial}. Then \thmref{difbicon} gives a $\gr$-graded
 differential bialgebra $H$.

\begin{thm} \label{Omega}
Let $H^{<0}$ denote $\oplus_{a\in\Z/2, b<0} H^{a,b} \subset H$. Then
\[ \Omega = H/ (H^{<0}, dH^{<0}) \]
is a $\gr_{\ge0}$-graded  differential bialgebra. The algebra
$\Omega$ is a universal bialgebra in the category of  differential 
$\gr_{\ge0}$-graded bialgebras coacting on 
$X^\bullet_1, X^\bullet_2, \dots, X^\bullet_r$. 
\end{thm}

\begin{pf}
The subspace $H^{<0} \oplus dH^{-1} \subset H$ is $d$-invariant 
(here $H^{-1} \overset{\text{def}}= H^{0,-1} \oplus H^{1,-1}$). Also this
subspace is a coideal. Hence, $\Omega = H/(H^{<0} \oplus dH^{-1})$ is 
a
 differential bialgebra. 

Having some $\gr_{\ge0}$-graded bialgebra $f\in \CH$ coacting on 
$X^\bullet_1, X^\bullet_2, \dots, X^\bullet_r$ we find by \thmref{constr1} 
the  morphism of bialgebras $\phi: H\to F$. If, in addition, $F$ is a 
 differential bialgebra, differentially coacting on
$X^\bullet_1, X^\bullet_2, \dots, X^\bullet_r$, we get that $\phi$ is 
a
differential bialgebra morphism. Since $H^{<0} \subset \Ker\phi$, this
implies $dH^{<0} \subset \Ker\phi$, whence the theorem is proven.
\end{pf}

\begin{defn}
We call $\Omega$ from \thmref{Omega} the algebra of differential forms 
on a
quantum (super)semigroup, corresponding to the function algebra
$\Omega^0 \overset{\text{def}}= \Omega^{0,0} \oplus \Omega^{1,0}$.
\end{defn}

This terminology is justified by examples.

\section{Examples of differential quantum supergroups}\label{sec2}
\subsection{Hecke $R$-matrices as construction data} \label{doubledR}
Let $V$ be a $\Z/2$-graded vector space with a graded basis 
$(x_i),1\le i\le n$ and let 
$\RB:V\tens V\to V\tens V,\quad x_i\tens x_j\mapsto\RB_{ij}^{kl}x_k\tens 
x_l$ 
be a solution to the Yang-Baxter equation. We assume that $\RB$ preserves 
the grading and is diagonalizable with two eigenvalues $q$ and
$-q^{-1},q\ne\pm i$. Denote by $dV$ another copy of $V$ with a 
basis $(dx_i)$. Set $X^1=V\oplus dV$ and 
\[ X^\bullet=T^\bullet(X^1)/(\RB(x_i\tens x_j)-qx_i\tens x_j,
\RB(dx_i\tens x_j)-q^{-1}dx_i\tens x_j,
\RB(dx_i\tens dx_j)+q^{-1}dx_i\tens dx_j)\]
It is understood that $\RB$ is extended to
\[ \RB:dV\tens V\to V\tens dV,\quad 
dx_i\tens x_j\mapsto \RB_{ij}^{kl} x_k\tens dx_l \]
\[ \RB:dV\tens dV\to dV\tens dV,\quad dx_i\tens dx_j
\mapsto \RB_{ij}^{kl} dx_k\tens dx_l \]
Make $X^\bullet$ into a $\Z/2\times\Z_{\ge0}\times\Z_{\ge0}$-graded 
algebra by setting
\[ \deg x_i = (p(x_i),0,1),\qquad \deg dx_i = (p(x_i),1,1) \]
where $p$ is the $\Z/2$-degree in $V$. Introduce a differential in $X^\bullet$ 
by $d(x_i) =dx_i$. 

Let $\Xi$ be a copy of $V$ with the opposite parity, that is, with a basis 
graded by
$(\xi_i)$, $p(\xi_i) =p(x_i) +1$. One more copy of $\Xi$ denoted $d\Xi$ 
has
a basis $(d\xi_i)$. Consider the algebra $Y$ generated by 
$Y^1= \Xi\oplus d\Xi$
\[ Y^\bullet=T^\bullet(Y^1)/(\RB(\xi_i\tens \xi_j)+ q^{-1}\xi_i\tens \xi_j,
\RB(d\xi_i\tens \xi_j)+qd\xi_i\tens \xi_j,
\RB(d\xi_i\tens d\xi_j)-qd\xi_i\tens d\xi_j) .\]
with degree 
\[ \deg \xi_i = (p(x_i)+1,0,1),\qquad \deg d\xi_i = (p(x_i)+1,1,1) ,\]
and differential $d(\xi_i) = d\xi_i$. 

The map $J:X^1\to Y^1$, $x_i\mapsto \xi_i$, $dx_i\mapsto d\xi_i$, of degree
$(1,0)\in \Z/2\times\Z$ commutes with the differential.

The data $(X^\bullet,Y^\bullet,J)$ determine a differential bialgebra 
$H$ as
\thmref{difbicon} claims. To find its structure explicitly we analyze 
the 
generators and relations proposed in the theorem. The matrices $T$ and 
$T'$ 
made of matrix elements of the spaces $X^1$ and $Y^1$ can be decomposed 
in blocks
\[ T= \begin{pmatrix} t&r \\ p&s \end{pmatrix} ,\qquad
   T'= \begin{pmatrix} t'&r' \\ p'&s' \end{pmatrix} ,\]
so the coaction on $X^1$ and $Y^1$ is 
\begin{align*}
\delta x_i  &= x_j\tens \ma tji +dx_j\tens \ma pji ,\\
\delta dx_i &= x_j\tens \ma rji +dx_j\tens \ma sji ,\\
\delta \xi_i  &= \xi_j\tens \ma {t'}ji +d\xi_j\tens \ma {p'}ji ,\\
\delta d\xi_i &= \xi_j\tens \ma {r'}ji +d\xi_j\tens \ma {s'}ji .
\end{align*}
The relations corresponding to $J$ say precisely that $T'=T$ in $H$. 
Clearly, the entries of $T$ generate the algebra $H$.The relations 
$\Rel(m:X^\bullet \tens X^\bullet\to X^\bullet)$ coincide with those 
asserting that $\Ker(m:X^\bullet \tens X^\bullet\to X^\bullet)$ is an 
$H$-subcomodule. A particular requirement is that 
\begin{multline*} 
\Ker(m_{11}:X^1\tens X^1 \to X^2) = \\ 
= \< \RB(x\tens y)-qx\tens y,
\RB(dx\tens y)-q^{-1}dx\tens y, \RB(dx\tens dy)+q^{-1}dx\tens dy \>
\end{multline*}
is an $H$-subcomodule. Similarly 
\[\Ker(m'_{11}:Y^1\tens Y^1\to Y^2) =\<\RB(\xi\tens\eta)+q^{-1}\xi\tens\eta,
\RB(d\xi\tens\eta)+qd\xi\tens\eta, \RB(d\xi\tens d\eta)-qd\xi\tens d\eta\>\]
is an $H$-subcomodule. The graded isomorphism of $H$-comodules
$J\tens J: X^1\tens X^1\to Y^1\tens Y^1$ sends the subspace
\[ K' = \< \RB(x\tens y)+q^{-1}x\tens y, 
\RB(dx\tens y)+qdx\tens y, \RB(dx\tens dy)-qdx\tens dy \> \]
to $\Ker m'_{11}$, whence $K'$ is an $H$-subcomodule.

Therefore, we have to impose on $H$ relations equivalent to the fact that
$\Ker m_{11}, K' \subset X^1\tens X^1$ are $H$-subcomodules. Remark that
\[ \Ker m_{11}\oplus K' = X^1\tens X^1 .\] 
Indeed, introduce a grading-preserving linear map
\[ \CR= \begin{pmatrix} \RB&0&0&0 \\ 0&0&\RB^{-1}&0 \\ 
0&\RB&q-q^{-1}&0 \\ 0&0&0&-\RB^{-1} \end{pmatrix} \]
\[ \CR: X^1\tens X^1 \to X^1\tens X^1= 
V\tens V\oplus V\tens dV\oplus dV\tens V\oplus dV\tens dV\]
(The matrix acts on a vector from the right.)
It satisfies the same quadratic equation as $\RB$
\[ (\CR-q)(\CR+q^{-1})=0 .\]
Its eigenspace with eigenvalue $q$ (resp. $-q^{-1}$) is $K'$ (resp. 
$\Ker m_{11}$).

We conclude that the equations to impose express the fact that $\CR$ is 
an
automorphism of the $H$-comodule $X^1\tens X^1$. Define accordingly
\be\label{(-)}
H= k\<\ma Tji \>/ ((-1)^{\hat l(\hat i-\hat k)} \CR^{ps}_{kl} \ma Tki 
\ma Tlj - 
(-1)^{\hat s(\hat m-\hat p)} \ma Tpm \ma Tsn \CR^{mn}_{ij}).
\end{equation}
The standard assertion is that $H$ is a bialgebra with the canonical 
coaction on $X^1,Y^1$. It coacts also on
\[ X^\bullet= T^\bullet(X^1)/(\Ker m_{11}) ,\qquad 
Y^\bullet= T^\bullet(Y^1)/(\Ker m'_{11}) \]
since the ideal of relations is a $H$-subcomodule by construction. The 
bialgebra $H$ was built so that any other bialgebra from $\CH$ is an
epimorphic image of it, hence it is universal.

The bialgebra $H$ has generators $\ma tji ,\ma rji ,\ma pji ,\ma sji$ 
of degree
\begin{alignat*}2
\deg\ma tji &= (p(x_i)-p(x_j),0), &\qquad\deg\ma rji &= (p(x_i)-p(x_j),1),\\
\deg\ma pji &= (p(x_i)-p(x_j),-1),&\qquad\deg\ma sji &= (p(x_i)-p(x_j),0).
\end{alignat*}
If $V$ is even (all $p(x_i)=0$), there is a presentation
\begin{align*}
\RB t_1t_2 &= t_1t_2\RB \\
\RB s_1s_2 &= s_1s_2\RB \\
r_1t_2 &= \RB t_1r_2\RB \\
r_1s_2 &= \RB s_1r_2\RB \\
p_1t_2 &= \RB t_1p_2\RB \\
p_1s_2 &= \RB s_1p_2\RB \\
r_1r_2 &=-\RB r_1r_2\RB \\
p_1p_2 &=-\RB p_1p_2\RB \\
r_1p_2 &=-\RB r_1p_2\RB \\
t_1s_2\RB-\RB s_1t_2 &= (q-q^{-1})r_1p_2
\end{align*}
in the usual notation, with $\RB$ standing for the matrix $(\RB_{ij}^{kl})$ 
and $t_1$, for example, standing for the Kronecker product matrix $t \otimes 
1 = (t_i^k\delta_j^l)$.
In  the general case the same presentation holds with additional 
minus signs from \eqref{(-)}.

It follows by \thmref{difbicon} that $H$ is a differential bialgebra with 
the differential 
\begin{alignat}2
dt &= r, &\qquad dr &= 0, \label{dtdr} \\
dp &=t-s,&\qquad ds &= r. \label{dpds}
\end{alignat}

In \thmref{Omega} the bialgebra $\Omega$ is constructed. Clearly, in our 
example it is generated by $\ma tji$ and $\ma rji =d\ma tji$ and has the
relations of $H$ plus the identifications $p=0, s=t$. So the defining 
relations of  $\Omega$ are
\begin{align}
\RB t_1t_2 &= t_1t_2\RB  \label{Rtt=}\\
dt_1\cdot t_2 &= \RB t_1\,dt_2\RB \\
dt_1\cdot dt_2&=-\RB dt_1\,dt_2\RB \label{dtdt=}
\end{align}
We recognize in $\Omega$ the algebra of differential forms related to 
$\RB$
\cite{SchWatZum,Sud:supcal,Tsy:dT,Zum:calcul}.

\subsection{The supersemigroup}
A straightforward check proves

\begin{prop}
$\CR$ satisfies the Yang--Baxter equation (in braid form).
\end{prop}

Therefore the bialgebra $H$ can be interpreted as the quantized algebra 
of
functions on a supersemigroup. If we interpret the algebra
$k\<t\>/(\RB t_1t_2-t_1t_2\RB)$ as the algebra of functions on quantum 
$Mat(n|m)$ related to $\RB$, then $H$ will be the algebra of functions 
on 
$Mat(n+m\mid n+m)$ related to $\CR$.

We can continue the procedure of constructing new solutions to the 
Yang--Baxter equation starting with $\CR$ instead of $\RB$.

\subsection{The supergroup}
Now we use the category $\CV$ of $\gr$-graded differential vector spaces
with the differential of degree $(0,1)\in \gr$. To make the differential
bialgebra from \secref{pracondifbial} into a Hopf algebra, we have to 
add 
the dual vector space $Z$ to the given $X=V\oplus dV$ with 
$\CR:X\tens X\to X\tens X$. The subspaces $V$ and $dV$ have degrees 0 
and 
$1\in\Z$. We define 
$Z=W\oplus U$, where $W=V\pti$ has degree $0\in\Z$ and the basis $(w^i)$
dual to $(x_i)$, and $U=(dV)\pti$ has degree $-1\in\Z$ and the basis $(u_i)$
dual to $(dx_i)$. The differential $d:Z\to Z$, $dw^i=0$, $du^i=-w^i$ makes
\[ \ev:X\tens Z\to k, \qqquad \coev:k\to Z\tens X \]
into morphisms of $\CV$. Also $\CR$ is a morphism of $\CV$ since it commutes
with $d:X\tens X \to X\tens X$, 
$d(x\tens y)= dx\tens y + \sigma(x) \tens dy$.

{\em Assume} now that $\RB^\sharp$ is invertible, so $\RB$ is Hecke.
By \thmref{invertible} $\RB^{-1\sharp} :W\tens W\pti \to W\pti\tens W$ 
is also invertible. A simple calculation gives
\[ \CR^\sharp= 
\begin{pmatrix} \RB^\sharp&0&0&(q-q^{-1})\X \\ 0&0&\RB^\sharp&0 \\ 
0&\RB^{-1\sharp}&0&0 \\ 0&0&0&-\RB^{-1\sharp} \end{pmatrix}
: X\tens X\pti \to X\pti\tens X ,\]
where $\X=1^\sharp$, $\X_{jl}^{ik} = \delta^{ik} \delta_{jl}$. Therefore,
$\CR^\sharp$ is also invertible and $\CR$ is Hecke. By \thmref{sharpsflats}
$\CR^\sharp$ and $\CR^{\sharp\sharp}$ are morphisms, and in particular
commute with the differential $d$.

Thus, we can consider the following data in $\CV$: the two objects 
$\CO=\{X,Z\}$ and the family $\CM$ of morphisms
\begin{align*}
\CR &: X\tens X \to X\tens X, \\
\CR^\sharp &: X\tens Z \to Z\tens X, \\
\CR^{\sharp\sharp} &: Z\tens Z\to Z\tens Z, \\
\ev &: X\tens Z \to k, \\
\coev &: k\to Z\tens X.
\end{align*}
These data are closed under duality in the sense of \thmref{constrmono}. 
The missing pairing and co-pairing are constructed as
\be\label{ev'}
Z\tens X @>\CR^{\sharp-1}>> X\tens Z @>\ev>> k,
\end{equation}
\be\label{coev'}
k @>\coev>> Z\tens X @>\CR^{-1\sharp-1}>> X\tens Z.
\end{equation}
They are non-degenerate as follows from the general theory \cite{Lyu:dis}.
Therefore the bialgebra $H$ constructed from the data $\CO,\CM$ is a 
differential Hopf algebra. It is generated by matrix elements $\ma Tij$ 
and
$\Tbar ij$ of $X$ and $Z$. The relations corresponding to $\CM$ are 
\begin{align}
(-1)^{\hat l(\hat i-\hat k)} \CR^{ps}_{kl} \ma Tki \ma Tlj &= 
(-1)^{\hat s(\hat m-\hat p)} \ma Tpm \ma Tsn \CR^{mn}_{ij} ,
\label{-RTT=-TTR} \\
(-1)^{\hat l(\hat i-\hat k)} \CR^{sl}_{pk} \ma Tki \Tbar lj &= 
(-1)^{\hat s(\hat m-\hat p)} \Tbar pm \ma Tsn \CR^{nj}_{mi} ,\\
(-1)^{\hat l(\hat i-\hat k)} \CR^{lk}_{sp} \Tbar ki \Tbar lj &= 
(-1)^{\hat s(\hat m-\hat p)} \Tbar pm \Tbar sn \CR^{ji}_{nm} ,\\
(-1)^{\hat k(\hat i-\hat k)} \ma Tki \Tbar kj &= \delta_i^j ,\\
(-1)^{\hat j(\hat k-\hat j)} \Tbar ik \ma Tjk &= \delta_i^j . 
\label{-TT=delta}
\end{align}
We know that the antipode $\gamma$ is invertible and the last two equations
give, in particular,
\[ \gamma(\Tbar ij ) = (-1)^{\hat j(\hat i-\hat j)} \ma Tji .\]
{}From \eqref{ev'}, \eqref{coev'} one can find also $\gamma(\ma Tij )$.

Divide the matrix $\bar T$ into four submatrices
\[ \bar T= \begin{pmatrix}\bar t&\bar r \\ \bar p&\bar s \end{pmatrix} 
\]
similarly to $T$. The degrees are
\begin{alignat*}2
\deg\ma{\bar t}ji &= (p(x_i)-p(x_j),0), 
&\qquad \deg\ma{\bar r}ji &= (p(x_i)-p(x_j),-1),\\
\deg\ma{\bar p}ji &= (p(x_i)-p(x_j),1),
&\qquad \deg\ma{\bar s}ji &= (p(x_i)-p(x_j),0).
\end{alignat*}
The  differential is given by \eqref{dtdr}, \eqref{dpds} and
\begin{alignat*}2
d\bar t &= \bar p, &\qquad d\bar r &= \bar s-\bar t, \\
d\bar p &= 0 ,&\qquad d\bar s &= \bar p. 
\end{alignat*}

\begin{rem}
The use of $\CR^\sharp$ and $\CR^{\sharp\sharp}$ is dictated by 
\thmref{sharpsflats}. 
\end{rem}

\subsubsection{Differential forms on quantum $GL(n)$} \label{DfoqGL}
The algebra 
\[ \Omega = H/ (H^{<0}, dH^{<0}) = H/(p,t-s,\bar r,\bar t-\bar s) \] 
is the algebra of differential forms on a quantum (super)group $GL(n)$ 
related to $\RB$. In the even case it is generated by $t,r,\bar t$, because
$\bar p= -\bar t\cdot r^t\cdot\bar t$. The defining relations are
\eqref{Rtt=}--\eqref{dtdt=} and 
\begin{align}
\RB^\sharp r_1\bar t_2 &= \bar t_1r_2 \RB^{-1\sharp}  \\
\RB^{\sharp\sharp} \bar t_1\bar t_2 &= \bar t_1\bar t_2 \RB^{\sharp\sharp} 
\label{tbar^2=} \\
t^t\, \bar t &=1= \bar t\, t^t \label{ttbar}
\end{align}
In  the general case there are similar relations with additional signs.

\section{Koszul complexes}\label{Koszul}
\subsection{The Berezinian}
In this section we temporarily forget about differentials $d$ and consider
arbitrary Hecke $\check R$-matrices. Recall that with each Hecke 
$\check R$-matrix $\RB:V\tens V\to V\tens V$ two $\Zp$-graded algebras
are associated. The symmetric algebra $\Sy^\bullet(V)$ is the quotient 
of  
$T^\bullet(V)$ by relations 
\[ \RB(x\tens y) = q x\tens y, \qquad x,y\in V \]
and the external algebra $\Ex^\bullet(V)$ has relations 
\[ \RB(x\tens y) = -q^{-1}x\tens y, \qquad x,y\in V. \]  

Our definition of a Koszul complex is a $q$-deformation of \cite{Lyu:ber},
where most results of this section were obtained for symmetric monoidal
categories ($q=1$). Also the following definition is a braided version
of one of the Koszul complexes for quadratic algebras \cite{Man:book}.
The Koszul complex for a Hecke symmetry possesses more structure;
 namely, it will have two differentials.

\begin{defn}
The {\em Koszul complex} of $V$ is the $\Zp\times\Zp$-graded algebra 
$\K(V)$, quotient of $T^{\bullet\bullet}(V\oplus V\pti)$ by relations
\begin{alignat*}2
\RB(v\tens\bar v) &= -q^{-1}v\tens\bar v &&\qquad\text{for }v,\bar v\in 
V,\\
\RB^\sharp(v\tens v') &= qv\tens v' &&\qquad\text{for }v\in V,v'\in V\pti,\\
\RB^{\sharp\sharp}(v'\tens\bar v') &= q v'\tens\bar v' && 
\qquad\text{for } v',\bar v'\in V\pti.
\end{alignat*}
\end{defn}

The middle commutation relation can be written also as
\be\label{qc} 
v'\tens v = qc_{V\pti,V}(v'\tens v) \qquad\text{for }v\in V,v'\in V\pti.
\end{equation}
Using the diamond lemma \cite{dia} we decompose $\K(V)$ into the tensor 
product of a symmetric and an external algebra. Precisely, the maps
\begin{align}
\Sy^m(V\pti) \tens \Ex^n(V) \to & 
K^{0m}(V) \tens K^{n0}(V) @>m>> K^{nm}(V), \label{fac} \\
\Ex^n(V) \tens \Sy^m(V\pti) \to &
K^{n0}(V) \tens K^{0m}(V) @>m>> K^{nm}(V) \notag
\end{align}
are isomorphisms of $\Zp\times\Zp$-graded vector spaces.

Let $\{v_i \mid i=1,\dots,n\}$ be a basis for $V$ and 
$\{v^i \mid i=1,\dots,n\}$ be the dual basis for $V\pti$. In the following
we use the summation convention on indices which occur once in an upper
and once in a lower position.

\begin{prop} \label{dv=vd}
The element $\di=v^iv_i =\coev(1) \in  K^{1,1}(V) \simeq V\pti\tens V$ 
obeys the following commutation relations in $K(V)$
\begin{alignat*}2
\di v &= -q^2 v\di &&\qquad \text{for }v\in V,\\
\di v' &= v'\di &&\qquad \text{for }v'\in V\pti.
\end{alignat*}
\end{prop}

\begin{pf}
Using graphical notation
\[
\unitlength=1mm
\begin{picture}(31,10)
\put(22,0.50){\oval(10,19)[t]}
\put(13,3){\makebox(0,0)[cc]{$v^i$}}
\put(31,3){\makebox(0,0)[cc]{$v_i$}}
\put(4,3){\makebox(0,0)[cc]{$\di=$}}
\end{picture}
\]
we have
\begin{align*} 
\di v &= v^iv_iv = -qv^i(-q^{-1}v_iv) = -qv^i\RB(v_i\tens v) = \\
&
\unitlength=0.80mm
\begin{picture}(110,28)
\put(6,13){\makebox(0,0)[cc]{$=-q$}}
\put(20,15){\oval(10,14)[t]}
\put(32,25){\line(-1,-2){12}}
\put(25,1){\line(0,1){7}}
\put(15,15){\line(0,-1){14}}
\put(13,24){\makebox(0,0)[cc]{$v^i$}}
\put(27,24){\makebox(0,0)[cc]{$v_i$}}
\put(35,24){\makebox(0,0)[cc]{$v$}}
\put(40,13){\makebox(0,0)[cc]{$=-q^2$}}
\put(55,15){\oval(10,14)[t]}
\put(48,24){\makebox(0,0)[cc]{$v^i$}}
\put(62,24){\makebox(0,0)[cc]{$v_i$}}
\put(46,1){\line(6,5){24}}
\put(50,15){\line(0,-1){9}}
\put(50,1){\line(0,1){2}}
\put(60,11){\line(0,-1){10}}
\put(70,24){\makebox(0,0)[cc]{$v$}}
\put(73,13){\makebox(0,0)[cc]{$=-q^2$}}
\put(82,21){\line(0,-1){20}}
\put(82,24){\makebox(0,0)[cc]{$v$}}
\put(94,15){\oval(10,14)[t]}
\put(87,24){\makebox(0,0)[cc]{$v^i$}}
\put(101,24){\makebox(0,0)[cc]{$v_i$}}
\put(89,15){\line(0,-1){14}}
\put(99,1){\line(0,1){14}}
\put(110,13){\makebox(0,0)[cc]{$=-q^2vd$,}}
\end{picture}
\end{align*}
and
\begin{align*} 
v'\di &= v'v^iv_i = q^{-1}(qv'v^i)v_i = 
q^{-1} \RB^{\sharp\sharp}(v'\tens v^i)v_i = \\
&
\unitlength=0.80mm
\begin{picture}(114,23)
\put(6,12){\makebox(0,0)[cc]{$=q^{-1}$}}
\put(21,20){\makebox(0,0)[cc]{$v^i$}}
\put(35,20){\makebox(0,0)[cc]{$v_i$}}
\put(28,1){\oval(10,36)[t]}
\put(29,1){\line(-5,6){5}}
\put(16,17){\line(5,-6){5.67}}
\put(15,20){\makebox(0,0)[cc]{$v'$}}
\put(41,12){\makebox(0,0)[cc]{$=$}}
\put(53,20){\makebox(0,0)[cc]{$v^i$}}
\put(67,20){\makebox(0,0)[cc]{$v_i$}}
\put(60,1){\oval(10,36)[t]}
\put(48,17){\line(3,-2){6}}
\put(72,1){\line(-3,2){6}}
\put(63,7){\line(-3,2){6}}
\put(47,20){\makebox(0,0)[cc]{$v'$}}
\put(77,12){\makebox(0,0)[cc]{$=$}}
\put(83,20){\makebox(0,0)[cc]{$v^i$}}
\put(97,20){\makebox(0,0)[cc]{$v_i$}}
\put(90,1){\oval(10,36)[t]}
\put(104,17){\line(0,-1){16}}
\put(104,20){\makebox(0,0)[cc]{$v'$}}
\put(114,12){\makebox(0,0)[cc]{$=\di v'$.}}
\end{picture}
\end{align*}
\end{pf}

\begin{cor}
$\di^2=0$
\end{cor}

Indeed,
\[ \di^2 = \di v^iv_i = v^i\di v_i = -q^2v^iv_i\di = -q^2\di^2 .\]

\begin{defn}
A differential of degree (1,1) $D=D_V :K(V)\to K(V)$ in the  
$\Zp\times\Zp$-graded vector space $\K(V)$ is defined by left 
multiplication by $\di$. Its $\Zp\times\Zp$-graded space of cohomologies
\[ \Ber(V) = H^*(K(V),D) \]
is called the \em{Berezinian} of $V$.
\end{defn}

\subsection{The Hecke sum}
The $\check R$-matrix $\CR$ from \secref{doubledR} is a special element 
of a more general family of $\check R$-matrices studied by Majid and Markl 
\cite{MaMa}. 

\begin{thm}\label{MajMar}
(a) Let $\RB:V\tens V\to V\tens V$, $\RB':U\tens U\to U\tens U$ be two
diagonalizable $\check R$-matrices with the eigenvalues $q$ and $-q^{-1}$,
and let $Q:U\tens V\to V\tens U$ be a bijective linear map satisfying
the equations
\begin{align*}
(1\tens Q)(Q\tens1)(1\tens \RB) &= (\RB\tens1)(1\tens Q)(Q\tens1) ,\\
(Q\tens1)(1\tens Q)(\RB'\tens1) &= (1\tens \RB')(Q\tens1)(1\tens Q). 
\end{align*}
Then the linear map $\CR:X\tens X\to X\tens X$, $X=V\oplus U$, given by
\[ \CR= \begin{pmatrix} 
\RB&0&0&0 \\ 0&0&Q^{-1}&0 \\ 0&Q&q-q^{-1}&0 \\ 0&0&0&\RB' \end{pmatrix} 
\]
is a diagonalizable $\check R$-matrix with eigenvalues $q$ and $-q^{-1}$.

(b) If in addition $\RB^\sharp$, $\RB^{\prime\sharp}$, $Q^\sharp$, 
$Q^{-1\sharp}$ are all invertible, then also $\CR^\sharp$, 
$\CR^{-1\sharp}$ are.
\end{thm}

\begin{pf}
(a) Reduces to a theorem from \cite{MaMa} and can be checked 
straightforwardly.

(b) The matrix
\be\label{Rsharp} 
\CR^\sharp= \begin{pmatrix} \RB^\sharp&0&0&(q-q^{-1})\X \\ 0&0&Q^\sharp&0 
\\ 0&Q^{-1\sharp}&0&0 \\ 0&0&0&\RB^{\prime\sharp} \end{pmatrix},
\end{equation}
where $\X=1^\sharp =(V\tens V\pti@>\ev>> k @>\coev>> U\pti\tens U)$, 
$\X_{jl}^{ik} = \delta^{ik} \delta_{jl}$, is obviously invertible. By
\thmref{invertible} $\CR^{-1\sharp}$ is also invertible.
\end{pf}

The space $X$ equipped with $\CR$ is denoted $V\oplus_QU$ and is called
the {\em Hecke sum}.

\subsection{Koszul complex of a Hecke sum}
Let us analyze in detail the structure of the algebra $\K(V\oplus_Q U)$.
The algebra $\Ex^\bullet(V\oplus_Q U)$ is a quotient of 
$T^\bullet(V\oplus U)$ by the relations
\begin{alignat*}2
\RB(v\tens\bar v) &= -q^{-1}v\tens\bar v &&\qquad\text{for }v,\bar v\in 
V,\\
Q(u\tens v) &= -qu\tens v && \qquad\text{for }v\in V,u\in U,\\
\RB'(u\tens\bar u) &= -q^{-1} u\tens\bar u &&\qquad\text{for }u,\bar u\in 
U.
\end{alignat*}
Since
\[ \CR^{\sharp\sharp} = \begin{pmatrix} 
\RB^{\sharp\sharp}&0&0&0 \\ 0&q-q^{-1}&Q^{-1\sharp\sharp}&0 \\ 
0&Q^{\sharp\sharp}&0&0 \\ 0&0&0&\RB^{\prime\sharp\sharp} \end{pmatrix} 
\]
the algebra $\Sy^\bullet((V\oplus_Q U)\pti)$ is a quotient of 
$T^\bullet(V\pti\oplus U\pti)$ by the relations
\begin{alignat*}2
\RB^{\sharp\sharp}(v'\tens\bar v') &= qv'\tens\bar v' 
&& \qquad\text{for }v',\bar v'\in V\pti,\\
Q^{\sharp\sharp}(u'\tens v') &= qu'\tens v' 
&& \qquad\text{for }v'\in V\pti,u'\in U\pti,\\
\RB^{\prime\sharp\sharp}(u'\tens\bar u') &= q u'\tens\bar u' 
&& \qquad\text{for }u',\bar u'\in U\pti.
\end{alignat*}
The commutation relations between $x\in V\oplus_Q U$ and 
$x'\in (V\oplus_Q U)\pti$ in $\K(V\oplus_Q U)$ are 
\[ \CR^\sharp(x\tens x') = qx\tens x' .\]
In particular, \eqref{Rsharp} implies that $\K(U)$ is a subalgebra of 
$\K(V\oplus_Q U)$, but $\K(V)$ is not. In fact, the following commutation
relation between $v\in V$ and $v'\in V\pti$ holds in $\K(V\oplus_Q U)$
\[ \RB^\sharp(v\tens v') + (q-q^{-1})\<v,v'\>\di_U = qv\tens v' ,\]
where $\di_U = u^ju_j$.

The factorization property \eqref{fac} applied to $V\oplus_Q U$ together
with factorization properties of $\Sy^\bullet((V\oplus_Q U)\pti)$ and
$\Ex^\bullet(V\oplus_Q U)$ imply that
\be\label{facVU}
\Sy(V\pti)\tens\Ex(V)\tens K(U) \to K(V\oplus_Q U)^{\tens3} 
@>m>> K(V\oplus_Q U)
\end{equation}
is an isomorphism of graded vector spaces. Proposition \ref{dv=vd} shows 
that \eqref{fac} is an isomorphism of complexes if the differential in
$\Sy^\bullet(V\pti)\tens\Ex^\bullet(V)$ is the insertion of 
$\di_V=v^i\tens v_i$ in the middle:
\be\label{DV}
D(y^1\dots y^m\tens y_1\dots y_n) = y^1\dots y^mv^i\tens v_iy_1\dots y_n.
\end{equation}
Composing \eqref{fac} and \eqref{facVU} we get an isomorphism of
$\Zp\times\Zp$-graded spaces
\[ \phi:\K(V)\tens\K(U) \to \Sy^\bullet(V\pti)\tens\Ex^\bullet(V)\tens\K(U)
\to \K(V\oplus_Q U) .\]

\begin{thm} \label{Bermultipl}
(a) The map $\phi$ is an isomorphism of complexes, that is,
\[ D_{V\oplus_Q U}\circ\phi = \phi\circ (D_V\tens1 + (-1)^n\tens D_U):
K^{n,m}(V)\tens K^{k,l}(U) \to K^{n+k+1,m+l+1}(V\oplus_Q U) .\]

(b) The isomorphism $\phi^{-1}$ induces an isomorphism
\[ \Ber(V\oplus_Q U) \simeq \Ber(V)\tens\Ber(U) .\]
\end{thm}

\begin{pf}
(a) From the commutation relations in $K(V\oplus_Q U)$
\[ Q(u\tens v) = -qu\tens v,\qquad Q^\sharp(v\tens u') = qv\tens u' \]
we find for $v\in V$, $u\in U$, $u'\in U\pti$
\[ u\tens v = -q^{-1}Q(u\tens v),\qquad 
u'\tens v = qQ^{\sharp-1}(u'\tens v).\]
Hence,
\[ \di_Uv = u^iu_iv = -q^{-1}u^iQ(u_i\tens v) = 
-(Q^{\sharp-1}\tens1) (u^i\tens Q(u_i\tens v)) = -vu^iu_i = -v\di_U .\]

Using \propref{dv=vd} for $X=V\oplus_Q U$, the above formula and \eqref{DV}
we compute for arbitrary $v_a'\in V\pti$, $y_b\in V$, $u_c'\in U\pti$,
$z_d\in U$
\begin{align*}
D_X(v_1'\dots v_m'y_1\dots y_n& u_1'\dots u_l'z_1\dots z_k) = \\
&= \di_X v_1'\dots v_m'y_1\dots y_n u_1'\dots u_l'z_1\dots z_k \\
&= v_1'\dots v_m'\di_X  y_1\dots y_n u_1'\dots u_l'z_1\dots z_k \\
&= v_1'\dots v_m'(\di_V+\di_U)y_1\dots y_n u_1'\dots u_l'z_1\dots z_k 
\\
&= v_1'\dots v_m'\di_V y_1\dots y_n u_1'\dots u_l'z_1\dots z_k \\
&\qqquad+(-1)^n v_1'\dots v_m'y_1\dots y_n\di_U u_1'\dots u_l'z_1\dots 
z_k\\
&= D_V(v_1'\dots v_m'y_1\dots y_n) u_1'\dots u_l'z_1\dots z_k \\
&\qqquad +(-1)^n v_1'\dots v_m'y_1\dots y_n D_U(u_1'\dots u_l'z_1\dots 
z_k).
\end{align*}

(b) Follows from (a) by K\"unneth's theorem \cite{Kun}.
\end{pf}

\subsection{Dual Koszul complex}
Consider now the Koszul complex $\K(\pti V)$ of the left dual $\pti V$,
which is generated by $\pti V\oplus V$ and has the relations
\begin{alignat*}2
\RB^{\flat\flat}({}'v\tens{}'\bar v) &= -q^{-1}{}'v\tens{}'\bar v && 
\qquad\text{for }{}'v,{}'\bar v\in \pti V,\\
\RB^\flat({}'v\tens v) &= q{\,}'v\tens v 
&&\qquad\text{for }v\in V,{}'v\in\pti V,\\
\RB(v\tens\bar v) &= q v\tens\bar v &&\qquad\text{for } v,\bar v\in V.
\end{alignat*}
The middle commutation relation is
\[ v\tens{}'v= qc_{V,\pti V} (v\tens{}'v) .\]
It has a special element 
\[ \di' = c_{V,\pti V}(v_i\tens{}^iv) = q^{-1}(v_i)({}^iv) = 
q^{-1}\di_{\pti V} \in K^{1,1}(\pti V) .\]
We shall be interested in the differential $D_{\pti V}^R$ given by the 
right multiplication by $\di'$ (it differs from $D_{\pti V}$ by a power 
of $q$ with a sign depending on the grading, as \propref{dv=vd} shows).

We want to extend the pairing $\<,\> = \ev: \pti V\tens V\to k$ to a
pairing between Koszul complexes of $\pti V$ and $V$. The natural pairings
$T^n(\pti V)\tens T^n(V)\to k$, $T^n(V)\tens T^n(V\pti)\to k$ are also
denoted $\<,\>$, 
$\<y_n\dots y_1 , z_1\dots z_n\> = \<y_1,z_1\>\dots\<y_n,z_n\>$. We use
Jimbo's symmetrizer and antisymmetrizer \cite{Jim:U(gl)}
\begin{align}
\Sym_m &= \sum_{\sigma\in\SSS_m} q^{l(\sigma)-m(m-1)/2} \RB_\sigma \notag 
\\
&= \sum_{\sigma\in\SSS_m} q^{m(m-1)/2-l(\sigma)} (\RB^{-1})_\sigma
:T^m(X) \to T^m(X) ,\label{Sym} \\
\Ant_n &= \sum_{\sigma\in\SSS_n} (-1)^{l(\sigma)} 
q^{n(n-1)/2-l(\sigma)} \RB_\sigma \notag \\
&= \sum_{\sigma\in\SSS_n} (-1)^{l(\sigma)}q^{l(\sigma)-n(n-1)/2}
(\RB^{-1})_\sigma : T^n(X) \to T^n(X) ,\label{Ant}
\end{align}
satisfying $(\Sym_m)^2= [m]_q!\Sym_m$, $(\Ant_n)^2 = [n]_q!\Ant_n$.

The pairing
\[ T^n(\pti V)\tens T^m(V)\tens T^m(V\pti)\tens T^n(V) \to k \]
\[ a\tens b\tens c\tens f \longmapsto \<a,\Ant_n(f)\> \<\Sym_m(b),c\> 
=
\<\Ant_n(a),f\> \<b,\Sym_m(c)\> \]
vanishes on ideals of relations of symmetric and exterior algebras.
Therefore, it induces the pairing
\[ \pi: \Ex^n(\pti V)\tens \Sy^m(V)\tens \Sy^m(V\pti)\tens \Ex^n(V) \to 
k\]
or
\be\label{piKK} 
\pi:K^{n,m}(\pti V) \tens K^{n,m}(V) \to k .
\end{equation}
When $q$ is not a root of unity 
this pairing is non-degenerate by the theory of Iwahori--Hecke algebras. 
In this case the natural map $\im\Sym_m\to\Sy^m(V)$ is an 
isomorphism and $\Sym_m$ is proportional to a projection. When $q$ is 
a 
root of unity the pairing is usually degenerate.

Similarly, the pairing
\[ T^m(V)\tens T^n(\pti V)\tens T^n(V) \tens T^m(V\pti) \to k \]
\[ b\tens a\tens f\tens c \longmapsto \<a,\Ant_n(f)\> \<\Sym_m(b),c\> 
\]
induces a pairing
\[ \Sy^m(V)\tens \Ex^n(\pti V)\tens \Ex^n(V) \tens \Sy^m(V\pti) \to k 
.\]
Interpreted as a pairing
\[ K^{n,m}(\pti V) \tens K^{n,m}(V) \to k \]
it coincides with $\pi$ from \eqref{piKK}. Indeed, using the commutation
relations \eqref{qc} and 
\[ {}'v\tens v = q^{-1} (c_{V,\pti V})^{-1} ({}'v\tens v) \]
for $'v\in\pti V$, $v\in V$, we can write both pairings as quotients of
equal maps 
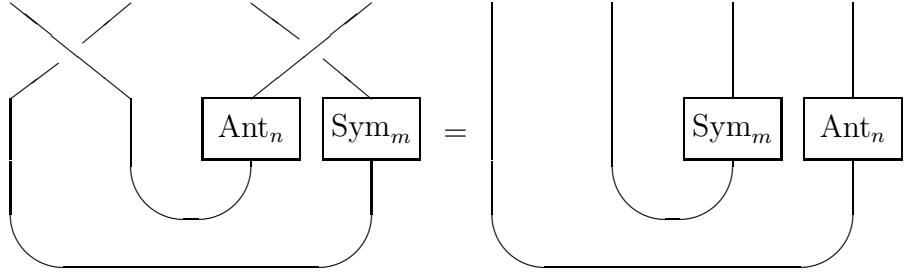
\[ T^n(\pti V)\tens T^m(V)\tens T^m(V\pti)\tens T^n(V) \to k \]
\[
\unitlength=0.8mm
\begin{picture}(148,45)
\put(112,18){\framebox(16,10)[cc]{$\Sym_m$}}
\put(32,18){\framebox(16,10)[cc]{$\Ant_n$}}
\put(30,18){\oval(20,20)[b]}
\put(30,17.50){\oval(60,35)[b]}
\put(20,18){\line(0,1){10}}
\put(20,28){\line(-5,4){20}}
\put(0,18){\line(0,1){10}}
\put(0,28){\line(4,3){8}}
\put(20,44){\line(-6,-5){8}}
\put(40,28){\line(5,4){20}}
\put(60,28){\line(-6,5){8}}
\put(40,44){\line(4,-3){8}}
\put(74,22){\makebox(0,0)[cc]{$=$}}
\put(110,18){\oval(20,20)[b]}
\put(110,17.50){\oval(60,35)[b]}
\put(52,18){\framebox(16,10)[cc]{$\Sym_m$}}
\put(132,18){\framebox(16,10)[cc]{$\Ant_n$}}
\put(140,44){\line(0,-1){16}}
\put(120,28){\line(0,1){16}}
\put(100,18){\line(0,1){26}}
\put(80,44){\line(0,-1){26}}
\end{picture}
\]

\begin{prop}\label{DptiVprimeR}
(a) The map
\begin{multline} 
D_{\pti V}^{\prime R}= \sum_{k=1}^m \sum_{l=1}^n (-1)^{n-l} q^{k+l-m-1} 
\unitlength=0.75mm
\makebox[60mm][l]{
\raisebox{-13.5mm}[15mm][15mm]{
\put(0,30){\line(0,-1){30}}
\put(20,0){\line(0,1){30}}
\put(30,30){\line(0,-1){30}}
\put(10,30){\line(6,-5){8}}
\put(40,5){\line(-6,5){7}}
\put(28,15){\line(-6,5){6}}
\put(40,5){\line(6,5){30}}
\put(80,30){\line(0,-1){30}}
\put(60,0){\line(0,1){19}}
\put(60,30){\line(0,-1){6}}
\put(50,30){\line(0,-1){14}}
\put(50,11){\line(0,-1){11}}
\put(0,34){\makebox(0,0)[cc]{$1$}}
\put(10,34){\makebox(0,0)[cc]{$l$}}
\put(30,34){\makebox(0,0)[cc]{$n$}}
\put(50,34){\makebox(0,0)[cc]{$1$}}
\put(70,34){\makebox(0,0)[cc]{$k$}}
\put(80,34){\makebox(0,0)[cc]{$m$}}
}}
 \label{DV'R} \\
: T^n(\pti V)\tens T^m(V) \to T^{n-1}(\pti V)\tens T^{m-1}(V) \notag
\end{multline}
composed with the projection 
$T^{n-1}(\pti V)\tens T^{m-1}(V) \to \Ex^{n-1}(\pti V) \tens \Sy^{m-1}(V)$
factorizes through a unique map
\[ D_{\pti V}^{\prime R}:\Ex^n(\pti V)\tens \Sy^m(V) \to 
\Ex^{n-1}(\pti V) \tens \Sy^{m-1}(V) .\]

(b) The map $D_{\pti V}^{\prime R}:K^{n,m}(\pti V) \to K^{n-1,m-1}(\pti 
V)$ 
so constructed is a transpose of $D_V$ in the following sense
\be\label{pipi}
\pi(D_{\pti V}^{\prime R}(a),b) = \pi(a,D_V(b))
\end{equation}
for $a\in K^{n,m}(\pti V)$, $b\in K^{n-1,m-1}(V)$.

(c) $(D_{\pti V}^{\prime R})^2 =0: K^{n,m}(\pti V) \to K^{n-2,m-2}(\pti 
V)$.
\end{prop}

\begin{pf}
(a) Using the identity $\RB-\RB^{-1}-q+q^{-1}=0$ one can show that the 
ideal of relations 
$[T^n(\pti V)\cap(\im(\RB^{\flat\flat}+q^{-1}))]\tens T^m(V)$ goes to 
$[T^{n-1}(\pti V)\cap(\im(\RB^{\flat\flat}+q^{-1}))]\tens T^{m-1}(V)$,
that is to itself, and $T^n(\pti V)\tens[(\im(\RB-q))\cap T^m(V)]$
goes to $T^{n-1}(\pti V)\tens[(\im(\RB-q))\cap T^{m-1}(V)]$.

(b) Equation \eqref{pipi} follows from the equation
\begin{multline*}
\left( T^n(\pti V)\tens T^m(V) @>D_{\pti V}^{\prime R}>> 
T^{n-1}(\pti V)\tens T^{m-1}(V) \to \right. \\ 
\left.@>\Ant_{n-1}\tens\Sym_{m-1}>> T^{n-1}(\pti V)\tens T^{m-1}(V)\right)=
\end{multline*}
\begin{multline*}
= \left( T^n(\pti V)\tens T^m(V) @>\Ant_n\tens\Sym_m>> 
T^n(\pti V)\tens T^m(V) = \right. \\
\left. = T^{n-1}(\pti V)\tens\pti V\tens V\tens T^{m-1}(V) 
@>1\tens\ev\tens1>> T^{n-1}(\pti V)\tens T^{m-1}(V) \right)
\end{multline*}
Substituting for $\Sym_m$ and $\Ant_n$ in the right hand side their 
expressions \eqref{Sym} and \eqref{Ant}, we reorder the sum, picking up
such $\sigma\in\SSS_m$ that $\sigma(k)=1$ and such $\sigma\in\SSS_n$ that 
$\sigma(l)=1$. The sum in the left hand side with fixed $k,l$ equals the
$(k,l)^{\text{th}}$ summand of $D_{\pti V}^{\prime R}$ multiplied by 
$\Ant_{n-1}\tens\Sym_{m-1}$.

(c) $D_V$ is a differential; this explains why $D_{\pti V}^{\prime R}$ 
is.
The proof consists of straightforward calculation.
\end{pf}

\begin{prop}
(a) The map
\begin{multline} 
\sum_{k=1}^m \sum_{l=1}^n (-1)^{l-1} q^{k+l-n-1} 
\Dprime  \\
: T^m(V\pti)\tens T^n(V) \to T^{m-1}(V\pti)\tens T^{n-1}(V) \to 
\Sy^{m-1}(V\pti)\tens \Ex^{n-1}(V) \label{D'V}
\end{multline}
factorizes through a unique map
\[ D'_V:\Sy^m(V\pti)\tens \Ex^n(V) \to \Sy^{m-1}(V\pti)\tens \Ex^{n-1}(V).\]

(b) The constructed map $D'_V:K^{n,m}(V) \to K^{n-1,m-1}(V)$ is a transpose
of $D_V^R$ in the following sense:
\[
\pi(D_V^R(a),b) = \pi(a,D'(b))
\]
for $a\in K^{n-1,m-1}(\pti V)$, $b\in K^{n,m}(V)$.

(c) $D^{\prime2} =0$.
\end{prop}

The proof is similar to that of \propref{DptiVprimeR}.

\begin{prop}\label{diffdiff}
The differentials are related as follows:
\begin{alignat*}2
D_V &= (-1)^n q^{2n+1} D_V^R &&: K^{n,m}(V) \to K^{n+1,m+1}(V) \\
D_V^{\prime R} &= (-1)^{n+1} q^{2n-1} D_V' &&: K^{n,m}(V) \to K^{n-1,m-1}(V)
\end{alignat*}
\end{prop}

\begin{pf} The first formula follows from \propref{dv=vd}. To get the 
second formula we represent $K^{n,m}(V)$ as $\Sy^m(V\pti) \tens\Ex^n(V)$.
Then $D_V^{\prime R}$ equals 
\begin{multline*}
\Sy^m(V\pti)\tens\Ex^n(V) @>q^{mn}c_{\tau(m,n)}>> 
\Ex^n(V)\tens\Sy^m(V\pti) @>D_V^{\prime R}>> \\
\to \Ex^{n-1}(V)\tens\Sy^{m-1}(V\pti) 
@>q^{-(n-1)(m-1)}(c^{-1})_{\tau(n-1,m-1)}>> 
\Sy^{m-1}(V\pti)\tens\Ex^{n-1}(V)
\end{multline*}
where the middle arrow is given by \eqref{DV'R}. Drawing this map, we 
reduce it to a factor map of
\begin{multline*} 
\sum_{k=1}^m \sum_{l=1}^n (-1)^{n-l} q^{k+l+n-2} 
\Dprime  \\
: T^m(V\pti)\tens T^n(V) \to T^{m-1}(V\pti)\tens T^{n-1}(V) \to 
\Sy^{m-1}(V\pti)\tens \Ex^{n-1}(V).
\end{multline*}
This differs from \eqref{D'V} by a power of $q$ with a sign.
\end{pf}

\subsection{Differentials on a Hecke sum}
It is possible to express the differentials $D^{\prime R}$, $D'$, $D_R$ 
on
a Hecke sum in terms of those on summands.

\begin{thm}\label{thmD'R}
The maps
\[ \psi: \Ex^p(\pti U)\tens\Sy^r(U) \tens K^{n,m}(\pti V) \to
\K(\pti(V\oplus_Q U))^{\tens3} @>m>> K^{p+n,r+m}(\pti(V\oplus_Q U)) \]
combine into an isomorphism of graded vector spaces. The differential 
$D^{\prime R}$ satisfies
\begin{multline*}
D_{\pti X}^{\prime R}\circ\psi = 
\psi\circ \big(q^{p+r}\tens D_{\pti V}^{\prime R}
+D_{\pti U}^{\prime R}\tens(-1)^nq^{n+m}\big) \\
: K^{p,r}(\pti U)\tens K^{n,m}(\pti V) @>>> K^{p+n-1,r+m-1}(\pti X ). 
\end{multline*}
\end{thm}

\begin{pf}
We have to compute the following expression for $D_{\pti X}^{\prime R}$:
\[\begin{CD}
\Ex^p(\pti U)\tens\Sy^r(U)\tens\Ex^n(\pti V)\tens\Sy^m(V) \\
@V1\tens q^{rn}c_{\tau(r,n)}\tens1VV \\
\Ex^p(\pti U)\tens\Ex^n(\pti V)\tens\Sy^r(U)\tens\Sy^m(V) \\
@VD_{\pti X}^{\prime R}VV \\
\Ex^{p-1}(\pti U)\tens\Ex^n(\pti V)\tens
\Sy^{r-1}(U)\tens\Sy^m(V)\oplus\hspace*{6cm}\\
\hspace*{6cm}\oplus\Ex^p(\pti U)\tens\Ex^{n-1}(\pti V)\tens
\Sy^r(U)\tens\Sy^{m-1}(V)\oplus\dots\\
@V1\tens q^{-n(r-1)}(c^{-1})_{\tau(n,r-1)}\tens1\oplus V
1\tens q^{-(n-1)r}(c^{-1})_{\tau(n-1,r)}\tens1\oplus\dots V \\
\Ex^{p-1}(\pti U)\tens\Sy^{r-1}(U)\tens
\Ex^n(\pti V)\tens\Sy^m(V)\oplus\hspace*{6cm}\\
\hspace*{6cm}\oplus\Ex^p(\pti U)\tens\Sy^r(U)\tens
\Ex^{n-1}(\pti V)\tens\Sy^{m-1}(V)\oplus\dots
\end{CD}\]
where we have singled out two summands from four (or even more) possible.
In fact these two suffice, as calculation shows:
{\allowdisplaybreaks
\begin{align*}
D_{\pti X}^{\prime R} &= \sum_{a=1}^p \sum_{b=1}^r (-1)^{p+n-a}
q^{b+a-r-m-1+rn-n(r-1)}
\makebox[60mm][l]{
\raisebox{-22mm}[23mm][23mm]{
\unitlength=0.75mm
\put(30,16){\line(0,1){20}}
\put(69,51){\line(0,-1){50}}
\put(79,1){\line(0,1){50}}
\put(30,36){\line(2,1){30}}
\put(60,1){\line(-2,1){30}}
\put(50,1){\line(-2,1){30}}
\put(20,16){\line(0,1){20}}
\put(20,36){\line(2,1){30}}
\put(20,51){\line(2,-1){14}}
\put(50,36){\line(-2,1){8}}
\put(60,36){\line(-2,1){14}}
\put(30,51){\line(2,-1){8}}
\put(50,16){\line(-2,-1){8}}
\put(60,36){\line(0,-1){20}}
\put(60,16){\line(-2,-1){14}}
\put(20,1){\line(2,1){14}}
\put(30,1){\line(2,1){8}}
\put(11,51){\line(0,-1){50}}
\put(1,1){\line(0,1){50}}
\put(1,56){\makebox(0,0)[cc]{$1$}}
\put(11,56){\makebox(0,0)[cc]{$p$}}
\put(20,56){\makebox(0,0)[cc]{$1$}}
\put(30,56){\makebox(0,0)[cc]{$r$}}
\put(50,56){\makebox(0,0)[cc]{$1$}}
\put(60,56){\makebox(0,0)[cc]{$n$}}
\put(69,56){\makebox(0,0)[cc]{$1$}}
\put(79,56){\makebox(0,0)[cc]{$m$}}
\put(25,51){\line(2,-1){12}}
\put(31,26){\line(1,0){24}}
\put(55,26){\line(0,1){10}}
\put(55,36){\line(-2,1){12}}
\put(50,36){\line(0,-1){9}}
\put(50,25){\line(0,-1){9}}
\put(19,26){\line(-1,0){7}}
\put(10,26){\line(-1,0){4}}
\put(6,26){\line(0,1){25}}
\put(6,56){\makebox(0,0)[cc]{$a$}}
\put(25,56){\makebox(0,0)[cc]{$b$}}
}}
\\
&+ \sum_{b=1}^r \sum_{c=1}^n (-1)^{n-c} q^{b+p+c-r-m-1+rn-(n-1)(r-1)}
\makebox[60mm][l]{
\raisebox{-22mm}[23mm][23mm]{
\unitlength=0.75mm
\put(30,16){\line(0,1){20}}
\put(69,51){\line(0,-1){50}}
\put(79,1){\line(0,1){50}}
\put(30,36){\line(2,1){30}}
\put(60,1){\line(-2,1){30}}
\put(50,1){\line(-2,1){30}}
\put(20,16){\line(0,1){20}}
\put(20,36){\line(2,1){30}}
\put(20,51){\line(2,-1){14}}
\put(50,36){\line(-2,1){8}}
\put(60,36){\line(-2,1){14}}
\put(30,51){\line(2,-1){8}}
\put(50,16){\line(-2,-1){8}}
\put(60,36){\line(0,-1){20}}
\put(60,16){\line(-2,-1){14}}
\put(20,1){\line(2,1){14}}
\put(30,1){\line(2,1){8}}
\put(11,51){\line(0,-1){50}}
\put(1,1){\line(0,1){50}}
\put(1,56){\makebox(0,0)[cc]{$1$}}
\put(11,56){\makebox(0,0)[cc]{$p$}}
\put(20,56){\makebox(0,0)[cc]{$1$}}
\put(30,56){\makebox(0,0)[cc]{$r$}}
\put(50,56){\makebox(0,0)[cc]{$1$}}
\put(60,56){\makebox(0,0)[cc]{$n$}}
\put(69,56){\makebox(0,0)[cc]{$1$}}
\put(79,56){\makebox(0,0)[cc]{$m$}}
\put(55,51){\line(-2,-1){30}}
\put(25,36){\line(0,-1){10}}
\put(25,26){\line(1,0){4}}
\put(31,26){\line(1,0){24}}
\put(55,26){\line(0,1){10}}
\put(55,36){\line(-2,1){12}}
\put(25,51){\line(2,-1){12}}
\put(50,36){\line(0,-1){9}}
\put(50,25){\line(0,-1){9}}
\put(25,56){\makebox(0,0)[cc]{$b$}}
\put(55,56){\makebox(0,0)[cc]{$c$}}
}}
\\
&+ \sum_{c=1}^n \sum_{d=1}^m (-1)^{n-c} q^{d+p+c-m-1+rn-(n-1)r}
\makebox[60mm][l]{
\raisebox{-22mm}[23mm][23mm]{
\unitlength=0.75mm
\put(30,16){\line(0,1){20}}
\put(79,1){\line(0,1){50}}
\put(30,36){\line(2,1){30}}
\put(60,1){\line(-2,1){30}}
\put(50,1){\line(-2,1){30}}
\put(20,16){\line(0,1){20}}
\put(20,36){\line(2,1){30}}
\put(20,51){\line(2,-1){14}}
\put(50,36){\line(-2,1){8}}
\put(60,36){\line(-2,1){14}}
\put(30,51){\line(2,-1){8}}
\put(50,16){\line(-2,-1){8}}
\put(60,16){\line(-2,-1){14}}
\put(20,1){\line(2,1){14}}
\put(30,1){\line(2,1){8}}
\put(11,51){\line(0,-1){50}}
\put(1,1){\line(0,1){50}}
\put(1,56){\makebox(0,0)[cc]{$1$}}
\put(11,56){\makebox(0,0)[cc]{$p$}}
\put(20,56){\makebox(0,0)[cc]{$1$}}
\put(30,56){\makebox(0,0)[cc]{$r$}}
\put(50,56){\makebox(0,0)[cc]{$1$}}
\put(60,56){\makebox(0,0)[cc]{$n$}}
\put(69,56){\makebox(0,0)[cc]{$1$}}
\put(79,56){\makebox(0,0)[cc]{$m$}}
\put(55,51){\line(-2,-1){30}}
\put(25,36){\line(0,-1){10}}
\put(25,26){\line(1,0){4}}
\put(31,26){\line(1,0){43}}
\put(74,26){\line(0,1){25}}
\put(69,51){\line(0,-1){24}}
\put(69,25){\line(0,-1){24}}
\put(60,16){\line(0,1){9}}
\put(60,27){\line(0,1){9}}
\put(50,36){\line(0,-1){9}}
\put(50,25){\line(0,-1){9}}
\put(55,56){\makebox(0,0)[cc]{$c$}}
\put(74,56){\makebox(0,0)[cc]{$d$}}
}}
\\
&+ \sum_{a=1}^p \sum_{d=1}^m (-1)^{n+p-a} q^{d+a-m-1+rn-nr}
\makebox[60mm][l]{
\raisebox{-22mm}[23mm][23mm]{
\unitlength=0.75mm
\put(30,16){\line(0,1){20}}
\put(79,1){\line(0,1){50}}
\put(30,36){\line(2,1){30}}
\put(60,1){\line(-2,1){30}}
\put(50,1){\line(-2,1){30}}
\put(20,16){\line(0,1){20}}
\put(20,36){\line(2,1){30}}
\put(20,51){\line(2,-1){14}}
\put(50,36){\line(-2,1){8}}
\put(60,36){\line(-2,1){14}}
\put(30,51){\line(2,-1){8}}
\put(50,16){\line(-2,-1){8}}
\put(60,16){\line(-2,-1){14}}
\put(20,1){\line(2,1){14}}
\put(30,1){\line(2,1){8}}
\put(11,51){\line(0,-1){50}}
\put(1,1){\line(0,1){50}}
\put(1,56){\makebox(0,0)[cc]{$1$}}
\put(11,56){\makebox(0,0)[cc]{$p$}}
\put(20,56){\makebox(0,0)[cc]{$1$}}
\put(30,56){\makebox(0,0)[cc]{$r$}}
\put(50,56){\makebox(0,0)[cc]{$1$}}
\put(60,56){\makebox(0,0)[cc]{$n$}}
\put(69,56){\makebox(0,0)[cc]{$1$}}
\put(79,56){\makebox(0,0)[cc]{$m$}}
\put(6,51){\line(0,-1){25}}
\put(6,26){\line(1,0){4}}
\put(12,26){\line(1,0){7}}
\put(31,26){\line(1,0){43}}
\put(74,26){\line(0,1){25}}
\put(69,51){\line(0,-1){24}}
\put(69,25){\line(0,-1){24}}
\put(60,16){\line(0,1){9}}
\put(60,27){\line(0,1){9}}
\put(50,36){\line(0,-1){9}}
\put(50,25){\line(0,-1){9}}
\put(74,56){\makebox(0,0)[cc]{$d$}}
\put(6,56){\makebox(0,0)[cc]{$a$}}
}}
\end{align*}
Here all crossings are interpreted via $\CR$ which reduces to $\RB$ and
$\RB'$ in the first and the third sum, giving 
$D_{\pti U}^{\prime R}\tens(-1)^nq^{n-m}$ and 
$q^{p+r}\tens D_{\pti V}^{\prime R}$ respectively. The second sum
vanishes, but the fourth does not. Indeed, using the equation
\[ \CR= Q\oplus(q-q^{-1}) : U\tens V \to V\tens U\oplus U\tens V \]
we find the fourth sum
\begin{multline*}
\sum_{a=1}^p \sum_{d=1}^m (-1)^{n+p-a} q^{d+a-m-1}
\unitlength=0.75mm
\makebox[59mm][l]{
\raisebox{-22mm}[23mm][23mm]{
\put(74,1){\line(0,1){50}}
\put(11,51){\line(0,-1){50}}
\put(1,1){\line(0,1){50}}
\put(1,56){\makebox(0,0)[cc]{$1$}}
\put(11,56){\makebox(0,0)[cc]{$p$}}
\put(20,56){\makebox(0,0)[cc]{$1$}}
\put(30,56){\makebox(0,0)[cc]{$r$}}
\put(45,56){\makebox(0,0)[cc]{$1$}}
\put(55,56){\makebox(0,0)[cc]{$n$}}
\put(64,56){\makebox(0,0)[cc]{$1$}}
\put(74,56){\makebox(0,0)[cc]{$m$}}
\put(45,51){\line(0,-1){50}}
\put(55,1){\line(0,1){50}}
\put(6,56){\makebox(0,0)[cc]{$a$}}
\put(69,56){\makebox(0,0)[cc]{$d$}}
\put(6,51){\line(0,-1){40}}
\put(6,11){\line(1,0){4}}
\put(12,11){\line(1,0){5}}
\put(17,11){\line(2,1){26}}
\put(57,31){\line(2,1){12}}
\put(69,37){\line(0,1){14}}
\put(64,51){\line(0,-1){15}}
\put(64,33){\line(0,-1){32}}
\put(30,51){\line(0,-1){32}}
\put(30,16){\line(0,-1){15}}
\put(20,1){\line(0,1){10}}
\put(20,14){\line(0,1){37}}
}}
= \\
= (q-q^{-1}) \sum_{a=1}^p \sum_{d=1}^m \sum_{t=1}^r (-1)^{n+p-a} q^{d+a-m-1}
\unitlength=0.75mm
\makebox[59mm][l]{
\raisebox{-22mm}[23mm][23mm]{
\put(74,1){\line(0,1){50}}
\put(11,51){\line(0,-1){50}}
\put(1,1){\line(0,1){50}}
\put(1,56){\makebox(0,0)[cc]{$1$}}
\put(11,56){\makebox(0,0)[cc]{$p$}}
\put(20,56){\makebox(0,0)[cc]{$1$}}
\put(32,56){\makebox(0,0)[cc]{$r$}}
\put(45,56){\makebox(0,0)[cc]{$1$}}
\put(55,56){\makebox(0,0)[cc]{$n$}}
\put(64,56){\makebox(0,0)[cc]{$1$}}
\put(74,56){\makebox(0,0)[cc]{$m$}}
\put(45,51){\line(0,-1){50}}
\put(55,1){\line(0,1){50}}
\put(6,56){\makebox(0,0)[cc]{$a$}}
\put(69,56){\makebox(0,0)[cc]{$d$}}
\put(26,56){\makebox(0,0)[cc]{$t$}}
\put(26,51){\line(0,-1){28}}
\put(26,23){\line(-2,-1){10}}
\put(16,18){\line(-1,0){4}}
\put(10,18){\line(-1,0){4}}
\put(6,18){\line(0,1){33}}
\put(26,1){\line(0,1){18}}
\put(26,19){\line(2,1){18}}
\put(56,34){\line(2,1){13}}
\put(69,40.50){\line(0,1){10.50}}
\put(64,51){\line(0,-1){12}}
\put(64,37){\line(0,-1){36}}
\put(20,51){\line(0,-1){29}}
\put(20,18){\line(0,-1){17}}
\put(23,1){\line(0,1){19}}
\put(23,23){\line(0,1){28}}
\put(29,51){\line(0,-1){29}}
\put(29,19){\line(0,-1){18}}
\put(32,1){\line(0,1){19}}
\put(32,24){\line(0,1){27}}
\put(32,22){\circle{2.83}}
\put(29,20){\circle{2.83}}
}}
\end{multline*}
where encircled crossings are $Q:U\tens V \to V\tens U$. Using 
commutation relations in $K(\pti X)$ we get rid of those and some other
crossings, changing them to powers of $q$ with a sign. So the fourth
sum equals
\begin{align*}
(q-q^{-1}) &\sum_{a=1}^p \sum_{t=1}^r \sum_{d=1}^m (-1)^{n+p-a} 
q^{d+a-m-1+t-r+n+d-1}
\unitlength=0.75mm
\makebox[59mm][l]{
\raisebox{-22mm}[23mm][23mm]{
\put(64,51){\line(0,-1){50}}
\put(74,1){\line(0,1){50}}
\put(11,51){\line(0,-1){50}}
\put(1,1){\line(0,1){50}}
\put(1,56){\makebox(0,0)[cc]{$1$}}
\put(11,56){\makebox(0,0)[cc]{$p$}}
\put(20,56){\makebox(0,0)[cc]{$1$}}
\put(30,56){\makebox(0,0)[cc]{$r$}}
\put(45,56){\makebox(0,0)[cc]{$1$}}
\put(55,56){\makebox(0,0)[cc]{$n$}}
\put(64,56){\makebox(0,0)[cc]{$1$}}
\put(74,56){\makebox(0,0)[cc]{$m$}}
\put(45,51){\line(0,-1){50}}
\put(55,1){\line(0,1){50}}
\put(30,51){\line(0,-1){50}}
\put(6,56){\makebox(0,0)[cc]{$a$}}
\put(69,56){\makebox(0,0)[cc]{$d$}}
\put(25,56){\makebox(0,0)[cc]{$t$}}
\put(25,51){\line(0,-1){20}}
\put(25,31){\line(-1,-1){8}}
\put(17,23){\line(-1,0){5}}
\put(10,23){\line(-1,0){4}}
\put(6,23){\line(0,1){28}}
\put(20,51){\line(0,-1){23}}
\put(20,24){\line(0,-1){23}}
\put(69,51){\line(0,-1){50}}
}}
\\
&= (q-q^{-1})D_{\pti U}^{\prime R}\tens(-1)^nq^{n-m} \sum_{d=1}^m q^{2d-1}\\
&= D_{\pti U}^{\prime R}\tens(-1)^nq^{n-m}(q^{2m}-1) .
\end{align*}
}
Summing up everything we get
\begin{align*}
D_{\pti X}^{\prime R} &= D_{\pti U}^{\prime R}\tens(-1)^nq^{n-m} +
q^{p+r}\tens D_{\pti V}^{\prime R} + 
D_{\pti U}^{\prime R}\tens(-1)^nq^{n-m}(q^{2m}-1) \\
&= D_{\pti U}^{\prime R}\tens(-1)^nq^{n+m} + 
q^{p+r}\tens D_{\pti V}^{\prime R},
\end{align*}
omitting $\psi$.
\end{pf}

The differential $D_{\pti X}^{\prime R}$ is a transpose of $D_X$, so it 
seems that Theorems~\ref{Bermultipl} and \ref{thmD'R} contradict  each 
other. The following Proposition shows that there is no contradiction, 
and,
moreover, \thmref{thmD'R} can be deduced from \thmref{Bermultipl} and 
vice versa on the assumption that $q$ is not a root of unity. We preferred 
to give independent proofs, which work for any $q$.

\begin{prop}\label{pi=qpipi}
Let $x\in K^{n,m}(V)$, $y\in K^{p,r}(U)$, $z\in K^{p,r}(\pti U)$,
$w\in K^{n,m}(\pti V)$. Then
\[ \pi_X(\psi(z\tens w),\phi(x\tens y)) = q^{np+mr} \pi_V(w,x) \pi_U(z,y).\]
\end{prop}

\begin{pf} 
Let $a\in T^m(V\pti)$, $b\in T^n(V)$, $x=ab$, $c\in T^r(U\pti)$,
$d\in T^p(U)$, $y=cd$, $e\in T^p(\pti U)$, $f\in T^r(U)$, $z=ef$, 
$g\in T^n(\pti V)$, $h\in T^m(V)$, $w=gh$. Then
\begin{align*}
\pi_V(w,x) \pi_U(z,y) &= \pi_V(gh,ab) \pi_U(ef,cd) \\
&= \<h,\Sym_ma\> \<g,\Ant_nb\> \<f,\Sym_rc\> \<e,\Ant_pd\> .
\end{align*}
Denote
\begin{alignat*}2
\tilde c\tens\tilde b &:= (c^{-1})_{\tau(n,r)} (b\tens c) 
&&\in T^r(U\pti) \tens T^n(V) ,\\
\tilde g\tens\tilde f &:= c_{\tau(n,r)} (f\tens g) 
&&\in T^n(\pti V) \tens T^r(U)
\end{alignat*}
Then
\begin{alignat*}2
\phi(x\tens y) &= \phi(ab\tens cd) &&= q^{-nr} a\tilde c\tilde bd, \\
\psi(z\tens w) &= \psi(ef\tens gh) &&= q^{rn} e\tilde g\tilde fh,
\end{alignat*}
and
\be\label{pipsiphi}
\nquad \pi_X(\psi(z\tens w),\phi(x\tens y)) = \<e\tens\tilde g, 
\<\Sym_{r+m}(\tilde f\tens h),a\tens\tilde c\>\Ant_{n+p}(\tilde b\tens 
d)\>.
\end{equation}

Introduce symmetric and antisymmetric shuffling operators
\begin{align*}
\Symshf_{r,m} &= \sum_{\sigma\in\Shuffles_{r,m}} 
q^{rm-l(\sigma)} (\CR^{-1})_\sigma ,\\
\Antshf_{n,p} &= \sum_{\sigma\in\Shuffles_{n,p}} 
(-1)^{l(\sigma)} q^{np-l(\sigma)} \CR_\sigma.
\end{align*}
Then we have
\begin{align*}
\Sym_{r+m} &= \Symshf_{r,m} \circ \Sym_r\tens\Sym_m , \\
\Ant_{n+p} &= \Antshf_{n,p} \circ \Ant_n\tens\Ant_p .
\end{align*}

Plugging this into \eqref{pipsiphi}, we see that only one term with 
$\sigma=1$ contributes from each shuffling operator. Therefore
\begin{align*}
\pi_X(\psi(z\tens w),\phi(x\tens y)) &= q^{np+mr} \<e\tens\tilde g,
\<\Sym_r\tilde f\tens\Sym_m h),a\tens\tilde c\> 
\Ant_n\tilde b\tens\Ant_p d)\> \\
&= q^{np+mr} \<\Sym_mh,a\> 
\<\tilde g\tens\Sym_r\tilde f, \tilde c\tens\Ant_n\tilde b \>
\<e,\Ant_pd\> \\
&= q^{np+mr} \<h,\Sym_ma\> 
\<\Sym_rf\tens g, \Ant_nb\tens c\> \<e,\Ant_pd\> \\
&= q^{np+mr} \<h,\Sym_ma\> \<g,\Ant_nb\> \<\Sym_rf,c\> \<e,\Ant_pd\> \\
&= q^{np+mr} \pi_V(w,x) \pi_U(z,y).
\end{align*}
Here we used the equation
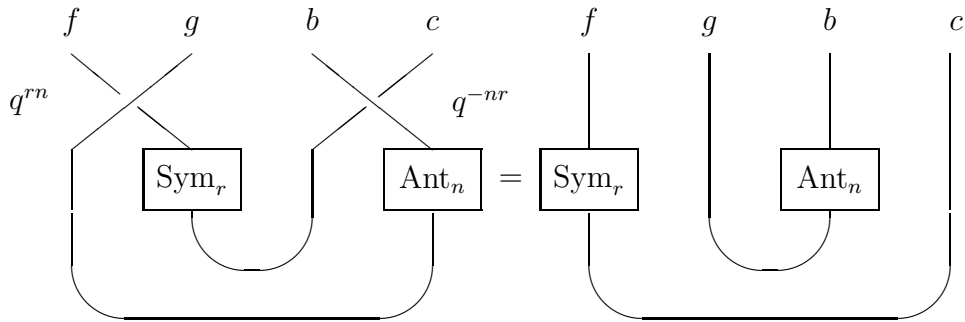
\[
\unitlength=0.8mm
\begin{picture}(156,51)
\put(127,18){\framebox(16,10)[cc]{$\Ant_n$}}
\put(61,18){\framebox(16,10)[cc]{$\Ant_n$}}
\put(39,18){\oval(20,20)[b]}
\put(39,17.50){\oval(60,35)[b]}
\put(9,18){\line(0,1){10}}
\put(125,18){\oval(20,20)[b]}
\put(125,17.50){\oval(60,35)[b]}
\put(21,18){\framebox(16,10)[cc]{$\Sym_r$}}
\put(87,18){\framebox(16,10)[cc]{$\Sym_r$}}
\put(135,28){\line(0,1){16}}
\put(115,18){\line(0,1){26}}
\put(9,48){\makebox(0,0)[cb]{$f$}}
\put(29,48){\makebox(0,0)[cb]{$g$}}
\put(49,48){\makebox(0,0)[cb]{$b$}}
\put(69,48){\makebox(0,0)[cb]{$c$}}
\put(95,48){\makebox(0,0)[cb]{$f$}}
\put(115,48){\makebox(0,0)[cb]{$g$}}
\put(135,48){\makebox(0,0)[cb]{$b$}}
\put(156,48){\makebox(0,0)[cb]{$c$}}
\put(2,36){\makebox(0,0)[cc]{$q^{rn}$}}
\put(77,36){\makebox(0,0)[cc]{$q^{-nr}$}}
\put(82,23){\makebox(0,0)[cc]{$=$}}
\put(155,18){\line(0,1){26}}
\put(95,28){\line(0,1){16}}
\put(69,28){\line(-5,4){20}}
\put(9,28){\line(5,4){20}}
\put(49,18){\line(0,1){10}}
\put(49,28){\line(5,4){9}}
\put(69,44){\line(-5,-4){9}}
\put(9,44){\line(6,-5){8}}
\put(29,28){\line(-5,4){9}}
\end{picture}
\]
\end{pf}

The result of this proposition can be stated as 
$\psi=q^{np+mr} {}^t\phi^{-1}$, and using this one can prove the 
equivalence of Theorems~\ref{Bermultipl} and \ref{thmD'R}, when the pairings 
$\pi$ are not degenerate, that is when $q$ is not a root of unity.

\subsection{8-dimensions}
The number which is called 8-dimension here was studied in
\cite{MajSoi:rank} under the name of rank. We believe that our term
is less confusing.

\begin{defn}
The 8-dimension of $V$ is the number
\[ \dim_8 V = (k @>\coev>> V\pti\tens V @>c>> V\tens V\pti @>\ev>> k) 
\equiv \dimeightV \in k \]
\end{defn}

The properties of 8-dimension for Hecke $R$-matrices are summarized 
in the following

\begin{prop}
(a) The addition formula for 8-dimensions
\[ \dim_8(V\oplus_QU) = \dim_8V + \dim_8U - (q-q^{-1})\dim_8V\,\dim_8U 
.\]

(b) The 8-dimension is related to the automorphism $V\to V$
\[ \nu_V^{-2}\equiv \nusquareVV = 1-(q-q^{-1})\dim_8V : V\to V .\]

(c) $\dim_8 V\pti = \dim_8 V$.
\end{prop}

\begin{pf}
(a) By definition
\[ \dim_8(V\oplus_QU) = (k @>\coev>> (V\oplus_QU)\pti\tens(V\oplus_QU)
@>\CR^{\sharp-1}>> (V\oplus_QU)\tens(V\oplus_QU)\pti @>\ev>> k) .\]
By \eqref{Rsharp} we get
\be\label{Rsharp-1}
\CR^{\sharp-1} = 
\begin{pmatrix} \RB^{\sharp-1}&0&0&A \\ 0&0&Q^{-1\sharp-1}&0 \\ 
0&Q^{\sharp-1}&0&0 \\ 0&0&0&\RB^{\prime\sharp-1} \end{pmatrix},
\end{equation}
where
\[ A = (q^{-1}-q) \left(V\pti\tens V @>\RB^{\sharp-1}>> V\tens V\pti @>\ev>>
k @>\coev>> U\pti\tens U @>\RB^{\prime\sharp-1}>> U\tens U\pti \right) 
.\]

Plugging this into the definition,
\begin{align*}
\dim_8(V\oplus_QU) 
&= (k @>\coev>> V\pti\tens V @>\RB^{\sharp-1}>> V\tens V\pti @>\ev>> k) 
\\
&\quad+ (q^{-1}-q)(k @>\coev>> V\pti\tens V @>\RB^{\sharp-1}>> 
V\tens V\pti @>\ev>> \\ 
&\qqquad\qqquad\qqquad @>\ev>> k @>\coev>> U\pti\tens U 
@>\RB^{\prime\sharp-1}>> U\tens U\pti @>\ev>> k) \\
&\quad+ (k @>\coev>> U\pti\tens U @>\RB^{\prime\sharp-1}>> 
U\tens U\pti @>\ev>> k) \\
&= \dim_8V + (q^{-1}-q)\dim_8V\,\dim_8U + \dim_8U .
\end{align*}

(b) Follows from the identity
\[
\unitlength=0.80mm
\begin{picture}(139,35)
\put(55,20){\oval(10,10)[t]}
\put(60,20){\line(0,-1){5}}
\put(65,15){\oval(10,10)[b]}
\put(50,20){\line(0,-1){5}}
\put(21,20){\line(0,-1){20}}
\put(22,21){\line(0,1){14}}
\put(26,33){\makebox(0,0)[cc]{$V$}}
\put(25,2){\makebox(0,0)[cc]{$V$}}
\put(25,17){\makebox(0,0)[lc]{$+(q-q^{-1})$}}
\put(54.50,15){\oval(9,10)[b]}
\put(5,20){\oval(10,10)[t]}
\put(10,20){\line(0,-1){5}}
\put(15,15){\oval(10,10)[b]}
\put(16,20){\oval(10,10)[t]}
\put(0,20){\line(0,-1){5}}
\put(4.50,15){\oval(9,10)[b]}
\put(100,20){\oval(10,10)[t]}
\put(105,20){\line(0,-1){5}}
\put(110,15){\oval(10,10)[b]}
\put(95,20){\line(0,-1){5}}
\put(99.50,15){\oval(9,10)[b]}
\put(65.50,20){\oval(9,10)[t]}
\put(110,20){\oval(8,10)[t]}
\put(70,20){\line(0,-1){5}}
\put(115,15){\line(0,1){20}}
\put(116,15){\line(0,-1){15}}
\put(73,24){\makebox(0,0)[cc]{$V$}}
\put(80,35){\line(0,-1){35}}
\put(84,33){\makebox(0,0)[cc]{$V$}}
\put(89,17){\makebox(0,0)[cc]{$=$}}
\put(135,35){\line(0,-1){35}}
\put(139,33){\makebox(0,0)[cc]{$V$}}
\put(119,33){\makebox(0,0)[cc]{$V$}}
\put(120,2){\makebox(0,0)[cc]{$V$}}
\put(127,17){\makebox(0,0)[cc]{$=$}}
\end{picture}
\]

(c) We have
\[ \dim_8\pti V =
\unitlength=0.50mm
\makebox[17mm][l]{
\raisebox{-10mm}[11mm][11mm]{
\put(15,30){\oval(20,20)[t]}
\put(15,10){\oval(20,20)[b]}
\put(25,30){\line(-1,-1){20}}
\put(25,10){\line(-1,1){9}}
\put(5,30){\line(1,-1){9}}
\put(31,30){\makebox(0,0)[cb]{$\pti V$}}
\put(0,30){\makebox(0,0)[cb]{$V$}}
}}
= V
\unitlength=0.50mm
\makebox[23mm][l]{
\raisebox{-5mm}{
\put(10,11){\oval(20,20)[l]}
\put(30,11){\oval(20,20)[r]}
\put(30,1){\line(-1,1){20}}
\put(10,1){\line(1,1){9}}
\put(30,21){\line(-1,-1){9}}
\put(9,22){\line(0,-1){2}}
\put(9,2){\line(0,-1){2}}
\put(31,0){\line(0,1){2}}
\put(31,20){\line(0,1){2}}
\put(15,21){\makebox(0,0)[cc]{$\pti V$}}
\put(25,21){\makebox(0,0)[cc]{$V\pti$}}
}}
V = \dimeightV = \dim_8 V .\]
\end{pf}

\begin{thm}\label{thmDR}
The differential $D^R$ satisfies
\begin{multline*}
D_{\pti X}^R\circ\psi = 
\psi\circ \big( \nu_U^{-2}q^{-2r}\tens D_{\pti V}^R
+D_{\pti U}^R\tens(-1)^nq^{-2n}\big) \\
: K^{p,r}(\pti U)\tens K^{n,m}(\pti V) @>>> K^{p+n+1,r+m+1}(\pti X ). 
\end{multline*}
\end{thm}

\begin{pf} The proof is quite analogous to that of \thmref{Bermultipl}. 
The only new point is the decomposition of $d_X'$ into  a $U$-part and
a $V$-part:
\begin{align*}
\di_X' &= c_{V,\pti V}(v_i\tens {}^iv + u_i\tens {}^iu) =
\CR^{\flat-1}(v_i\tens {}^iv + u_i\tens {}^iu) \\
&= \RB^{\flat-1}(v_i\tens {}^iv) + \RB^{\prime\flat-1} (u_i\tens {}^iu)
+ (q^{-1}-q) \dim_8\pti U\ \RB^{\flat-1}(v_i\tens {}^iv) \\
&= \RB^{\prime\flat-1} (u_i\tens {}^iu)
+ (1 + (q^{-1}-q) \dim_8\pti U) \RB^{\flat-1}(v_i\tens {}^iv) \\
&= \di_U' + \nu_U^{-2} \di_V'
\end{align*}
Here we used the explicit form of 
\[ \CR^{\flat-1} = 
\begin{pmatrix} \RB^{\flat-1}&0&0&0 \\ 0&0&Q^{-1\flat-1}&0 \\ 
0&Q^{\flat-1}&0&0 \\ F&0&0&\RB^{\prime\flat-1} \end{pmatrix},\]
where
\[ F = (q^{-1}-q) \big(U\tens\pti U @>\RB^{\prime\flat-1}>> \pti U\tens 
U 
@>\ev>> k @>\coev>> V\tens\pti V @>\RB^{\flat-1}>> \pti V\tens V \big) 
.\]
So the factor $\nu_U^{-2}$ appears in the statement.
\end{pf}

\begin{thm}\label{thmD'}
(a) The differential $D'$ satisfies
\begin{multline*}
D_{V\oplus_Q U}'\circ\phi = 
\phi\circ (D_V'\tens q^{p-r}\nu_U^{-2} + (-1)^n q^{m-n} \tens D_U') \\
: K^{n,m}(V)\tens K^{p,r}(U) \to K^{n+p-1,m+r-1}(V\oplus_Q U) .
\end{multline*}

(b) The map induced by $\phi$ 
\[ H^*(K(V), D_V') \tens H^*(K(U), D_U') 
\to H^*(K(V\oplus_Q U), D_{V\oplus_QU}') \]
is an isomorphism.
\end{thm}

\begin{pf}
(a) The proof is similar to that of \thmref{thmD'R}. When calculating
$D_X'$ one uses the identity
\[ \big(V\pti\tens V @>c_{X\pti,X}>> X\tens X\pti @>\ev>> k \big) =
\nu_U^{-2}\big(V\pti\tens V @>c_{V\pti,V}>> V\tens V\pti @>\ev>> k \big).\]
It is proven using \eqref{Rsharp-1}
\begin{align*}
&\big(V\pti\tens V @>\CR^{\sharp-1}>> X\tens X\pti @>\ev>> k \big) = \\
&= \big(V\pti\tens V @>\RB^{\sharp-1}>> V\tens V\pti @>\ev>> k\big) \\
&\quad+ (q^{-1}-q)\big(V\pti\tens V @>\RB^{\sharp-1}>> 
V\tens V\pti @>\ev>>  k @>\coev>> U\pti\tens U 
@>\RB^{\prime\sharp-1}>> U\tens U\pti @>\ev>> k\big) \\
&= \big(V\pti\tens V @>\RB^{\sharp-1}>> V\tens V\pti @>\ev>> k\big)
(1+(q^{-1}-q) \dim_8U) \\
&=\nu_U^{-2}\big(V\pti\tens V @>\RB^{\sharp-1}>> V\tens V\pti @>\ev>>k\big).
\end{align*}

(b) Renormalize the differentials introducing
\begin{alignat}2
\bar D_V &= \nu_U^{-2}q^{n-m}D_V' &&: K^{n,m}(V)\to K^{n-1.m-1}(V) ,\notag\\
\bar D_U &= q^{r-p}D_U' &&: K^{p,r}(U) \to K^{p-1.r-1}(U) ,\label{Dbars}\\
\bar D_X &= q^{r-p+n-m} \phi^{-1}D_X'\phi &&: 
K^{n,m}(V)\tens K^{p,r}(U) \to \K(V)\tens \K(U) .\notag
\end{alignat}
These are also differentials and $H^*(K(V),\bar D_V) = H^*(K(V), D_V')$,
$H^*(K(U),\bar D_U) = H^*(K(U), D_U')$, 
$H^*(K(V)\tens K(U),\bar D_X) \simeq H^*(K(X), D_X')$. But 
$\bar D_X=\bar D_V\tens1 + (-1)^n\tens\bar D_U$ and the K\"unneth 
theorem says that the tensor product of the two first spaces is 
isomorphic to the third one.
\end{pf}

When $q$ is not a root of unity, Theorems \ref{thmDR} and \ref{thmD'} 
can be deduced from each other, using duality and \propref{pi=qpipi}.

\subsection{The Laplacian}
\begin{defn}
The anticommutator
\[ L= D'D+DD' : K^{m,n}(V) \to K^{m,n}(V) \]
is called the \em{Laplacian} for the Koszul bidifferential complex.
\end{defn}

Theorems \ref{Bermultipl} and \ref{thmD'} have a straightforward 

\begin{cor}
The Laplacian for a Hecke sum can be calculated on
$K^{n,m}(V)\tens K^{p,r}(U)$ as
\[ L_X = L_V\tens q^{p-r} \nu_U^{-2} + q^{m-n}\tens L_U .\]
\end{cor}

The Laplacian is calculated for an arbitrary Hecke 
$\check R$-matrix in

\begin{thm} \label{Laplace=}
The Laplacian in $K(V)$ is multiplication by the number
\[ L\big|_{K^{m,n}(V)} = q^{n-m} \dim_8 V + [m-n]_q .\]
\end{thm}

\begin{pf}
We compute the lifting of $L$ to the space $T^m(V\pti)\tens T^n(V)$:
{\allowdisplaybreaks
\begin{align*}
L &= D'D+DD' \\
&= q^{-n-m} 
\unitlength=0.5mm
\makebox[39mm][l]{
\raisebox{-16mm}[21mm][17mm]{
\put(74,60){\line(0,-1){60}}
\put(54,60){\line(0,-1){60}}
\put(14,65){\makebox(0,0)[cb]{$1$}}
\put(34,65){\makebox(0,0)[cb]{$m$}}
\put(54,65){\makebox(0,0)[cb]{$1$}}
\put(74,65){\makebox(0,0)[cb]{$n$}}
\put(43,14.50){\oval(10,9)[r]}
\put(5,45){\oval(10,10)[t]}
\put(24,29){\line(0,1){9}}
\put(34,60){\line(0,-1){23}}
\put(34,34){\line(0,-1){9}}
\put(34,23){\line(0,-1){23}}
\put(24,41){\line(0,1){19}}
\put(14,60){\line(0,-1){15}}
\put(41,21){\line(-5,2){41}}
\put(0,37.33){\line(0,1){7.67}}
\put(14,42){\line(0,-1){9}}
\put(14,31){\line(0,-1){31}}
\put(24,0){\line(0,1){26}}
\put(43,10){\line(-1,0){1}}
\put(10,45){\line(5,-2){32}}
\put(42,32.17){\line(0,-1){22.17}}
\put(64,60){\line(0,-1){60}}
}} 
+ \sum_{l=1}^n (-1)^l q^{l-n-m} 
\unitlength=0.5mm
\makebox[39mm][l]{
\raisebox{-16mm}[21mm][17mm]{
\put(74,60){\line(0,-1){60}}
\put(54,60){\line(0,-1){60}}
\put(14,65){\makebox(0,0)[cb]{$1$}}
\put(34,65){\makebox(0,0)[cb]{$m$}}
\put(54,65){\makebox(0,0)[cb]{$1$}}
\put(64,65){\makebox(0,0)[cb]{$l$}}
\put(74,65){\makebox(0,0)[cb]{$n$}}
\put(43,15){\oval(10,10)[l]}
\put(43,14.50){\oval(10,9)[r]}
\put(5,45){\oval(10,10)[t]}
\put(10,45){\line(5,-2){41}}
\put(51,28.67){\line(0,-1){28.67}}
\put(43,20){\line(1,0){7}}
\put(52,20){\line(1,0){1}}
\put(55,20){\line(3,2){9}}
\put(64,26){\line(0,1){34}}
\put(24,29){\line(0,1){9}}
\put(34,60){\line(0,-1){23}}
\put(34,34){\line(0,-1){9}}
\put(34,23){\line(0,-1){23}}
\put(24,41){\line(0,1){19}}
\put(14,60){\line(0,-1){15}}
\put(41,21){\line(-5,2){41}}
\put(0,37.33){\line(0,1){7.67}}
\put(14,42){\line(0,-1){9}}
\put(14,31){\line(0,-1){31}}
\put(24,0){\line(0,1){26}}
}} 
\\
&\quad+ \sum_{k=1}^m q^{k-n-m} 
\unitlength=0.50mm
\makebox[39mm][l]{
\raisebox{-16mm}[21mm][17mm]{
\put(74,60){\line(0,-1){60}}
\put(54,60){\line(0,-1){60}}
\put(14,65){\makebox(0,0)[cb]{$1$}}
\put(24,65){\makebox(0,0)[cb]{$k$}}
\put(34,65){\makebox(0,0)[cb]{$m$}}
\put(54,65){\makebox(0,0)[cb]{$1$}}
\put(74,65){\makebox(0,0)[cb]{$n$}}
\put(43,14.50){\oval(10,9)[r]}
\put(5,45){\oval(10,10)[t]}
\put(0,45){\line(0,-1){45}}
\put(40,21){\line(-2,1){16}}
\put(24,29){\line(0,1){9}}
\put(34,60){\line(0,-1){23}}
\put(34,34){\line(0,-1){9}}
\put(34,23){\line(0,-1){23}}
\put(24,41){\line(0,1){19}}
\put(14,60){\line(0,-1){15}}
\put(14,42){\line(0,-1){42}}
\put(10,45){\line(5,-2){31}}
\put(41,32.67){\line(0,-1){22.67}}
\put(41,10){\line(1,0){2}}
\put(64,0){\line(0,1){60}}
}} 
+ \sum\begin{Sb}1\le k\le m\\1\le l\le n\end{Sb} (-1)^l q^{k+l-n-m}
\unitlength=0.5mm
\makebox[39mm][l]{
\raisebox{-16mm}[21mm][17mm]{
\put(74,60){\line(0,-1){60}}
\put(54,60){\line(0,-1){60}}
\put(14,65){\makebox(0,0)[cb]{$1$}}
\put(24,65){\makebox(0,0)[cb]{$k$}}
\put(34,65){\makebox(0,0)[cb]{$m$}}
\put(54,65){\makebox(0,0)[cb]{$1$}}
\put(64,65){\makebox(0,0)[cb]{$l$}}
\put(74,65){\makebox(0,0)[cb]{$n$}}
\put(43,15){\oval(10,10)[l]}
\put(43,14.50){\oval(10,9)[r]}
\put(5,45){\oval(10,10)[t]}
\put(10,45){\line(5,-2){41}}
\put(51,28.67){\line(0,-1){28.67}}
\put(0,45){\line(0,-1){45}}
\put(43,20){\line(1,0){7}}
\put(52,20){\line(1,0){1}}
\put(55,20){\line(3,2){9}}
\put(64,26){\line(0,1){34}}
\put(40,21){\line(-2,1){16}}
\put(24,29){\line(0,1){9}}
\put(34,60){\line(0,-1){23}}
\put(34,34){\line(0,-1){9}}
\put(34,23){\line(0,-1){23}}
\put(24,41){\line(0,1){19}}
\put(14,60){\line(0,-1){15}}
\put(14,42){\line(0,-1){42}}
}}
\\
&\quad+\sum\begin{Sb}1\le k\le m\\1\le l\le n\end{Sb} (-1)^{l-1}q^{k+l-n-m}
\unitlength=0.50mm
\makebox[39mm][l]{
\raisebox{-16mm}[21mm][17mm]{
\put(74,60){\line(0,-1){60}}
\put(54,60){\line(0,-1){60}}
\put(14,65){\makebox(0,0)[cb]{$1$}}
\put(24,65){\makebox(0,0)[cb]{$k$}}
\put(34,65){\makebox(0,0)[cb]{$m$}}
\put(54,65){\makebox(0,0)[cb]{$1$}}
\put(64,65){\makebox(0,0)[cb]{$l$}}
\put(74,65){\makebox(0,0)[cb]{$n$}}
\put(43,37){\oval(10,10)[l]}
\put(43,36.50){\oval(10,9)[r]}
\put(5,25){\oval(10,10)[t]}
\put(55,42){\line(3,2){9}}
\put(40,43){\line(-2,1){16}}
\put(24,51){\line(0,1){9}}
\put(64,48){\line(0,1){12}}
\put(43,42){\line(1,0){10}}
\put(34,60){\line(0,-1){13}}
\put(10,25){\line(5,-2){34}}
\put(44,11.33){\line(0,-1){11.33}}
\put(34,45){\line(0,-1){28}}
\put(34,14){\line(0,-1){14}}
\put(14,0){\line(0,1){22}}
\put(14,25){\line(0,1){35}}
\put(0,25){\line(0,-1){25}}
}}
\\
&= q^{-n-m}\dim_8 V + \sum_{l=1}^n (-1)^l q^{l-n-m} \nu_V^{-2} 
\unitlength=0.50mm
\makebox[32mm][l]{
\raisebox{-16mm}[21mm][17mm]{
\put(60,60){\line(0,-1){60}}
\put(40,60){\line(0,-1){60}}
\put(20,0){\line(0,1){60}}
\put(10,60){\line(0,-1){60}}
\put(0,0){\line(0,1){60}}
\put(0,65){\makebox(0,0)[cb]{$1$}}
\put(20,65){\makebox(0,0)[cb]{$m$}}
\put(40,65){\makebox(0,0)[cb]{$1$}}
\put(50,65){\makebox(0,0)[cb]{$l$}}
\put(60,65){\makebox(0,0)[cb]{$n$}}
\put(50,60){\line(-1,-3){9}}
\put(30,0){\line(1,3){9}}
}} 
\\
&\quad+ \sum_{k=1}^m q^{k-n-m} 
\unitlength=0.50mm
\makebox[37mm][l]{
\raisebox{-16mm}[21mm][17mm]{
\put(70,60){\line(0,-1){60}}
\put(60,0){\line(0,1){60}}
\put(50,60){\line(0,-1){60}}
\put(30,0){\line(0,1){60}}
\put(10,65){\makebox(0,0)[cb]{$1$}}
\put(20,65){\makebox(0,0)[cb]{$k$}}
\put(30,65){\makebox(0,0)[cb]{$m$}}
\put(50,65){\makebox(0,0)[cb]{$1$}}
\put(70,65){\makebox(0,0)[cb]{$n$}}
\put(20,60){\line(-1,-3){20}}
\put(10,0){\line(0,1){27}}
\put(10,33){\line(0,1){27}}
}}
\end{align*}
}
When this map projects to $\Sy^m(V\pti)\tens\Ex^n(V)$ the braiding 
can be replaced by its eigenvalue $q$ or $-q^{-1}$
\begin{align*}
L\big|_{K^{n,m}(V)} &= q^{-n-m}\dim_8 V - 
\nu_V^{-2} \sum_{l=1}^nq^{2l-1-n-m} + \sum_{k=1}^m q^{2k-1-n-m} \\
&= q^{-n-m}\big\{ \dim_8 V + (-1+(q-q^{-1})\dim_8 V) (q^{2n}-1)/(q-q^{-1})\\ 
&\qqquad\qqquad\qqquad + (q^{2m}-1)/(q-q^{-1}) \big\} \\
&= q^{-n-m}\big\{ q^{2n}\dim_8 V + (q^{2m}-q^{2n})/(q-q^{-1}) \big\} \\
&= q^{n-m} \dim_8 V + [m-n]_q .
\end{align*}
\end{pf}

Assume that the cohomology space $H^{n,m}(K(V),D) \subset \Ber V$ is
non-trivial. Since the restriction of $L$ to $H^{n,m}(K(V),D)$ vanishes,
this can happen only if $L\big|_{K^{n,m}(V)} = q^{n-m}\dim_8V + [m-n]_q 
=0$.
So we find
\begin{align*}
\dim_8V &= q^{m-n} [n-m]_q ,\\
\nu_V^2 &= q^{2(n-m)} .
\end{align*}
If, additionally, the category is a ribbon category with the ribbon twist
$\nu_V=q^{n-m}: V\to V$ (square root of $\nu_V^2$) we find a categorical
dimension of $V$ as
\[ \dim_\CC V \equiv 
\unitlength=0.70mm
\makebox[23mm][l]{
\raisebox{-14mm}[15mm][15mm]{
\put(15,30){\oval(20,20)[t]}
\put(15,10){\oval(20,20)[b]}
\put(30,35){\makebox(0,0)[cb]{$V$}}
\put(0,30){\makebox(0,0)[cb]{$V\pti$}}
\put(5,10){\line(5,3){20}}
\put(21,22){\framebox(8,8)[cc]{$\nu$}}
\put(5,30){\line(1,-1){11}}
\put(25,10){\line(-1,1){6}}
}}
= [n-m]_q .\]
We conclude that 8-dimension is a sort of categorical dimension (a 
$q$-integer) multiplied by a power of $q$.

\begin{rem}
The construction of the Berezinian applies to an object $V$ in an abelian
$k$-linear braided rigid monoidal category $\CC$ with the braiding
$B=c_{V,V}:V\tens V \to V\tens V$, satisfying $(B-q)(B+q^{-1})=0$.
As an example take a $\C$-linear category $\CC_1$ generated over $\C$ 
by
tangles with two colours denoted $V$ and $V\pti$ with the relations
\[
\unitlength=0.80mm
\begin{picture}(113,27)
\put(20,20){\line(-1,-1){20}}
\put(0,20){\line(1,-1){9}}
\put(11,9){\line(1,-1){9}}
\put(28,10){\makebox(0,0)[cc]{$-$}}
\put(36,20){\line(1,-1){20}}
\put(36,0){\line(1,1){9}}
\put(47,11){\line(1,1){9}}
\put(60,10){\makebox(0,0)[lc]{$=(q-q^{-1})$}}
\put(90,20){\line(0,-1){20}}
\put(105,0){\line(0,1){20}}
\put(105,25){\makebox(0,0)[cc]{$V$}}
\put(90,25){\makebox(0,0)[cc]{$V$}}
\put(56,25){\makebox(0,0)[cc]{$V$}}
\put(36,25){\makebox(0,0)[cc]{$V$}}
\put(20,25){\makebox(0,0)[cc]{$V$}}
\put(0,25){\makebox(0,0)[cc]{$V$}}
\put(113,10){\makebox(0,0)[cc]{,}}
\end{picture}
\]
\[ \nusquareVV = \alpha ,\]
where $\alpha\in\C-0$ is a parameter (see e.g. \cite{Lyu:tan}). Denote 
by 
$\CC$ its Karoubi envelope, whose objects are idempotents of $\CC_1$. 
The
category $\CC$ is a $\C$-linear braided rigid monoidal category. For
generic values of the parameter $q$ (all except a countable number) this
category is semisimple abelian. The above discussion shows that if 
$\alpha$ is not a power of $q^2$ the cohomology $\Ber V$ vanishes.
Therefore non-vanishing of the Berezinian implies a sort of integrality
condition on structure constants of $\CC$.
\end{rem}

\begin{conjecture}
Let $\CC$ be a $k$-linear abelian braided rigid monoidal category with
$\End_\CC I=k$, $\dim_k\Hom_\CC(A,B) <\infty$, generated by an object 
$V$
and its dual $V\pti$ such that $(c_{V,V}-q)(c_{V,V}+q^{-1}) =0$. If
$\Ber V$ is not null, it is an invertible object, and the pair 
$(n,m)\in\Z^2$ such that $\Ber V = H^{n,m}(K(V),D)$ is called the 
superdimension of $V$, $\sdim V =n|m$. 
\end{conjecture}

The object $V$ is called {\em even} if there exists $n\in\Z_{>0}$ such 
that $\Ex^n(V)$ is
invertible and $\Ex^{n+1}(V)=0$. The object $V$ is called {\em odd} if 
there exists $m\in\Z_{>0}$ such that $\Sy^m(V)$ is invertible and $\Sy^{m+1}(V)=0$.

\begin{rem}
If $\CC$ consists of $k$-vector spaces (it is equipped with a faithful
exact monoidal functor $\CC\to k$-vect), the number $n$ such that 
$\Ex^n(V)\ne0$, $\Ex^{n+1}(V)=0$, is not necessarily $\dim_k V$. There
are examples constructed by Gurevich \cite{Gur:DAN,Gur:AiA} in which 
$n<\dim_kV$.
\end{rem}

\begin{prop}
Let $q$ be not a root of unity and let $\CC$ be as in the above conjecture.

(a) Assume that $V\in\Ob\CC$ is even, $\Ex^{n+1}(V)=0$, $\Ex^n(V)\ne0$.
Then $\Ber V =\Ex^n(V)$ and $\sdim V =n|0$.

(b) Assume that $V\in\Ob\CC$ is odd, $\Sy^{m+1}(V)=0$, $\Sy^m(V)\ne0$.
Then also $\Sy^{m+1}(V\pti)=0$, $\Sy^m(V\pti)$ is invertible and 
$\Ber V =\Sy^m(V\pti)$, $\sdim V =0|m$.
\end{prop}

\begin{pf}
(a) We have $K^{n,0}(V)= \Ex^n(V)$ and $K^{n+1,1}(V)=0$. Thus
$K^{n,0}(V) \subset \Ker D$ and $\im D\cap K^{n,0}(V)=0$, wherefore
$\Ex^n(V) \subset \Ber V$. This implies 
$\dim_8V = q^{-n}[n]_q = (1-q^{-2n})/(q-q^{-1})$. Since $q$ is not a root
of unity, \thmref{Laplace=} claims that $L\big|_{K^{a,b}(V)} \ne0$ if 
$a-b\ne n$. Therefore the subcomplexes $(K^{a,b}(V))_{a-b=p}$ are acyclic 
except $p=n$. The last subcomplex has only one non-zero term $\Ex^n(V)$ 
and the assertion follows.

(b) Similarly.
\end{pf}

In many examples the vector space $V$ with
$(\RB-q)(\RB+q^{-1}) =0$ can be represented in the form
\be\label{XVVV}
V= V_1\oplus_{Q_1}V_2\oplus_{Q_2}\dots \oplus_{Q_{k-1}}V_k
\end{equation}
with some order in which the operations $\oplus_{Q_i}$ are performed and
any space $V_i$ is even or odd. If $q$ is not a root of unity, from the
above Proposition and \thmref{Bermultipl} we deduce that $\Ber V$ is 
one dimensional.

The case of a root $q$  of unity does not follow from the above, but at 
least we know that if $V$ is one-dimensional, so is $\Ber V\simeq V$ or 
$V\pti$. Hence, $V$ given by \eqref{XVVV} with $\dim_kV_i=1$ has 
one-dimensional $\Ber V$. In particular, it is true for the vector 
representation $V$ of $U_q(\esel(n|m))$, $\sdim V=n|m$. 

\subsubsection{}
Using the renormalized differentials $\bar D$ from \eqref{Dbars} we can 
write \thmref{Laplace=} as
\begin{align*}
D_V \bar D_V + \bar D_V D_V &= \nu_U^{-2} q^{2(n-m)} 
{q^{2(m-n)}-\nu_V^{-2}\over q-q^{-1}} : K^{n,m}(V) \to K^{n,m}(V) ,\\
D_U \bar D_U + \bar D_U D_U &= 
{q^{2(r-p)}-\nu_U^{-2}\over q-q^{-1}} : K^{p,r}(U) \to K^{p,r}(U).
\end{align*}
Write $\K(V)$ as a direct sum of subcomplexes 
$\oplus_{k\in\Z} C_V^\bullet(k)$ with $C_V^n(k)=K^{n,n-k}(V)$. Then
\[ K(X) = \oplus_{k,l\in\Z} C_V^\bullet(k) \tens C_U^\bullet(l) .\]
Clearly, $H^*(C_V^\bullet(k),D_V)=0$, $H^*(C_V^\bullet(k),\bar D_V)=0$
unless $q^{2k}=\nu_V^2$, and $H^*(C_U^\bullet(l),D_U)=0$,
$H^*(C_U^\bullet(l),\bar D_U)=0$ unless $q^{2l}=\nu_U^2$. When 
$q^{2k}=\nu_V^2$ (resp. $q^{2l}=\nu_U^2$) the differentials $D_V$, 
$\bar D_V$ (resp. $D_U$, $\bar D_U$) anticommute. 

\begin{thm}\label{thmiiiiii}
The following conditions are equivalent for any $V$ with a Hecke 
$\check R$-matrix:

\begin{enumerate}
\renewcommand{\labelenumi}{(\roman{enumi})}
\item $\K(V) = I_{st}\oplus M$ is a direct sum of bidifferential 
subcomplexes $M$ and $I_{st}$ for some $s,t\in\Zp$, where $H^*(M,D_V)=0$
and $I_{st}^{n,m} = \C$ if $n=s$, $m=t$ and $0$ otherwise.
\item All the complexes $C_V^\bullet(k)$ with $q^{2k}=\nu_V^2$ are 
$D_V$-acyclic, except one which decomposes as $I_s\oplus N$ for some
$s\in\Zp$, where $H^*(N,D_V)=0$ and $I_s^n=\C$ if $n=s$ and $0$ otherwise.
\item The natural embedding $i$ and projection $j$ in
\[ \frac{\Ker D_V\cap\Ker D_V'}{\Ker D_V\cap\Ker D_V'\cap(\im D_V+\im 
D_V')}
@<j<< \frac{\Ker D_V\cap\Ker D_V'}{\im D_V\cap\Ker D_V'}
@>i>> \frac{\Ker D_V}{\im D_V} \]
are isomorphisms of one dimensional spaces.
\end{enumerate}
\end{thm}

\begin{pf}
(i)$\,\Leftrightarrow\,$(ii): Clear from the above reasoning.

(i)$\,\Rightarrow\,$(iii): The restriction of $i$ (resp. $j$) to the subcomplex
$M$ is an embedding into (resp. surjection from) the zero space.

The proof of (iii)$\,\Rightarrow\,$(ii) is omitted. It follows by a 
classification theorem for indecomposable bidifferential complexes
with anticommuting $D$, $\bar D$, which hopefully will be published elsewhere.
We do not use this implication in this paper.
\end{pf}

\begin{prop}\label{proiii}
If $V$ and $U$ satisfies conditions $(\mathrm i)$ or $(\mathrm i\mathrm 
i)$ 
of the above theorem, then so does $V\oplus_QU$.
\end{prop}

\begin{pf} The category of bidifferential complexes decomposable as in 
(i) is closed under tensor multiplication by K\"unneth's theorem.
\end{pf}

\begin{conjecture}
Any Hecke $\check R$-matrix satisfies the conditions (i)-(iii) of 
\thmref{thmiiiiii}.
\end{conjecture}

\subsubsection{Duality for Berezinians}
Assume that $q$ is not a root of unity, so that the pairing
$\pi:K^{n,m}(\pti V)\tens K^{n,m}(V) \to k$ is non-degenerate. Since
$D_{\pti V}^{\prime R}$ is the transpose of $D_V$, the pairing
\[ (\Ker D_{\pti V}^{\prime R}/\im D_{\pti V}^{\prime R}) \tens
(\Ker D_V/\im D_V) \to k \]
is non-degenerate. Hence, 
$H^*(K(\pti V),D_{\pti V}') = H^*(D_{\pti V}^{\prime R})$ is naturally 
a dual space to $\Ber V = H^*(K(V), D_V)$. This duality breaks down
when $q$ is a root of unity.

We shall denote
\[ \Ber' V = \frac{\Ker D_V\cap\Ker D_V'}
{\Ker D_V\cap\Ker D_V'\cap(\im D_V+\im D_V')} .\]
For any $q$ $\pi$ induces a pairing 
\[ \frac{\im D_{\pti V}^{\prime R} + \im D_{\pti V}^R + 
(\Ker D_{\pti V}^{\prime R}\cap\Ker D_{\pti V}^R)}
{\im D_{\pti V}^{\prime R} + \im D_{\pti V}^R} \tens \Ber'V \to k ,\]
which is non-degenerate if $q$ is not a root of unity. By \propref{diffdiff}
the left differentials $D_{\pti V}$, $D_{\pti V}'$ can be used above
instead of the right ones $D_{\pti V}^R$, $D_{\pti V}^{\prime R}$. 
The first multiplicand can be written also as
\[ \frac{\Ker D_{\pti V}'\cap\Ker D_{\pti V}}
{(\im D_{\pti V}' + \im D_{\pti V}) \cap 
(\Ker D_{\pti V}'\cap\Ker D_{\pti V})} = \Ber'\pti V .\]
Therefore, we obtain a pairing
\be\label{piB'B'}
\pi: \Ber'\pti V \tens \Ber'V \to k ,
\end{equation}
non-degenerate if $q$ is not a root of unity.

\begin{conjecture}
The above pairing is always non-degenerate.
\end{conjecture}

Theorems \ref{Bermultipl}, \ref{thmD'R}, \ref{thmDR}, \ref{thmD'} 
imply the existence of external products, coherent with $i$'s and 
$j$'s from \thmref{thmiiiiii}(iii),
\begin{alignat*}4
& \Ber'V\tens\Ber'U &\to& \Ber'(V\oplus_QU) ,&\qquad &
[\omega_V]\tens[\omega_U] &\mapsto& [\phi(\omega_V\tens\omega_U)] ,\\
& \Ber'\pti U\tens\Ber'\pti V &\to& \Ber'\pti(V\oplus_QU) ,&\qquad&
[\omega_{\pti U}]\tens[\omega_{\pti V}] &\mapsto &
[\psi(\omega_{\pti U}\tens\omega_{\pti V})] ,
\end{alignat*}
where 
\begin{eqnarray}
\omega_V&\in& K^{n,m}(V)\cap\Ker D_V\cap\Ker D_V',\\ 
\omega_U&\in& K^{p,r}(U)\cap\Ker D_U\cap\Ker D_U', \\
\omega_{\pti U}&\in &K^{p,r}(\pti U)\cap\Ker D_{\pti U}\cap\Ker D_{\pti 
U}',\\
\omega_{\pti V}&\in &K^{n,m}(\pti V)\cap\Ker D_{\pti V}\cap\Ker D_{\pti 
V}'.
\end{eqnarray}
\propref{pi=qpipi} shows that
\be\label{q^np+mr}
\nqquad  \pi_X( [\psi(\omega_{\pti U}\tens\omega_{\pti V})], 
[\phi(\omega_V\tens\omega_U)] ) = q^{np+mr} 
\pi_V([\omega_{\pti V}], [\omega_V]) \pi_U([\omega_{\pti U}], [\omega_U])
\end{equation}
This gives some evidence in favour of the conjecture.

\section{Calculation of Berezinians}\label{CalBer}
\subsection{The quantum superdeterminant}
We calculate Berezinians in several examples and show that these spaces
are one dimensional. Let a non-zero vector $\omega\in \Ber V$ constitute
a basis. The coaction $\delta: \Ber V \to\Ber V\tens H$, 
$\omega\mapsto\omega\tens\tau$ determines a group-like element $\tau\in 
H$,
where $H$ is the Hopf superalgebra determined by 
\eqref{-RTT=-TTR}--\eqref{-TT=delta}. Thus $\eps(\tau)=1$, 
$\Delta(\tau) = \tau\tens\tau$. The element $\tau$
can be called $\Ber T$, the Berezinian of $(\ma Tij)$, or
the quantum superdeterminant $\sdet_qT$.

The commutation properties of $\tau$ are determined by the braiding 
properties of $\omega$. Introduce linear bijections $\alpha:V\to V$, 
$v_i\mapsto v_j\ma\alpha ji$, and $\beta:V\to V$, 
$v_i\mapsto v_j\ma\beta ji$ by
\begin{align*}
c(v\tens\omega) &= \omega\tens\alpha(v) ,\\
c(\omega\tens v)&= \beta(v)\tens\omega.
\end{align*}

\begin{prop}
(a) The maps $\alpha$, $\beta$ are symmetries of $\RB:V\tens V \to V\tens 
V$
in the following sense
\[ (\alpha\tens\alpha)\RB = \RB(\alpha\tens\alpha), \qqquad
(\beta\tens\beta)\RB = \RB(\beta\tens\beta) .\]

(b) The maps $\alpha\beta$, $\beta\alpha$ are automorphisms of the
$H$-comodule $V$.

(c) The element $\omega$ has a well defined degree $p(\omega)$ and
\begin{align*}
\tau\ma Tlj \tau^{-1} &= \sum_{i,k} (-1)^{p(\omega)(p(v_i)-p(v_k))}
\ma\alpha lk \ma Tki \ma\alpha{-1i}j \\
&= \sum_{i,k} (-1)^{p(\omega)(p(v_i)-p(v_k))}
\ma\beta{-1l}k \ma Tki \ma\beta ij \\
\tau\Tbar lj \tau^{-1} &= \sum_{i,k} (-1)^{p(\omega)(p(v_i)-p(v_k))}
\ma\alpha{-1k}l \Tbar ki \ma\alpha ji \\
&= \sum_{i,k} (-1)^{p(\omega)(p(v_i)-p(v_k))}
\ma\beta kl \Tbar ki \ma\beta{-1j}i
\end{align*}
\end{prop}

\begin{pf}
(a) Follows from the Yang--Baxter equation applied to $V\tens V\tens \Ber 
V$
and to $\Ber V\tens V\tens V$.

(b) Indeed, $c^2=1\tens\beta\alpha : \Ber V\tens V \to \Ber V\tens V$
and $c^2=\alpha\beta\tens1 : V\tens \Ber V \to V\tens \Ber V$ are
automorphisms.

(c) $\Ber V$ is a $\Z/2$- or $\gr$-graded space. The formulae express
the fact that $c:V\tens\Ber V \to \Ber V\tens V$, 
$c: \Ber V\tens V \to V\tens\Ber V$ are morphisms.
\end{pf}

\begin{cor}
If for some $a,b\in k^\times$
\begin{align}
\alpha(v_i) &= (-1)^{p(v_i)p(\omega)} av_i \label{alphaa} \\
\intertext{or}
\beta(v_i) &= (-1)^{p(v_i)p(\omega)} bv_i, \label{betab}
\end{align}
then $\tau=\Ber T$ is a central element.
\end{cor}

In the case of \eqref{alphaa} or \eqref{betab},  the Hopf algebra $H'=H/(\tau-1)$, 
referred to as a {\em semispecial quantum linear group},
is big enough. In the particular case of a {\em special quantum linear 
group}, 
when $a=b=1$, $p(\omega)=0$ (resp. $p(\omega)=1$), the algebra 
$H'=H/(\tau-1)$ (resp. $H'=H/(\tau^2-1)$) is a coquasitriangular Hopf 
algebra. Here we can add to the defining system of morphisms an 
isomorphism  to the trivial comodule $\Ber V\simeq k$ (resp. 
$(\Ber V)^{\tens2}\simeq k$).

\subsection{One dimensional example}
Take $V=\C^{1|0}$ with $\RB=q$. Let $v\in V$, $v\ne0$ and let $w\in V\pti$
be its dual vector. $K(V) = \C[w,v]/(v^2)$ contains the element $\di=wv$.
The differentials are given by
\begin{alignat*}2
D(w^k) &= w^{k+1}v &,\qqquad D(w^kv) &= 0, \\
D'(w^k) &= 0 &,\qqquad D'(w^kv) &= q^{-1} [k]_qw^{k-1} .
\end{alignat*}
The Laplacian is
\[ L\big|_{K^{0,m}} =q^{-1}[m+1]_q,\qqquad L\big|_{K^{1,m}} =q^{-1}[m]_q.\]
The cohomologies $H^*(K(V),D')$  are infinite dimensional if $q$ is a 
root of unity and one dimensional otherwise. The Berezinian 
$\Ber V\simeq \Ber'V$ is always one dimensional and 
$\omega_V=v\in K^{1,0}(V)$, $v\in\Ker D\cap\Ker D'$, is a cocycle giving 
its basis. Remark that $K(V)= K^{1,0}(V)\oplus M$, 
$M=\oplus \begin{Sb}(n.m)\ne(1,0)\\n,m\ge0\end{Sb} K^{n,m}(V)$ is a
decomposition into bidifferential subcomplexes and $M$ is $D$-acyclic.
A basis of $\Ber\pti V \simeq \Ber'\pti V$ is given by the dual vector
$\omega_{\pti V}= {}'v\in\pti V=K^{1,0}(\pti V)$ and 
$\pi(\omega_{\pti V}, \omega_V) =1$, so the conjectures are verified in 
this
example.

\subsection{Standard $GL(n|m)$ $R$-matrices}
The standard $R$-matrices for $GL(n|m)$ are obtained by iterating the
following construction.

\begin{prop}
Let $\RB:V\tens V \to V\tens V$ be a Hecke $\check R$-matrix in a 
$\Z/2$-graded space $V$. Let $\phi:V\to V$ be a symmetry of $\RB$, 
a bijective linear map of degree $0$, satisfying 
$\phi\tens\phi\circ\RB = \RB\circ\phi\tens\phi$. Consider a one dimensional
$\Z/2$-graded vector space $U$ with an $\check R$-matrix 
$\RB':U\tens U \to U\tens U$, such that either $U$ is even and $\RB'=q$,
or $U$ is odd and $\RB'=-q^{-1}$. The map $Q:U\tens V\to V\tens U$,
$Q(u\tens v) = \phi(v)\tens u$, satisfies the conditions of \thmref{MajMar},
defining a Hecke $\check R$-matrix in $V\oplus_QU$.
\end{prop}

\begin{pf} Clear.
\end{pf}

We start with one dimensional space $V_1$, add up one dimensional spaces
$V_i$ and obtain 
$V= V_1\oplus_{Q_1}V_2\oplus_{Q_2}\dots \oplus_{Q_{k-1}}V_k$, using
diagonal matrices
\[ \phi_i: V_1\oplus V_2\oplus\dots \oplus V_i \to 
V_1\oplus V_2\oplus\dots \oplus V_i, \qquad \phi_i(v_j)=q_{i+1,j}v_j ,\]
which are symmetries of $\check R$-matrices on
$V_1\oplus_{Q_1}V_2\oplus_{Q_2}\dots \oplus_{Q_{i-1}}V_i$. Here 
$q_{ab}\in k^\times$, $a>b$ are parameters. We introduce additional
parameters $q_{ab}$, $a\le b$ satisfying
\[ q_{ab} q_{ba} =1 \qqquad\text{if } a\ne b, \]
\[ q_{ii} = \begin{cases} q & \qquad\text{if $v_i$ is even,} \\
-q^{-1} & \qquad\text{if $v_i$ is odd.} \end{cases} \]
The resulting $\check R$-matrix in the $\Z/2$-graded space $V$ is described 
by
\begin{alignat*}2
\RB(v_i\tens v_j) &= q_{ij} v_j\tens v_i &\qquad& \text{for } i\le j ,\\
\RB(v_i\tens v_j) &= q_{ij} v_j\tens v_i + (q-q^{-1})v_i\tens v_j
&\qquad& \text{for } i>j .
\end{alignat*}
The set of indices is divided in two parts: even indices
$\{e_1<e_2<\dots<e_n\}$ and odd indices $\{o_1<o_2<\dots<o_m\}$.

Theorem~\ref{Bermultipl} and \propref{proiii} say that 
$\Ber V\simeq \Ber'V$ is one dimensional and has a basic vector
\[ \omega_V = v^{o_1}\dots v^{o_m} v_{e_1}\dots v_{e_n} \in \Ber V \]
(including the case of roots of unity). The ordering of this product
can be changed due to commutation relations based on the expression
for $\RB^\sharp$
\begin{alignat*}2
\RB^\sharp(v_i\tens v^j) &= q_{ji} v^j\tens v_i 
&\quad& \text{for } i\ne j ,\\
\RB^\sharp(v_j\tens v^j) &= q_{jj} v^j\tens v_j + 
(q-q^{-1}) \sum_{i>j} v_i\tens v_i &\quad& .
\end{alignat*}

The element
\[ \omega_{\pti V} = 
{}^{e_n}v\dots{}^{e_1}v v_{o_m}\dots v_{o_1} \in K(\pti V) \]
gives a basis of $\Ber\pti V \simeq\Ber'\pti V$. Lifting these elements 
to
\begin{align*} 
\tilde\omega_V &= v^{o_1}\tens\dots\tens v^{o_m}\tens 
v_{e_1}\tens\dots\tens v_{e_n} \in T^m(V\pti)\tens T^n(V) ,\\
\tilde\omega_{\pti V} &= {}^{e_n}v\tens\dots\tens{}^{e_1}v\tens 
v_{o_m}\tens\dots\tens v_{o_1} \in T^n(\pti V)\tens T^m(V) ,
\end{align*}
we find by \eqref{q^np+mr} 
\[ \pi(\omega_{\pti V},\omega_V) = q^{n(n-1)/2+m(m-1)/2} .\]
Therefore, $\pi(\tilde\omega_{\pti V},-)$ is a nontrivial linear 
functional on $\Ker D_V$, vanishing on $\im D_V$. So we conclude that
{\allowdisplaybreaks
\begin{align}
&\delta(\omega_V) = \omega_V\tens 
\frac1{\pi(\tilde\omega_{\pti V},\tilde\omega_V)}
\pi(\tilde\omega_{\pti V},\tilde\omega_{V(0)}) \tilde\omega_{V(1)} \notag 
\\
&= \omega_V\tens q^{-n(n-1)/2-m(m-1)/2} \sum_{a,b} (\sign)
\<\Ant({}^{e_n}v\tens\dots\tens{}^{e_1}v)\tens
\Sym(v_{o_m}\tens\dots\tens v_{o_1}), \notag  \\
&\qqquad v^{a_1}\tens\dots\tens v^{a_m}\tens 
v_{b_1}\tens\dots\tens v_{b_n}\>  \tbar{a_1}{o_1} \dots\tbar{a_m}{o_m} 
\ma t{b_1}{e_1} \dots\ma t{b_n}{e_n}\notag\\
&= \omega_V\tens \sum_{a,b} (\sign) \sum_{\mu\in\SSS_m} q^{-l(\mu)}
\<(\RB^{-1})_\mu(v_{o_m}\tens\dots\tens v_{o_1}), 
v^{a_1}\tens\dots\tens v^{a_m} \> 
\tbar{a_1}{o_1} \dots\tbar{a_m}{o_m}\notag \\
&\qqquad\times \sum_{\lambda\in\SSS_n} (-1)^{l(\lambda)} q^{-l(\lambda)}
\< (\RB^{\flat\flat})_\lambda ({}^{e_n}v\tens\dots\tens{}^{e_1}v),
v_{b_1}\tens\dots\tens v_{b_n} \> 
\ma t{b_1}{e_1} \dots\ma t{b_n}{e_n}\notag\\ 
&= \omega_V\tens \sum_{\mu\in\SSS_m} q^{-l(\mu)}
\Big(\prod\begin{Sb}1\le i<j\le m\\ \mu(i)>\mu(j)\end{Sb} 
q_{o_io_j}^{-1}\Big) \tbar{o_\mu(1)}{o_1} \dots\tbar{o_\mu(m)}{o_m}\notag 
\\
&\qqquad\times \sum_{\lambda\in\SSS_n} (-1)^{l(\lambda)} q^{-l(\lambda)}
\Big(\prod\begin{Sb}1\le i<j\le n\\ \lambda(i)>\lambda(j)\end{Sb} 
q_{e_ie_j}\Big) 
\ma t{e_{\lambda(1)}}{e_1} \dots\ma t{e_{\lambda(n)}}{e_n}\notag\\
&\equiv \omega_V\tens\detq\bar t_{\text{odd}}\,\detq t_{\text{even}}\notag\\
&\equiv \omega_V\tens \sdet_q t \label{sdetqt}.
\end{align}
}
Here (sign) comes from the graded coproduct and it disappears in the
final answer because only even elements $\tbar{o_i}{o_j}$ and 
$\ma t{e_i}{e_j}$ contribute to the final formula. When $m=0$ the 
superdeterminant coincides with the usual quantum determinant $\detq t$ 
from \cite{Schir}.

In the formulae
\begin{align}
c(v\tens\omega_V) &= \omega_V\tens \frac1{\pi(\omega_{\pti V},\omega_V)} 
(\pi\tens1)(1\tens c) (\tilde\omega_{\pti V}\tens v\tens\tilde\omega_V) 
\label{pi11c} \\
c(\omega_V\tens v) &= \omega_V\tens \frac1{\pi(\omega_V,\omega_{V\pti})} 
(1\tens\pi)(c\tens1) (\tilde\omega_V\tens v\tens\tilde\omega_{V\pti})
\label{1pic1}
\end{align}
the contribution of terms proportional to $(q-q^{-1})$ is  nil.
Therefore we find
\begin{align*}
\alpha(v_i) &= q_{ie_1}\dots q_{ie_n} q_{io_1}^{-1}\dots q_{io_m}^{-1}v_i,\\
\beta(v_i) &= q_{e_1i}\dots q_{e_ni} q_{o_1i}^{-1}\dots q_{o_mi}^{-1}v_i.
\end{align*}
Their composition
\[ \alpha\beta(v_i) = \prod_{j=1}^nq_{ie_j}\ \prod_{k=1}^mq_{io_k}^{-1}\ 
\prod_{j=1}^nq_{e_ji}\ \prod_{k=1}^mq_{o_ki}^{-1}\ v_i = q^2 v_i \]
because
\[ q_{ij} q_{ji} = \begin{cases} 1 &\ \text{if } i\ne j,\\
q^2 &\ \text{if $i=j$ is even},\\ q^{-2} &\ \text{if $i=j$ is odd}.
\end{cases} \]

If for some constant $a$ and all $i$
\[ q_{ie_1}\dots q_{ie_n}q_{io_1}^{-1}\dots q_{io_m}^{-1} =(-1)^{mp(v_i)}a\]
we have semispecial linear group and $\tau$ is a central element. If 
$n\ne m$, we can rescale $\RB$, multiplying it by $q^{1/(m-n)}$. Then
$\alpha$ and $\beta$ multiply by $q^{-1}$ and their product becomes 1.
So if additionally $a=q$, we have a special linear group.
When $n=m$ rescaling will not help to construct a special linear group.

\subsection{Berezinians for differential supergroups obtained from
standard quantum $GL(n|m)$}
We note that provided $q$ is not a sixth root of unity, the algebra $H$
of functions on the differential quantum supergroup has the same growth
properties as the corresponding supercommutative algebra: specifically,
it has a linear basis consisting of all alphabetically ordered
monomials in $\ma pij$, $\ma rij$, $\ma sij$, $\ma tij$ with the powers
of $p$ and $r$ not exceeding 1 (\cite{Sud:matel}, Theorem 3).
Again \thmref{Bermultipl} tells us that $\Ber X\simeq\Ber'X$ is one 
dimensional. The basis $(u_i)=(dv_i) \subset U=dV$ has the changed
degree $p(u_i)=(p(v_i),1)\in\gr$; hence a basic vector of $\Ber U$ is
similar to that of $\Ber V$ with swapped even and odd indices
\[ \omega_U = u^{e_n}\dots u^{e_1}u_{o_m}\dots u_{o_1} \in \Ber U .\]
This gives a basic vector of $\Ber X$
\be\label{omegaX} 
\omega_X = v^{o_1}\dots v^{o_m} v_{e_1}\dots v_{e_n} 
u^{e_n}\dots u^{e_1}u_{o_m}\dots u_{o_1} \in \Ker D_X\cap\Ker D_X' .
\end{equation}
Using the commutation relations in $K(X)$ 
\begin{alignat*}2
qv_i u^j &= q_{ji} u^j v_i 
&\quad& \text{if } i\ne j ,\\
qv_j u^j &= q_{jj} u^j v_j + (q-q^{-1}) \sum_{i>j} u^i v_i &\quad&
\end{alignat*}
we represent this cocycle in another form
\[ \bar\omega_X = v^{o_1}\dots v^{o_m} 
\big( \sum_{k_i,l_i\ge e_i} c_{k_1\dots k_n}^{l_1\dots l_n} 
u^{k_n}\dots u^{k_1} v_{l_1}\dots v_{l_n} \big) u_{o_m}\dots u_{o_1} .\]

An element 
\[ \omega_{\pti X} = {}^{o_1}u\dots{}^{o_m}u u_{e_1}\dots u_{e_n}
{}^{e_n}v\dots{}^{e_1}v v_{o_m}\dots v_{o_1} \in K(\pti X) \]
or in another form
\[ \bar\omega_{\pti X} = {}^{o_1}u\dots{}^{o_m}u 
\big( \sum_{r_i,p_i\le e_i} b_{p_1\dots p_n}^{r_1\dots r_m} ({}^{p_n}v)
\dots({}^{p_1}v) u_{r_1}\dots u_{r_n} \big) v_{o_m}\dots v_{o_1} \]
is a cocycle from $\Ker D_{\pti X}\cap\Ker D_{\pti X}'$ giving a basis
of $\Ber'\pti X\simeq \Ber\pti X$. The pairing between these cocycles 
is
found from \eqref{q^np+mr}
\[ \pi(\omega_{\pti X},\omega_X) = q^{(n+m)(n+m-1)} .\]
The superdeterminant $\tau_X$ is 
\[ \tau_X = 
q^{-(n+m)(n+m-1)} \pi(\omega_{\pti X},\omega_{X(0)}) \omega_{X(1)}.\]
The explicit formula is rather complicated. At least $\tau_X$ is 
grouplike and $d\tau_X=0$.

Lift $\omega_X$ to an element of the tensor product
\[ \tilde\omega_X \in T^m(V\pti)\tens T^n(V)\tens T^n(U\pti)\tens T^m(U) 
\]
inserting $\tens$ between elements of \eqref{omegaX}. We know in 
principle how to calculate $c(x\tens\tilde\omega_X)$, 
$c(\tilde\omega_X\tens x)$ for $x\in X$. To find $c(x\tens[\omega_X])$
and $c([\omega_X]\tens x)$ we use \eqref{pi11c} and \eqref{1pic1} applied
to $X$. All terms proportional to $(q-q^{-1})$ do not contribute to
\begin{align*}
\alpha_X(x) &= q^{-(n+m)(n+m-1)}
(\pi\tens1)(1\tens c) (\omega_{\pti X}\tens x\tens\tilde\omega_X) \\
\intertext{and}
\beta_X(x) &= q^{-(n+m)(n+m-1)}
(1\tens\pi)(c\tens1) (\tilde\omega_X\tens x\tens\omega_{X\pti}) .
\end{align*}
The principal terms give
\begin{align*}
\alpha_X(v_i) &= q_{ie_1}\dots q_{ie_n} q_{io_1}^{-1}\dots q_{io_m}^{-1}
q_{o_1i}^{-1}\dots q_{o_mi}^{-1} q_{e_1i}\dots q_{e_ni} v_i \\
&= \prod_{k=1}^n q^{2\delta_{ie_k}}\ \prod_{l=1}^m q^{2\delta_{io_l}} 
v_i \\
&= q^2 v_i \\
\intertext{and similarly}
\alpha_X(u_i) &= (-1)^{n+m} q^2 u_i ,\\
\beta_X(v_i) &= v_i ,\\
\beta_X(u_i) &= (-1)^{n+m} u_i .
\end{align*}
The $\gr$-gradings of the elements involved  are
\begin{align*}
\deg(v_i) &= (p(v_i),0) ,\\
\deg(u_i) &= (p(v_i),1) ,\\
\deg(\omega_X) &= (0,m-n) ;
\end{align*}
hence
\begin{align*}
(-1)^{\deg(\omega_X)\deg(v_i)} &= 1,\\
(-1)^{\deg(\omega_X)\deg(u_i)} &= (-1)^{n+m} .
\end{align*}
Finally,
\begin{align*}
\alpha(x) &= (-1)^{\deg(\omega_X)\deg(x)}q^2x ,\\
\beta(x) &= (-1)^{\deg(\omega_X)\deg(x)}x ,
\end{align*}
so $\tau_X$ is a central element and we are in a semispecial situation.
The quotient $H/(\tau_X-1)$ is a differential Hopf algebra which is no
longer coquasitriangular.

\subsubsection{The superdeterminant for differential forms}
Considering another quotient of $H$, the algebra of differential forms
$\Omega$, we can ask about the image of $\tau_X$ in $\Omega$. Unlike
$\tau_X$ its image can be calculated by similar reasoning to the
proof of \propref{pi=qpipi}. We get in $\Omega$
{\allowdisplaybreaks
\begin{align*}
\tau_X &= q^{-n(n-1)-m(m-1)} (\sign) 
\<\Ant_m({}^{o_1}u\tens\dots\tens{}^{o_m}u) 
\tens\Sym(u_{e_1}\tens\dots\tens u_{e_n})\tens \\
&\quad\tens\Ant_n({}^{e_n}v\tens\dots\tens{}^{e_1}v)
\tens\Sym_m(v_{o_m}\tens\dots\tens v_{o_1}), \\
&\quad\sum_{a,b,c,d} v^{a_1}\tens\dots\tens v^{a_m}\tens 
v_{b_1}\tens\dots\tens v_{b_n}\tens 
u^{c_n}\tens\dots\tens u^{c_1}\tens u_{d_m}\tens\dots\tens u_{d_1} \> 
\\
&\quad \tbar{a_1}{o_1} \dots\tbar{a_m}{o_m} 
\ma t{b_1}{e_1} \dots\ma t{b_n}{e_n}
\tbar{c_n}{e_n} \dots\tbar{c_1}{e_1} \ma t{d_m}{o_m} \dots\ma t{d_1}{o_1} 
\\
&= \sum_a \sum_{\mu\in\SSS_m} q^{-l(\mu)}
\<(\RB^{-1})_\mu(v_{o_m}\tens\dots\tens v_{o_1}), 
v^{a_1}\tens\dots\tens v^{a_m} \> \tbar{a_1}{o_1} \dots\tbar{a_m}{o_m} 
\\
&\quad\times\sum_b\sum_{\lambda\in\SSS_n} (-1)^{l(\lambda)} q^{-l(\lambda)}
\< (\RB^{\flat\flat})_\lambda ({}^{e_n}v\tens\dots\tens{}^{e_1}v),
v_{b_1}\tens\dots\tens v_{b_n} \> \ma t{b_1}{e_1} \dots\ma t{b_n}{e_n}\\ 
&\quad\times\sum_c\sum_{\nu\in\SSS_n} (-1)^{l(\nu)} q^{-l(\nu)}
\<\RB_\nu(u_{e_1}\tens\dots\tens u_{e_n}), 
u^{c_n}\tens\dots\tens u^{c_1} \> \tbar{c_n}{e_n} \dots\tbar{c_1}{e_1} 
\\
&\quad\times \sum_d \sum_{\kappa\in\SSS_m} q^{-l(\kappa)}
\< (\RB^{-1\flat\flat})_\kappa ({}^{o_1}u\tens\dots\tens{}^{o_m}u),
u_{d_m}\tens\dots\tens u_{d_1} \> \ma t{d_m}{o_m} \dots\ma t{d_1}{o_1} 
\\
&\equiv \detq\bar t_{\text{odd}}\,\detq t_{\text{even}}\ 
\detq\bar t_{\text{even}}\,\detq t_{\text{odd}} \\
&\equiv \sdet_q t\ \sdet_q\bar t \\
&= 1
\end{align*}
}
because
\begin{alignat*}2
\delta([\omega_V]) &= [\omega_V]\tens \sdet_qt &\qqquad& 
\text{by \eqref{sdetqt}}\\
\delta([\omega_{V\pti}]) &= [\omega_{V\pti}]\tens \sdet_q\bar t 
&\qqquad& \text{similarly}\\
\Ber V\tens\Ber(V\pti) &\simeq k \in\CC &\qqquad& \text{by \eqref{piB'B'}}
\end{alignat*}
imply
\begin{align*}
\delta([\omega_V]\tens[\omega_{V\pti}]) &= 
[\omega_{V}]\tens [\omega_{V\pti}]\tens \sdet_qt \sdet_q\bar t \\
&= [\omega_{V}]\tens [\omega_{V\pti}]\tens1
\end{align*}
and, finally,
\[ \sdet_qt \sdet_q\bar t =1 .\]

Therefore, for the standard $\RB$-matrix of $GL(n|m)$ type there are 
epimorphisms of differential Hopf algebras
\[ H\to H/(\tau_X-1) \to \Omega \]
and the image of the superdeterminant $\tau_X$ is 1 in $\Omega$. So it
is unreasonable to look for an $SL$-version of the algebra $\Omega$ of
differential forms on quantum $GL(n|m)$, because it is, in a sense, already 
of $SL$ type!

\section{Hopf bimodules}\label{bimodules}
\subsection{Recollection of basic facts}
There is a well known definition of a Hopf module over
a Hopf algebra.  We recall it in a $\Z/2$-graded version.

A {\em left Hopf module} $(M,a_L,\delta_L)$ over a $\Z/2$-graded Hopf 
algebra $F$ is a left $F$-module $(M,a_L:F\tens M\to M)$ with a coaction
$\delta_L:M \to F\tens M$, $m\mapsto m\mone \tens m\nul$ such that
\[ \delta_L(fm) = 
(-1)^{\hat f\two \hat m\mone} f\one m\mone \tens f\two m\nul .\]
Here $\hat x\in\Z/2$ is the parity of a homogeneous element $x$.
Similarly a right Hopf module $(M,a_R,\delta_R)$ is defined with
$a_R:M\tens F\to M$ and $\delta_R:M \to M\tens F$, 
$m\mapsto m\nul \tens m\one$. The mixed notions $(M,a_L,\delta_R)$ 
and $(M,a_R,\delta_L)$ are required to satisfy
\[ \delta_R(fm) = 
(-1)^{\hat f\two \hat m\nul} f\one m\nul \tens f\two m\one \]
and
\[ \delta_L(mf) = 
(-1)^{\hat  m\nul \hat f\one}  m\mone f\one \tens  m\nul  f\two\]
respectively.

The results of this section were also independently obtained by 
Schauenburg~\cite{Sch:HYD} in greater generality.

\begin{defn}
A Hopf $F$-bimodule $(M,a_L,a_R,\delta_L,\delta_R)$ is a vector space
$M$ with left and right actions of $F$ and left and right coactions of 
$F$ such that $(M,a_L,\delta_L)$, $(M,a_L,\delta_R)$, $(M,a_R,\delta_L)$,
$(M,a_R,\delta_R)$ are Hopf modules, $(M,a_L,a_R)$ is an $F$-bimodule
(i.e. $(fm)g=f(mg)$) and $(M,\delta_L,\delta_R)$ is an $F$-bicomodule
(i.e. $(1\tens \delta_R) \delta_L = (\delta_L\tens1) \delta_R$). Schauenburg 
calls such $M$ two-sided two-cosided Hopf modules \cite{Sch:HYD}.
\end{defn}

\begin{examples}
1) The Hopf algebra $F$ is a Hopf $F$-bimodule, when equipped with
regular actions and coactions.

2) Differential forms of the first order $\Omega^1$ (see \secref{DfoqGL})
make a Hopf module over the algebra of functions $F=\Omega^0$ with
coactions determined by the decomposition $\Delta = \delta_L \oplus \delta_R 
: \Omega^1 \to \Omega^0\tens\Omega^1 \oplus \Omega^1 \tens\Omega^0$.

3) The same for an arbitrary $\grp$-graded differential Hopf algebra.
\end{examples}

Let $M$ be a left Hopf $F$-module and let
\[ M^l =\{ m\in M \mid \delta_L(m) = 1\tens m \} .\]
Then $M$ is isomorphic to a direct sum of $\dim M^l$ copies of the
regular Hopf module, $M\simeq F\tens M^l$ (see e.g. \cite{Swe:book}). 
This
gives an equivalence between the category of left Hopf $F$-modules
and the category of vector spaces. 

Let $M$ be a Hopf $F$-bimodule. Then $\delta_R(M^l)\subset M^l\tens F$
by the bicomodule property, so $M^l$ is a right $F$-comodule. It can be
viewed also as a left $F^*$-module. The right coadjoint action of $F$
\[ m\tl f = (-1)^{\hat m\hat f\one} \gamma(f\one) m f\two \]
preserves $M^l$, so $M^l$ is a right $F$-module.

\begin{thm}[See also Schauenburg \cite{Sch:HYD}]\label{Hbimstruc}
The module $M^l$ satisfies
\be\label{ntrig}
\delta_R(n\tl g) = 
(-1)^{\hat g\one \hat g\two + \hat n\one(\hat g\one+\hat g\two)} 
n\nul\tl g\two \tens \gamma(g\one) n\one g\tre
\end{equation}
for any $n\in M^l$, $g\in F$. 

Given a right $F$-module and $F$-comodule $(N,\tl,\delta_R)$ which 
satisfies~\eqref{ntrig}, we make $M=F\tens N$ into a Hopf bimodule by 
setting
\begin{align}
g.(f\tens n) &= (gf)\tens n ,\label{g.(fn)} \\
(f\tens n).g &= (-1)^{\hat n\hat g\one} (fg\one) \tens (n\tl g\two) ,\\
\delta_L(f\tens n) &= f\one \tens (f\two \tens n) ,\\
\delta_R(f\tens n) &= (-1)^{\hat f\two \hat n\nul} 
(f\one\tens n\nul) \tens (f\two n\one) .\label{deltaR(fn)}
\end{align}
We have $M^l = k\tens N \simeq N$, and every Hopf bimodule $M$ can be
constructed in this way.
\end{thm}

\begin{cor}\label{abeve}
The abelian category of Hopf $F$-bimodules is equivalent to the abelian
category $\DY_F$ of right $F$-modules and right $F$-comodules satisfying
~\eqref{ntrig}.
\end{cor}

The last category was introduced by Yetter \cite{Yet:cros}, who called
such $N$ right crossed $F$-bimodules.

\begin{pf*}{Proof of \thmref{Hbimstruc}}
Clearly, $a_L$, $a_R$, $\delta_L$, $\delta_R$ are uniquely determined
by $\delta_R\big|_{M^l}$ and $\tl\big|_{M^l}$ via 
\eqref{g.(fn)}--\eqref{deltaR(fn)}. On the other hand, for any right
crossed $F$-bimodule $N$ these formulae define operations such that
$(M,a_L,\delta_L)$, $(M,a_L,\delta_R)$, $(M,a_R,\delta_L)$ are Hopf 
bimodules, $(M,a_L,a_R)$ is  a bimodule, and $(M,\delta_L,\delta_R)$ 
is a bicomodule. Only the Hopf relation between $a_R$ and $\delta_R$
is left. It is equivalent to the identity~\eqref{ntrig}. Indeed, to
deduce \eqref{ntrig} we remark that 
\[ \delta_R( (-1)^{\hat n\hat g\one} (\gamma(g\one)\tens n).g\two ) =
\delta_R(1\tens(n\tl g)) = 1\tens\delta_R(n\tl g) \]
is equal in a Hopf bimodule to
\begin{align*}
& (-1)^{\hat n\hat g\one}\delta_R(\gamma(g\one)\tens n).g\two\tens g\tre 
=\\
& = (-1)^{\hat n\one\hat g + \hat n\nul \hat g\one 
+ \hat g\nul(\hat g\one + \hat g\two)}
(\gamma(g\one)\tens n\nul).g\two \tens \gamma(g\nul)n\one g\tre \\
&= (-1)^{\hat n\one \hat g\one + \hat g\two(\hat g\one +\hat n\one)}
\tens n\nul\tl g\two \tens \gamma(g\one) n\one g\tre .
\end{align*}
Vice versa, \eqref{ntrig} implies the Hopf relation between $a_R$
and $\delta_R$.
\end{pf*}

\begin{rem} When the pairing $\<,\>:  F^\circ\tens F \to k$ of $F$ with 
its 
dual Hopf algebra $F^\circ$ is non-degenerate,  the category of finite
dimensional left modules over the Drinfeld double $D(F^\circ)$ 
\cite{Dri:qua} coincides with the subcategory of finite dimensional 
objects of $\DY_F$. The double  $D(F^\circ)$ is defined as a $\Z/2$-graded 
Hopf algebra generated by its Hopf subalgebras $F^\circ$ and $F^\op$ 
with the commutation relations
\[ (-1)^{\hat x\one \hat y} x\one y\one \<y\two,x\two\> =
(-1)^{\hat x\one \hat y\two} \<y\one,x\one\> y\two x\two \]
for any $x\in F^\op$ and $y\in F^\circ$.
\end{rem}

\begin{rem} If $F$ is coquasitriangular, which means the existence of
a Hopf pairing $\rho: F\tens F^\op \to k$ with the property
\[ (-1)^{\hat f\hat g\one} g\one f\one \rho(f\two,g\two) =
(-1)^{\hat f\two \hat g\one} \rho(f\one,g\one) f\two g\two \]
for $f,g \in F$, the category comod-$F$ is embedded in $\DY_F$. The right
action of $F$ on a right $F$-comodule $N$ is chosen as
\[ n.f = n\nul \rho(n\one, f) \]
for $n\in N$, $f\in F$. Dually, we can say that some representations of
the double $D(F^\circ)$ come from representations of the
quasitriangular algebra $F^\circ$.
\end{rem}

\subsubsection{Tensor product of Hopf bimodules}
Given two Hopf $F$-bimodules $M$ and $N$ we make their tensor product
$M\tens_F N$ into a Hopf bimodule by setting
\begin{align*}
f.(m\tens_F n) &= (f.m)\tens_F n ,\\
(m\tens_F n).f &= m\tens_F (n.f) ,\\
\delta_L(m\tens_F n) &= 
(-1)^{\hat m\nul \hat n\mone} m\mone n\mone \tens (m\nul\tens_F n\nul) 
,\\
\delta_R(m\tens_F n) &= 
(-1)^{\hat m\one \hat n\nul} (m\nul\tens_F n\nul) \tens m\one n\one .
\end{align*}
All necessary checks are left to the reader. Since $N$ is a free left
$F$-module, we have an isomorphism of left $F$-modules 
$M\tens_F N @>\sim>> M\tens N^l$. This implies that the map
\[ M^l \tens N^l \to  (M\tens_FN)^l \]
is an isomorphism. Indeed, for $n\in N^l$ clearly $m\tens n\in(M\tens_FN)^l$
iff $m\in M^l$. The right $F$-module and comodule structure induced
on $M^l \tens N^l$ by this isomorphism is 
\begin{align}
\delta_R(m\tens n) &= (-1)^{\hat m\one \hat n\nul} 
(m\nul\tens n\nul) \tens m\one n\one , \label{mnmnmn}\\
(m\tens n)\tl f &= (-1)^{(\hat m+\hat n)\hat f\one} 
\gamma(f\one)(m\tens_F n)f\two \notag\\
&= (-1)^{(\hat m+\hat n)\hat f\one} \gamma(f\one) m\tens_F n f\two \notag\\
&= (-1)^{\hat m\hat f\nul +\hat n(\hat f\nul+\hat f\one+\hat f\two)} 
\gamma(f\nul)mf\one \tens_F \gamma(f\two)nf\tre \notag\\
&= (-1)^{\hat n\hat f\one} (m\tl f\one) \tens (n\tl f\two).\label{mtlfntlf}
\end{align}
Therefore we can strengthen Corollary~\ref{abeve}.

\begin{prop}[See also Schauenburg \cite{Sch:HYD}]
The category of Hopf $F$-bimodules is tensor equivalent to the category
$\DY_F$ with the tensor product determined by \eqref{mnmnmn} and 
\eqref{mtlfntlf} (for $D(F^\circ)$-modules this is the usual
tensor product).
\end{prop}

\begin{cor}
The category of Hopf $F$-bimodules is a braided tensor category.
\end{cor}

The braiding was discovered by Woronowicz \cite{Wor:calcul}. The explanation 
is quite simple: $\DY_F$ is braided with the braiding~\cite{Yet:cros}
\[ c: X\tens Y \to Y\tens X, \qquad 
x\tens y \mapsto (-1)^{\hat x\hat y\nul} y\nul \tens x\tl y\one \]
where $x\in X$, $y\in Y$, $X,Y\in \DY_F$. On $D(F^\circ)$-modules
this braiding is $PR$, where $R$ is the universal $R$-matrix of the 
double \cite{Dri:qua}. The induced braiding for Hopf $F$-bimodules is
\be\label{brbim}
c: M\tens_F N \to N\tens_F M, \qquad fm\tens_F n \mapsto
(-1)^{\hat m\hat n\nul} fn\nul \tens _F m\tl n\one ,
\end{equation}
where $m\in M^l$, $n\in N^l$, $f\in F$. Another presentation of the 
braiding is
\[ c(m\tens n) = (-1)^{(\hat m\mone+\hat m\nul)(\hat n\nul+\hat n\one)} 
m\mtwo n\nul \gamma(n\one) \tens \gamma(m\mone) m\nul n\two \]
for $m\in M$, $n\in N$.

\subsection{Differential Hopf algebras determined by Hopf bimodules}
\begin{example}
The algebra $F$ considered as a regular right $F$-module equipped with
the right coadjoint coaction
\[ \nabla f = (-1)^{\hat f\one \hat f\two} f\two\tens\gamma(f\one)f\tre 
\]
becomes itself a right crossed bimodule. The subobject 
$K=\Ker(\e:F\to k)$ is also in $\DY_F$.
\end{example}

{}From the results of Woronowicz \cite{Wor:calcul} one can conclude that 
{\em a bicovariant first order differential calculus} is precisely a Hopf 
bimodule $M$ together with an epimorphism $\omega:K \to M^l \in\DY_F$. 
The differential $d:F \to M$ is recovered from $\omega$ as 
$df= f\one \omega(f\two-\e(f\two))$. If $d:F \to M$ is given, we construct
a map $\omega:F \to M^l$, $\omega(f)=\gamma(f\one) df\two$, whose 
restriction to $K$ is a morphism from $\DY_F$.

Suppose that such $d:F \to M$ constitute a part of $\grp$-graded
differential Hopf algebra $A$, so $A^0=F$, $A^1=M$. The Cartan--Maurer
formula tells us that
\begin{align}
d\omega(f) &= (d\gamma(f\one)) df\two \notag\\
&= -\gamma(f\one) (df\two) \gamma(f\tre) df\four \notag\\
&= - \omega(f\one) \omega(f\two) \label{CarMaur}
\end{align}
for any $f\in F$. 

\begin{thm}\label{MtoEF(M)}
Let $M$ be a Hopf $F$-bimodule,  and let $\omega:K \to M^l$ be an
epimorphism in $\DY_F$. There exists a universal $\grp$-graded algebra
$E^\bullet\in F$-Hopf-bimod, with a differential which is a bicomodule
map, generated by $E^0=F$, $E^1=M$. It is
\[ E^\bullet = E_F^\bullet(M) = 
T_F^\bullet(M)/(\omega(a\one)\tens\omega(a\two))_{a\in J} ,\]
where $J=\Ker(\omega:K\to M^l)$. Moreover, $E_F^\bullet(M)$ is a 
$\grp$-graded differential Hopf $k$-algebra.
\end{thm}

\begin{pf}
If we drop the differential, the algebra $T_F^\bullet(M)$ will be the
universal $\grp$-graded algebra and a Hopf bimodule such that 
$T_F^0(M)=F$, $T_F^1(M)=M$. When the differential is considered,
\eqref{CarMaur} shows that $\omega(a\one) \omega(a\two)=0$ in $E$
for $a\in J$. On the other hand,  the ideal
$(\omega(a\one)\tens\omega(a\two))_{a\in J}$ is a Hopf subbimodule, and
one can check  that the differential $d:E^0\to E^1$
extends to the whole of $E^\bullet$ uniquely.

The algebra $T_F^\bullet(M)$ has a comultiplication
$\Delta: T_F(M) \to T_F(M)\tens T_F(M)$ coinciding with 
$\Delta:F\to F\tens F$ and 
$\Delta = \delta_L\oplus\delta_R: M\to F\tens M \oplus M\tens F$ in the
lowest degrees. It extends to all $T_F^n(M)$ making $T_F^\bullet(M)$
into a Hopf algebra with the antipode $\gamma$, satisfying
$\gamma(df)= d\gamma(f)$ for $f\in F$. The ideal
$(\omega(a\one)\tens\omega(a\two))_{a\in J}$ is a $\gamma$-invariant
coideal, therefore $E^\bullet$ is a Hopf algebra.
\end{pf}

\begin{cor}
$E_F^\bullet(M)$ is a universal $\grp$-graded differential Hopf algebra 
generated by $E_F^0(M)=F$, $E_F^1(M)=M$.
\end{cor}

\begin{rem}\label{remBrz}
This theorem should be compared with a result of Brzezi\'nski~\cite{Brz}.
With the same assumptions he proves the existence of a graded differential
Hopf algebra $(M^{\wedge},d)$ generated by $F$ and $M$, namely
\[ M^{\wedge} = T_F^\bullet(M)/(\Ker(\sigma-1)) ,\]
where $\sigma=c: M^l\tens M^l \to M^l\tens M^l$ is the 
braiding~\eqref{brbim}. By \thmref{MtoEF(M)} $M^{\wedge}$ is a quotient
of $E_F(M)$ and, indeed, one can check straightforwardly that
\[ \{\omega(a\one)\tens\omega(a\two) \mid a\in J\} \subset \Ker(\sigma-1).\]
Indeed,
\[ (\omega\tens\omega)\nabla = (1-\sigma) (\omega\tens\omega) \Delta \]
and $\nabla(J)\subset J\tens F$. In the Hecke case  both algebras 
coincide (see discussion in \secref{relOmegaGLn}). 
\end{rem}

The {\em universal differential calculus} \cite{Wor:calcul} is worth 
mentioning as a particular case. This is the Hopf bimodule 
$U=\Ker(m:F\tens F\to F)$ with the operations
\begin{align*}
f.(g\tens h) &= fg\tens h ,\\
(g\tens h).f &= g\tens hf ,\\
\delta_L(g\tens h) &= 
(-1)^{\hat g\nul \hat h\mone} g\mone h\mone \tens (g\nul\tens h\nul) ,\\
\delta_R(g\tens h) &= 
(-1)^{\hat g\one \hat h\nul} (g\nul\tens h\nul) \tens g\one h\one 
\end{align*}
and the differential $d:F\to U$, $df=1\tens f- f\tens1$. The map
$\omega':K\to U$,  $b\to \gamma(b\one)\tens b\two$ is an embedding
and $\omega'(K)=U^l$. Therefore, $J=0$ and $E_F^\bullet(U)=T_F^\bullet(U)$.

\begin{prop}\label{proEFM}
Let $E^\bullet$ be a $\grp$-graded algebra and a Hopf $F$-bimodule with
a differential which is a bicomodule map, generated by $E^0=F$, $E^1=M$.
If a system of defining relations of $E^\bullet$ is obtained by
differentiating a system of defining relations of the bimodule $M$, 
the algebra $E^\bullet$ is isomorphic to $E_F^\bullet(M)$.
\end{prop}

\begin{pf}
Cover $M$ by the universal Hopf bimodule $U$ as in the commutative diagram
with exact rows
\[ \begin{CD}
0 @>>> J @>>> K @>\omega>> M^l @>>> 0 \\
@.   @VVV  @V\omega'VV    @VVV    @.  \\
0 @>>> N @>>> U @>p>> M   @>>> 0 
\end{CD} \]
The upper exact sequence is in $\DY_F$ and $N^l=J$. 

A {\em system of defining relations} of $M$ means a collection 
$\{r_i\} \subset N\subset U$ such that $F\{r_i\}F =N$. Differentiating
it we get a system of relations of $E^\bullet$ 
$\{(p\tens p)(dr_i)\} \subset M\tens_F M$, where $dr_i\in U\tens_FU$.
We have to prove that
\[ F\{(p\tens p)(dr_i)\}F = 
F\{\omega(b\one)\tens\omega(b\two) \mid b\in J\}F .\]

\smallskip
{\sl The $\subset$ inclusion.} Since $r_i\in N$, it can be represented
as $r_i= \sum_k a_i^k \omega'(b_k)$, where $a_i^k\in F$ and $(b_k)$ is
a $k$-basis of $J$. Hence,
\begin{align*}
dr_i &= \sum_k da_i^k \omega'(b_k) + \sum_k a_i^k d\omega'(b_k) \\
&= \sum_k da_i^k \omega'(b_k) - 
\sum_k a_i^k \omega'(b_{k(1)}) \omega'(b_{k(2)})  \\
&\in U\tens_FN + F\{\omega'(b\one)\tens_F\omega'(b\two) \mid b\in J\}
\end{align*}
which implies
\[ (p\tens p)(dr_i) \in F\{\omega(b\one)\tens\omega(b\two) \mid b\in J\}. 
\]

\smallskip
{\sl The $\supset$ inclusion.} Represent an arbitrary element of $N^l$
in the form $\omega'(b) = \sum_i f_i r_i g_i$, where $f_i,g_i\in F$,
$b \in J$. Then
\begin{align*}
- \omega'(b\one)\tens_F\omega'(b\two) &= d\omega'(b) = d(f_ir_ig_i) \\
&= \sum_i (df_i)r_ig_i + \sum_i f_i(dr_i)g_i - \sum_i f_ir_idg_i \\
&\in U\tens_FN + F\{dr_i\}F + N\tens_FU
\end{align*}
which implies
\[ \omega(b\one)\tens\omega(b\two) \in F\{(p\tens p)(dr_i)\}F .\]
\end{pf}

Together with \thmref{MtoEF(M)} this proposition states that given a 
bimodule with a differential $d:F\to M$ one constructs a graded 
differential Hopf algebra $E_F^\bullet(M)$ simply by differentiating 
the relations of $M$. However, this method is too universal to single 
out interesting cases.

\subsection{The quantum $GL(n|m)$ case}\label{relOmegaGLn}
The algebra of differential forms $\Omega$ constructed from an arbitrary
Hecke $\check R$-matrix satisfies the hypotheses of \propref{proEFM}. 
Therefore $\Omega^\bullet \simeq E_F^\bullet(\Omega^1)$ are isomorphic 
$\grp$-graded differential Hopf algebras. It is possible to find 
the kernel $J$ explicitly. We will do it in the purely even case; the 
general case differs only by signs.

Let $\RB:V\tens V \to V\tens V$ be a Hecke $\check R$-matrix, and let
$\ma tab = \ma {t_V}ab$, $\tpri ab = t_{\pti V\,a}{@!}^b$ be matrix
elements from $\Omega^\bullet$. Multiply equation~\eqref{Rtt=}, or
\[ (d\mt ab )\mt cf = \RB_{gh}^{ac} \mt gi d\mt hj \RB_{bf}^{ij}, \]
by $\gamma(\mt kc )\gamma(\mt la ) = \tpri ck \tpri al$ on the left:
\[ \gamma(\mt kc )\gamma(\mt la ) (d\mt ab )\mt cf = 
\RB_{gh}^{ac} \tpri ck \tpri al \mt gi d\mt hj \RB_{bf}^{ij} .\]
Applying \eqref{tbar^2=} in the form
\[ \RB_{gh}^{ac} \tpri al \mt gi = \mt cm \tpri np \RB_{ip}^{lm} \]
we get
\[ \gamma(\mt la \mt kc ) d(\mt ab \mt cf ) - 
\gamma(\mt kc )\gamma(\mt la ) \mt ab d\mt cf =
\gamma(\mt kc) \mt cm \tpri np d\mt hj \RB_{ip}^{lm} \RB_{bf}^{ij} .\]
This simplifies to
\[ \omega(\mt lb \mt kf ) - \delta^l_b \omega(\mt kf ) =
\RB_{ip}^{lk} \RB_{bf}^{ij} \omega(\mt pj ) ,\]
so we have
\be\label{ttinJ}
\mt lb \mt kf - \delta^l_b \mt kf - \RB_{ip}^{lk} \RB_{bf}^{ij} \mt pj
+ \RB_{ij}^{lk} \RB_{bf}^{ij} \in J .
\end{equation}
This equation is  equivalent to \eqref{Rtt=} modulo the other relations. 
Similarly, equation~\eqref{tbar^2=} is equivalent to the relation
\be\label{tbartbarinJ}
\tbar lh \tbar kg - \delta_l^h \tbar kg - 
{\RB^{-1}}_{ac}^{gh} {\RB^{-1}}_{kl}^{pc} \tbar pa +
{\RB^{-1}}_{ac}^{gh} {\RB^{-1}}_{kl}^{ac} \in J .
\end{equation}
Equation \eqref{ttbar} is equivalent to the relation
\be\label{tbartinJ} 
\tbar gh + {{u^2_{-1}}^i}_k {\RB^{-1}}_{ig}^{jh} \mt kj -
(1+\nu_V^{-2}) \delta_g^h \in J ,
\end{equation}
where ${{u^2_{-1}}^i}_k = \sum_c {\RB^{\sharp-1}}_{ik}^{cc}$, 
and also to the relation
\be\label{ttbarinJ}
\mt ab + {{u^2_1}_m}^k \RB_{bk}^{al}\tbar lm - \delta_b^a(1+\nu_V^2) \in 
J,
\end{equation}
where ${{u^2_1}_m}^k = \sum_c {\RB^{-1\flat-1}}_{mk}^{cc}$. The relations
\eqref{ttinJ}--\eqref{tbartinJ} make a complete list of relations
of $M=F\tens K/J$. Therefore, the right ideal in $F$ generated by
\eqref{ttinJ}--\eqref{tbartinJ} coincides with $J$. 

In calculating $M^l$ we can use the following remarks. There is an isomorphism
\[ \frac{T(\mt ab,\tbar ab)_{a,b}} {\{\eqref{ttinJ}, \eqref{tbartbarinJ}, 
\eqref{tbartinJ}\}T(\mt ab,\tbar ab)_{a,b}}\: @>j>\sim> \:k\{1,\mt ab\}_{a,b}.\]
Indeed, any word in $t,\bar t$ starting with $tt\dots$ can be shortened
using \eqref{ttinJ}, a word starting with $\bar t\bar t\dots$ shortens
by \eqref{tbartbarinJ}, and \eqref{tbartinJ}, \eqref{ttbarinJ}  reduce 
$\bar tt\dots$ and $t\bar t\dots$ to previous cases. One can show that 
the ideal of relations of $F$ projects to 0 by $j$. It is sufficient to 
check that $j$ projects \eqref{ttinJ}, $\mt ab$\eqref{ttinJ}, 
\eqref{tbartbarinJ}, $\tbar ab$\eqref{tbartbarinJ}, \eqref{tbartinJ}, 
$\mt ab$\eqref{tbartinJ} to 0. This implies that $\dim M^l= (\dim V)^2$ 
and $M^l$ is spanned by $\omega(\mt ab)$. 

Relations in the differential algebra of left invariant differential 
forms, which is an algebra in the category $\DY_F$ corresponding to
$\Omega$, are found by Tsygan~\cite{Tsy:dT}. They can be also obtained 
in the form $\omega(b\one)\omega(b\two)$, where $b$ is given 
by \eqref{ttinJ}:
\be\label{R2R3}
\omega(\mt kc) \RB^{lc}_{bz} \omega(\mt zf) + \RB^{lk}_{ip} 
\omega(\mt pj) \RB^{ij}_{xz} \omega(\mt zy) \RB^{xy}_{bf} =0.
\end{equation}
Equations \eqref{tbartbarinJ}, \eqref{tbartinJ} also give some relations
which follow from the above due to the identification \eqref{tbartinJ}
\[ \omega(\tbar gh) = 
- {{u^2_{-1}}^i}_k {\RB^{-1}}_{ig}^{jh} \omega(\mt kj) .\]

The algebra $\Omega$ coincides in the Hecke case with the differential
graded Hopf algebra $M^{\wedge} = T_F^\bullet(M)/(\Ker(\sigma-1))$
constructed by Brzezi\'nski~\cite{Brz} after Woronowicz's 
ideas~\cite{Wor:calcul}. To prove this we have to show that the set of
relations~\eqref{R2R3} coincides with $\Ker(\sigma-1)$. By definition~
\eqref{brbim} the braiding $\sigma$ is
\[ \sigma(\omega(\mt ab) \tens \omega(\mt cf)) = \omega(\mt gh)
\tens \omega((\mt ab - {\delta^a}_b) \gamma(\mt cg) \mt hf ) .\]
The substitution 
$\gamma(\mt cg) = {{(u_1^2)^{-1}}_g}^i\, \tbar ij \,{{u_1^2}_j}^c$ together
with relations \eqref{ttinJ}--\eqref{ttbarinJ} reduces this expression 
to
\[ {\RB^{\flat-1}}_{bc}^{ij} {\RB^{-1}}_{lg}^{ia} \RB_{kp}^{lh} 
\RB_{jf}^{km} \omega(\mt gh) \tens \omega(\mt pm) .\]
This formula as well as the identity
\[ \sigma(\RB_{jc}^{ib} \omega(\mt ab) \tens \omega(\mt cf) ) =
\RB_{jf}^{km}\, \RB_{kp}^{lh} \omega(\mt gh) \tens \omega(\mt pm)\,
{\RB^{-1}}_{lg}^{ia} \]
were obtained by Sudbery~\cite{Sud:supcal}. In the basis
\[ X_{jf}^{ia} = \RB_{jc}^{ib} \omega(\mt ab)\tens \omega(\mt cf) \]
of $M^l\tens M^l$ the braiding is expressed as
\[ \sigma(X_{jf}^{ia}) = \RB_{jf}^{km} X_{km}^{lg} {\RB^{-1}}_{lg}^{ia} 
.\]

The relations \eqref{R2R3} form the subspace
\[ I = \text{span}\, \{ X_{bf}^{lk} + 
\RB_{ip}^{lk} X_{xy}^{ip} \RB_{bf}^{xy} \} \]
which is contained in
\[ \Ker(\sigma-1) = \{ \Tr(AX)\equiv A_{ia}^{jf}X_{jf}^{ia} \mid RA=AR 
\} \]
by Remark~\ref{remBrz}. All matrices $A$ commuting with $\RB$ have the 
form $P_+BP_+ + P_-CP_-$, where $\RB = qP_+ - q^{-1} P_-$ is the spectral
decomposition. Since $\Tr(A(X+RXR)) = (1+q^{\pm2}) \Tr(AX)$ if 
$A=P_\pm B P_\pm$, we conclude that $\Ker(\sigma-1) \subset I$, proving
the claim.

\bibliographystyle{plain}

\end{document}